\documentclass[twocolumn,aps,prl,superscriptaddress,amsfonts,longbibliography]{revtex4-1}
\usepackage[tight]{subfigure}
\usepackage{physics,tensor,siunitx}
\usepackage{amsmath,dsfont}
\usepackage{mathtools}
\usepackage{amssymb}
\usepackage{booktabs}
\usepackage{mathrsfs}
\usepackage{varwidth}
\usepackage{graphicx,color,caption}
\usepackage{bm}
\usepackage{cases}
\usepackage{array}
\usepackage{dcolumn}
\usepackage{chngcntr}
\usepackage{hepparticles}
\usepackage{heppennames}
\usepackage{hepnicenames}
\usepackage{epstopdf}
\usepackage{newfloat}
\usepackage{nicefrac}
\usepackage{comment}
\usepackage{svg}
\usepackage{etex}

\DeclareFloatingEnvironment[name={Extended Data Figure}]{suppfigure}
\DeclareUnicodeCharacter{2212}{-}
\usepackage{hyperref}
\hypersetup{
    colorlinks=true,
    linkcolor=blue,     
    citecolor=blue,    
    urlcolor={}         
}
\usepackage[capitalise]{cleveref}
\usepackage{tikz}
\usetikzlibrary{decorations.pathmorphing} 
\usetikzlibrary{fit}					
\usetikzlibrary{backgrounds}	
\usetikzlibrary{shapes.geometric} %
\usetikzlibrary{positioning}
\usetikzlibrary{arrows.meta}
\usetikzlibrary{quantikz}
\usepackage{yquant}

\begin{document}

\title{Quantum nuclear dynamics on a distributed set of ion-trap quantum computing systems}

\author{Anurag Dwivedi}
\affiliation{Department of Chemistry, Indiana University, Bloomington, Indiana 47405, USA}
\affiliation{Indiana University Quantum Science and Engineering Center, Bloomington, Indiana 47405, USA}
\author{A. J. Rasmusson}
\affiliation{Department of Physics, Indiana University, Bloomington, Indiana 47405, USA}
\affiliation{Indiana University Quantum Science and Engineering Center, Bloomington, Indiana 47405, USA}
\author{Philip Richerme}
\affiliation{Department of Physics, Indiana University, Bloomington, Indiana 47405, USA}
\affiliation{Indiana University Quantum Science and Engineering Center, Bloomington, Indiana 47405, USA}
\author{Srinivasan S. Iyengar}
\affiliation{Indiana University Quantum Science and Engineering Center, Bloomington, Indiana 47405, USA}
\affiliation{Department of Chemistry, Indiana University, Bloomington, Indiana 47405, USA}
\date{\today}

\begin{abstract}
Quantum nuclear dynamics with wavepacket time-evolution is classically intractable and viewed as a promising avenue for quantum information processing. 
Here, we use an IonQ 11-qubit trapped-ion quantum computer, Harmony, to study the quantum wavepacket dynamics of a shared-proton within a short-strong hydrogen-bonded system. We also provide the first application of distributed quantum computing for chemical dynamics problems, where the distributed set of quantum processes is constructed using a tensor network formalism. For a range of initial states, we experimentally drive the ion-trap system to emulate the quantum nuclear wavepacket as it evolves along the potential surface generated from electronic structure. Following the experimental creation of the nuclear wavepacket, we extract measurement observables such as its time-dependent spatial projection and its characteristic vibrational frequencies to good agreement with classical results. Vibrational eigenenergies obtained from quantum computational are in agreement with those obtained from classical simulations to within a fraction of a kcal/mol, thus suggesting chemical accuracy. Our approach opens a new paradigm for studying the quantum chemical dynamics and vibrational spectra of molecules and also provides the first demonstration for parallel quantum computation on a distributed set of ion-trap quantum computers.
\end{abstract}
\maketitle
Quantum nuclear dynamics has critical impact on a range of problems of interest in biological, materials, and atmospheric systems. For example, the rate limiting step in the oxidation of fatty acids by the enzyme, soybean lipoxygenase-1, is a hydrogen transfer step that is thought to be deeply impacted by quantum mechanical tunneling \cite{htrans,Klinman-ACR-2015,SHS-PCET-ACR-2009,Klinman-SLO-1}. Protonated and hydroxide-rich water clusters that are thought to catalyze atmospheric reactions and are also of fundamental interest in condensed phase chemistry \cite{protonjpcb,protonIJMS,HO2-qwaimd}, have spectral features influenced by the quantum mechanical nature of the hydrated proton \cite{Johnson-Jordan-21mer,Johnson-Jordan-Zundel-Science,admp-21mer,admp-21mer-2}. Several problems in materials chemistry including the study of  nitrogen fixation \cite{Hoffman-N2-H2-redelim, Thorneley-Lowe-PCET-N2, Schrock-N2-NCET} and artificial photosynthesis \cite{Meyer-2005_IC_44_6802} are strongly controlled by such hydrogen transfer processes \cite{htrans}. In all these cases, there is a critical interplay between the electronic and nuclear degrees of freedom that influences chemical properties including reactivity. 

Quantum chemical reaction dynamical study
\cite{Wyatt-Zhang, VSCF, McCoy, McCoy2, IyengarParker, MCTDH-Meyer1, Wyatt-Bohm-1,Schatz-Kupperman,Delos,feit1,feit2,feit3,dkos1,dkos2,kosloff2,Leforestier-Lanczos,dev,jl1,stuart-clary,stuart-Nature,youhong,Miller-rate-PI,Makri-PI-Review,talezer,bh1,TIW,lill1,jl2,SincDVR,Matsunaga,Gerber,DAFprop,Jung} 
 of such complex problems entails the ab initio investigation of the time evolution of nuclear wavepackets on carefully constructed multi-dimensional potential energy surfaces. The study of these problems leads naturally to detailed examination of chemical reaction dynamics 
and is considered to be an exponential scaling problem with nuclear dimension \cite{Wyatt-Zhang,Feynman-Comp,Nielsen-Chuang-QuantComp,Feynman1982,Berman-Comp-complexity}. 
Indeed, such studies are deeply confounded by the following primary challenges. (a) Accurate electron correlation methods have a computational scaling \cite{Berman-Comp-complexity} that is steeply algebraic \cite{schlegel1991computational}, where, for example, the well-known and extremely accurate CCSD(T) approach\cite{raghavachari1989fifth}, scales as $O(N^{6-7})$, for $N$  electrons in the system. (b) Quantum nuclear dynamics is, as noted above, thought to scale in an exponential manner with increasing number of nuclear degree of freedom \cite{Wyatt-Zhang,Feynman-Comp,Nielsen-Chuang-QuantComp,Feynman1982,Ohrn,kuppermann,talezer,monte}. 

This paper deals with the development of quantum algorithms for the study of quantum nuclear wavepacket dynamics and implementation of these on quantum hardware. Our work here follows on our previous studies of quantum nuclear wavepacket dynamics on ion-trap systems \cite{Sandia-1D-3Qubits,Debadrita-Mapping-1D-3Qubits} and provides one approach to extend these ideas to multi-dimensional quantum dynamics conducted here on an ion-trap quantum hardware device. We perform our ion-trap experiments on IonQ's 11-qubit trapped-ion quantum computer, Harmony. Our focus here is on obtaining the vibrational properties, beyond the harmonic approximation, for  specifically chosen nuclear degrees of freedom in a hydrogen bonded system, through the simulation of the time-evolution of a quantum nuclear wavepacket on an ion-trap quantum device. 
This paper tackles this challenge by providing the first quantum computed results for a multidimensional nuclear dynamics problem. The specific problem of choice is a hydrogen bonded system where, for simplicity, the hydrogen bond coordinate and a specific  gating mode which modulates the strength of the hydrogen bond are together treated as a single multi-dimensional quantum mechanical wavepacket. The evolution of this system is studied using an ion trap quantum computer. At the end, we find that the molecular vibrations captured from the time-evolution of the wavepacket are in good agreement 
with quantum wavepacket dynamics studies on a classical computer.

This paper builds on Ref. \cite{Sandia-1D-3Qubits}, where we chose a single hydrogen bonded system, with the hydrogen bonded direction as the quantum mechanical degree of freedom (that is, one dimensional quantum system) and this quantized dimension was directly treated on Sandia National lab's, QSCOUT ion trap quantum computer\cite{Sandia-1D-3Qubits}. Good agreement 
was found between the classically studied wavepacket dynamics problem and that studied on the ion trap quantum system. Here, not only do we provide a natural generalization of this application to more quantum dimensions, we also provide a circuit based implementation on a distributed set of ion-trap quantum computing systems, thus opening the doors for parallel quantum computation. Given that we are in the NISQ era of quantum computing\cite{Nai-Hui-Hybrid-QC,Preskill2018-NISQ}, with deep quantum circuits yielding large errors in computation, a distributed approach substantially reduces the circuit depth for these quantum operations and also provides a first  parallel implementation of a chemical dynamics problem on quantum computers.  

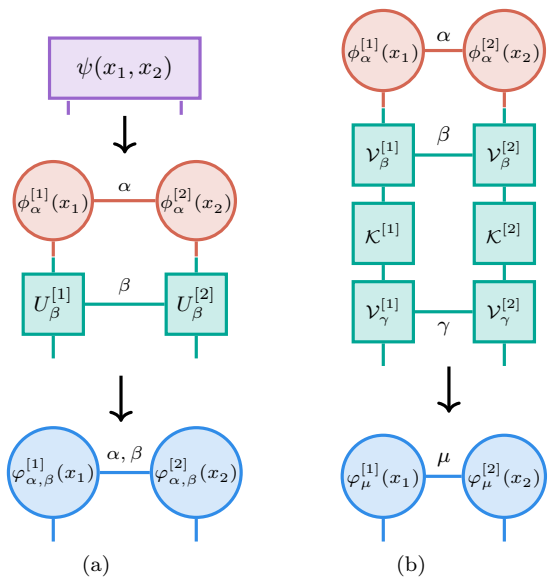
\begin{figure}[tbp]
    \subfigure[]{
\definecolor{bblue}{rgb}{0.19, 0.55, 0.91}
\definecolor{green}{rgb}{0.0, 0.65, 0.58}
\definecolor{purple}{rgb}{0.6, 0.4, 0.8}
\definecolor{orange}{rgb}{0.83, 0.4, 0.32}

\tikzstyle{psi} = [rectangle, very thick, minimum width=2cm, minimum height = 0.8cm, inner sep=0.1cm,draw=purple, fill=purple!20]

\tikzstyle{phi} = [circle, very thick, minimum size=0.8cm, inner sep=0.1mm, draw=orange, fill=orange!20]
    
\tikzstyle{phinew} = [circle, very thick, minimum size=0.7cm, inner sep=0.1mm, draw=bblue, fill=bblue!20]
    
\tikzstyle{U} = [rectangle, very thick, minimum size=0.8cm, inner sep=0.1cm,draw=green, fill=green!20]
    
\tikzset{arw/.style={-Stealth, thick}}
    \begin{tikzpicture}[>=latex, node distance=1cm,
        every edge quote/.style={font=\fontsize{7}{1}},scale=0.5]    
        \node (psi1) [psi, font=\fontsize{9}{1}]{$\psi(x_1,x_2)$};
      
        \node (psix01) [below=0mm of psi1, xshift = -7.5mm, minimum size=0mm, inner sep=0mm]{};
        \node (psix02) [below=0mm of psi1, xshift = +7.5mm, minimum size=0mm, inner sep=0mm]{};

        \node (psix00) [below=2mm of psi1, minimum size=0mm, inner sep=0mm]{};
        \node (ppx01) [below=5mm of psix00, minimum size=0mm, inner sep=0mm]{};
        
        \node (psix1) [below=2mm of psi1, xshift = -7.5mm, minimum size=0mm, inner sep=0mm]{};
        \node (psix2) [below=2mm of psi1, xshift = +7.5mm, minimum size=0mm, inner sep=0mm]{};

        \node (pphi1) [phi,below=8mm of psi1,xshift = -9.5mm, font=\fontsize{7}{1}]{$\phi^{[1]}_\alpha(x_1)$};
        \node (pphi2) [phi, below=8mm of psi1,xshift = +9.5mm, font=\fontsize{7}{1}]{$\phi^{[2]}_\alpha(x_2)$};
               
        \node (ppx1) [below=2mm of pphi1, minimum size=0mm, inner sep=0mm]{};
        \node (ppx2) [below=2mm of pphi2, minimum size=0mm, inner sep=0mm]{};
        
        \node (u1) [U, below=2mm of ppx1, font=\fontsize{8}{1}]{$U^{[1]}_\beta$};
        \node (u2) [U, below=2mm of ppx2, font=\fontsize{8}{1}]{$U^{[2]}_\beta$};
              
        \node (ux1) [below=3mm of u1, minimum size=0mm, inner sep=0mm]{};
        \node (ux2) [below=3mm of u2, minimum size=0mm, inner sep=0mm]{};
        
        \node (eql0sgn) [below=1mm of ux1, xshift=9mm]{};
        \node (eqlsgn) [below=5mm of eql0sgn]{};  
        
        \node (phit1) [phinew, below=12mm of u1, font=\fontsize{7}{1}]{$\varphi^{[1]}_{\alpha,\beta}(x_1)$};
        \node (phit2) [phinew, below=12mm of u2, font=\fontsize{7}{1}]{$\varphi^{[2]}_{\alpha,\beta}(x_2)$};
             
        \node (phitx1) [below=3mm of phit1, minimum size=0mm, inner sep=0mm]{};
        \node (phitx2) [below=3mm of phit2, minimum size=0mm, inner sep=0mm]{};

        \draw[-To, very thick, black, to path={-| (\tikztotarget)}] (psix00) -- (ppx01);
        \draw[-To, very thick, black, to path={-| (\tikztotarget)}] (eql0sgn) edge (eqlsgn);
        
        \path[-]
            
           
            
            (psix01) edge[very thick, purple] node[left,black, yshift=-1mm, font=\fontsize{7}{1}] {} (psix1)
            (psix02) edge[very thick, purple] 
            node[left,black, yshift=-1mm,font=\fontsize{10}{1}] {} (psix2)

            (pphi1) edge[very thick, orange] node[above,black,font=\fontsize{7}{1}] {$\alpha$} (pphi2)

            (pphi1) edge[very thick, orange] node[left,black,yshift=-1mm,font=\fontsize{8}{1}] {} (ppx1)
            (pphi2) edge[very thick, orange] node[left,black,yshift=-1mm,font=\fontsize{8}{1}] {} (ppx2)

            (u1) edge[very thick, green] node[above,black,font=\fontsize{7}{1}] {$\beta$}  node[below,black,font=\fontsize{7}{1}] {} (u2)
            
            (u1) edge[very thick, green] (ppx1)
            (u1) edge[very thick, green] node[left,black,yshift=-0.5mm,font=\fontsize{8}{1}] {}(ux1)
            (u2) edge[very thick, green] (ppx2)
            (u2) edge[very thick, green] node[left,black,yshift=-0.5mm,font=\fontsize{8}{1}] {}(ux2)
            
            (phit1) edge[very thick, bblue] node[above,black,font=\fontsize{7}{1}] {${\alpha,\beta}$} node[below,black] {} 
            (phit2)
            
            (phit1) edge[very thick, bblue] node[left,black,yshift=-0.5mm,font=\fontsize{8}{1}] {}(phitx1)
            (phit2) edge[very thick, bblue] node[left,black,yshift=-0.5mm,font=\fontsize{8}{1}] {}(phitx2)
            ;

    \end{tikzpicture}
        \label{Fig:MPS_tEvo_trotter-a}}
    \subfigure[]{
\definecolor{bblue}{rgb}{0.19, 0.55, 0.91}
\definecolor{green}{rgb}{0.0, 0.65, 0.58}
\definecolor{purple}{rgb}{0.6, 0.4, 0.8}
\definecolor{orange}{rgb}{0.83, 0.4, 0.32}

    \tikzstyle{psi} = [rectangle, very thick, minimum width=2cm, minimum height = 0.8cm, inner sep=0.1cm,draw=purple, fill=purple!20]
    
    \tikzstyle{phi} = [circle, very thick, minimum size=0.8cm, inner sep=0.2mm, draw=orange, fill=orange!20]
    
    \tikzstyle{phinew} = [circle, very thick, minimum size=0.8cm, inner sep=0.2mm, draw=bblue, fill=bblue!20]
    
    \tikzstyle{U} = [rectangle, very thick, minimum size=0.8cm, inner sep=0.1cm,draw=green, fill=green!20]
    
    \tikzset{arw/.style={-Stealth, thick}}

    \begin{tikzpicture}[>=latex,node distance=3cm,
        every edge quote/.style={font=\fontsize{7}{1}},
        ]
      

        
        
		\node (pphi1) [phi, xshift = -8mm, font=\fontsize{7}{1}]{$\phi^{[1]}_\alpha(x_1)$};
        \node (pphi2) [phi, right=5mm of pphi1, font=\fontsize{7}{1}]{$\phi^{[2]}_\alpha(x_2)$};

        \node (ppx01) [below=5mm of psix00, minimum size=0mm, inner sep=0mm]{}; 
        
        \node (ppx1) [below=2mm of pphi1, minimum size=0mm, inner sep=0mm]{};
        \node (ppx2) [below=2mm of pphi2, minimum size=0mm, inner sep=0mm]{};
        
        \node (V1) [U, below=2mm of ppx1, font=\fontsize{7}{1}]{$\mathcal{V}^{[1]}_\beta$};
        \node (V2) [U, below=2mm of ppx2, font=\fontsize{7}{1}]{$\mathcal{V}^{[2]}_\beta$};
                
        \node (K1) [U, below=2mm of V1, font=\fontsize{7}{1}]{$\mathcal{K}^{[1]}$};
        \node (K2) [U, below=2mm of V2, font=\fontsize{7}{1}]{$\mathcal{K}^{[2]}$};

        \node (u1) [U, below=2mm of K1, font=\fontsize{7}{1}]{$\mathcal{V}^{[1]}_\gamma$};
        \node (u2) [U, below=2mm of K2, font=\fontsize{7}{1}]{$\mathcal{V}^{[2]}_\gamma$};
             
        \node (ux1) [below=3mm of u1, minimum size=0mm, inner sep=0mm]{};
        \node (ux2) [below=3mm of u2, minimum size=0mm, inner sep=0mm]{};
        
        \node (eql0sgn) [below=10mm of ux1, right=7.5mm of ux1]{};
        \node (eqlsgn) [below=5mm of eql0sgn]{};  
        
        \node (phit1) [phinew, below=12mm of u1, font=\fontsize{7}{1}]{$\varphi^{[1]}_{\mu}(x_1)$};
        \node (phit2) [phinew, below=12mm of u2, font=\fontsize{7}{1}]{$\varphi^{[2]}_{\mu}(x_2)$};
        
        \node (phitx1) [below=3mm of phit1, minimum size=0mm, inner sep=0mm]{};
        \node (phitx2) [below=3mm of phit2, minimum size=0mm, inner sep=0mm]{};

        \draw[-To, very thick, black, to path={-| (\tikztotarget)}] (eql0sgn) edge (eqlsgn);
        
        \path[-]
            
           

            
            (pphi1) edge[very thick, orange] node[above,black,font=\fontsize{8}{1}] {$\alpha$} (pphi2)
           
            (pphi1) edge[very thick, orange] node[left,black,yshift=-1mm,font=\fontsize{8}{1}] {} (ppx1)
            (pphi2) edge[very thick, orange] node[left,black,yshift=-1mm,font=\fontsize{8}{1}] {} (ppx2)
            
            (V1) edge[very thick, green] node[above,black,font=\fontsize{8}{1}] {$\beta$} (V2)
            
            (u1) edge[very thick, green] node[below,black,font=\fontsize{8}{1}] {$\gamma$} (u2)

            (V1) edge[very thick, green] (ppx1)
            (K1) edge[very thick, green] node[left,black,yshift=-0.5mm,font=\fontsize{8}{1}] {}(u1)
            (V2) edge[very thick, green] (ppx2)
            (K2) edge[very thick, green] node[left,black,yshift=-0.5mm,font=\fontsize{8}{1}] {}(u2)
            
            (V1) edge[very thick, green] (K1)
            (V2) edge[very thick, green] (K2)

            (u1) edge[very thick, green] (ux1)
            (u2) edge[very thick, green] (ux2)

            (phit1) edge[very thick , bblue] node[above,black,font=\fontsize{8}{1}] {${\mu}$} node[below,black] {} 
            (phit2)
           
            (phit1) edge[very thick, bblue] node[left,black,yshift=-0.5mm,font=\fontsize{8}{1}] {}(phitx1)
            (phit2) edge[very thick, bblue] node[left,black,yshift=-0.5mm,font=\fontsize{8}{1}] {}(phitx2)
            ;

    \end{tikzpicture}\label{Fig:MPS_tEvo_trotter-c}}
    \caption{Figure \ref{Fig:MPS_tEvo_trotter-a} depicts  the time evolution of the tensor network representation of a wavepacket. 
    Figure \ref{Fig:MPS_tEvo_trotter-c} enumerates this action when the time-evolution is written in Trotterized form. Here, the initial vector is given by entanglement variables $\alpha$ while the Trotterized propagator has two entangled variables, $\beta$ and $\gamma$, which combine to determine the index $\mu \equiv \left( \alpha, \beta, \gamma\right)$ for the propagated system. }
    \label{Fig:MPS_tEvo_trotter}
\end{figure}
Our nuclear 
Hamiltonian 
comprises the nuclear kinetic energy operator, 
and the multidimensional potential energy surface 
obtained from electronic structure calculations (see SI). 
%
The initial quantum nuclear wavepacket state is chosen as a matrix product state (MPS), where the quantum state is expressed as a tensor product of lower-rank tensors each of which pertain to the individual physical dimensions of the nuclear vibrational problem. Figure \ref{Fig:MPS_tEvo_trotter} complements our discussion and more details can be found in SI. The matrix product state for the initial wavepacket is represented as orange circles at the top of \cref{Fig:MPS_tEvo_trotter-a} with entanglement, or bond, dimension $\alpha$. This is a critical step here, as we will see in arriving at a distributed quantum computing protocol. Concurrently, the time-evolution operator, is concisely represented in \cref{Fig:MPS_tEvo_trotter-a} (green boxes) with entanglement, or bond, dimension $\beta$. Given that the time-evolution operator here arrives from electronic structure, this is further written out using the kinetic and electronic structure-based potential operators, as a Trotter factorization \cite{Nelson-Trotter} in \cref{Fig:MPS_tEvo_trotter-c}. Additionally, in \cref{Fig:MPS_tEvo_trotter-c}, the potential propagator is also represented as a matrix product state, yielding for example, $\left\{ \mathcal{V}^{[1]}\right\};\left\{\mathcal{V}^{[2]}\right\}$, with bond dimensions $\beta$ and $\gamma$. 
The bond dimensions signify the extent of entanglement, or the number of such one-dimensional propagators and serves directly in creating quantum replicas to be processed on a distributed architecture. The final number of bond dimensions is explicitly written out in \cref{Fig:MPS_tEvo_trotter-a}, at the bottom, and is concisely represented as $\mu$, in Figure \ref{Fig:MPS_tEvo_trotter-c}. 

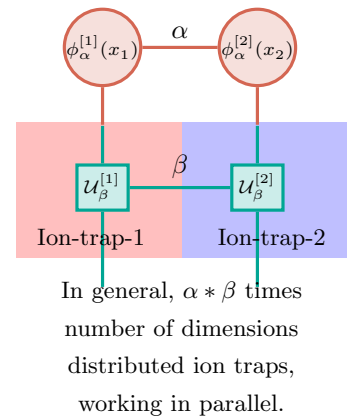
\begin{figure}[tbp]
    \centering
{
\definecolor{bblue}{rgb}{0.19, 0.55, 0.91}
\definecolor{green}{rgb}{0.0, 0.65, 0.58}
\definecolor{purple}{rgb}{0.6, 0.4, 0.8}
\definecolor{orange}{rgb}{0.83, 0.4, 0.32}

    \tikzstyle{psi} = [rectangle, very thick, minimum width=2cm, minimum height = 0.6cm, inner sep=0.0cm,draw=purple, fill=purple!20]
    
    \tikzstyle{phi} = [circle, very thick, minimum size=0.6cm, inner sep=0.0mm, draw=orange, fill=orange!20]
    
    \tikzstyle{phinew} = [circle, very thick, minimum size=0.6cm, inner sep=0.0mm, draw=bblue, fill=bblue!20]
    
    \tikzstyle{phiempty} = [circle, very thick, minimum size=0.cm, inner sep=0.0mm, draw=bblue, fill=bblue!20]

    \tikzstyle{U} = [rectangle, very thick, minimum size=0.6cm, inner sep=0.1cm,draw=green, fill=green!20]
    
    \tikzset{arw/.style={-Stealth, thick}}

    \begin{tikzpicture}[>=latex, node distance=0.7cm,
        every edge quote/.style={font=\fontsize{7}{1}}]
      

        
        
        \fill[fill=blue!100!white!25](0,-2.8)--(0,-1)--(2.2,-1)--(2.2,-2.8)--(0,-2.8);
        \fill[fill=red!100!white!25](0,-2.8)--(0,-1)--(-2.2,-1)--(-2.2,-2.8)--(0,-2.8);

        \node (pphiempty) [phiempty, font=\fontsize{7}{1}]{};

        \node (pphi1) [phi, left=5mm of pphiempty, font=\fontsize{7}{1}]{$\phi^{[1]}_\alpha(x_1)$};
        \node (pphi2) [phi, right=10mm of pphi1, font=\fontsize{7}{1}]{$\phi^{[2]}_\alpha(x_2)$};

        \node (ppx01) [below=5mm of psix00, minimum size=0mm, inner sep=0mm]{}; 
        
        \node (ppx1) [below=5mm of pphi1, minimum size=0mm, inner sep=0mm]{};
        \node (ppx2) [below=5mm of pphi2, minimum size=0mm, inner sep=0mm]{};
        
        \node (V1) [U, below=5mm of ppx1, font=\fontsize{7}{1}]{$\mathcal{U}^{[1]}_\beta$};
        \node (V2) [U, below=5mm of ppx2, font=\fontsize{7}{1}]{$\mathcal{U}^{[2]}_\beta$};         

             
        \node (ux1) [below=10mm of V1, minimum size=0mm, inner sep=0mm]{};
        \node (ux2) [below=10mm of V2, minimum size=0mm, inner sep=0mm]{};

        
        \path[-]
            
           

            
            (pphi1) edge[very thick, orange] node[above,black,font=\fontsize{10}{1}] {$\alpha$} (pphi2)
           
            (pphi1) edge[very thick, orange] node[left,black,yshift=-1mm,font=\fontsize{7}{1}] {} (ppx1)
            (pphi2) edge[very thick, orange] node[left,black,yshift=-1mm,font=\fontsize{7}{1}] {} (ppx2)
            
            (V1) edge[very thick, green] node[above,black,font=\fontsize{10}{1}] {$\beta$} (V2)
            

            (V1) edge[very thick, green] (ppx1)
            (V1) edge[very thick, green] node[left,black,yshift=-0.5mm,font=\fontsize{7}{1}] {}(ux1)
            (V2) edge[very thick, green] (ppx2)
            (V2) edge[very thick, green] node[left,black,yshift=-0.5mm,font=\fontsize{7}{1}] {}(ux2)
            
           
            ;
        \filldraw[black] (1.2,-2.3) circle (0pt)node[below]{Ion-trap-2};
        \filldraw[black] (-1.2,-2.3) circle (0pt)node[below]{Ion-trap-1};
        \filldraw[black] (0.0,-3) circle (0pt)node[below]{In general, $\alpha*\beta$ times};
        \filldraw[black] (0.0,-3.5) circle (0pt)node[below]{number of dimensions};
        \filldraw[black] (0.0,-4) circle (0pt)node[below]{distributed ion traps,};        
        \filldraw[black] (0.0,-4.5) circle (0pt)node[below]{working in parallel.};        
           
    \end{tikzpicture}

}
    \caption{\label{Fig:MPS_tEvo_trotter-b} Derived from \cref{Fig:MPS_tEvo_trotter-a} and depicts the simultaneous study of quantum dynamics for individual one-dimensional wavefunctions on distinct ion-trap quantum computers. }
\end{figure}
\begin{figure*}[tbp]
    \centering
    \input{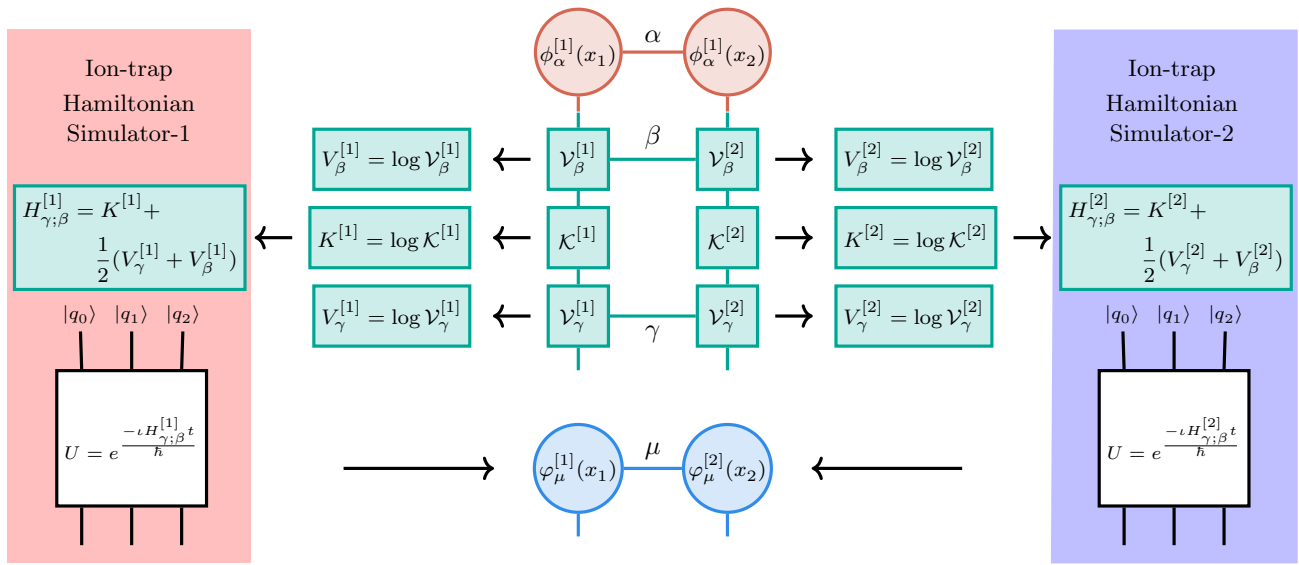}
    \caption{A detailed exposition of \cref{Fig:MPS_tEvo_trotter-a} that precisely outlines how the distributed simulations are carried out. The individual one-dimensional potentials are derived from each one-dimensional propagator as shown, yielding a family of effective one-dimensional Hamiltonians for the system. These individual one-dimensional Hamiltonians are then be independently and concurrently simulated on separate ion-trap quantum computers.}
    \label{Fig:MPS_tEvo_trotter-d}
\end{figure*}
Classically, we would follow a procedure similar to that illustrated in \cref{Fig:MPS_tEvo_trotter-c}, which involves the parallel application of one-dimensional potential propagators on each matrix product state (MPS) dimension. That is, the equivalent of $U^{[i]}_\beta\phi^{[i]}_\alpha(x_i) \rightarrow \varphi^{[i]}_{\alpha,\beta}(x_i)$. In this paper, we examine a  general quantum approach for such higher dimension problems, and this is described in \cref{Fig:MPS_tEvo_trotter-b} with more details in \cref{Fig:MPS_tEvo_trotter-d}. As can be seen from \cref{Fig:MPS_tEvo_trotter-b}, we may create $\alpha*\beta$ copies of unitary evolutions per dimension, each conducted on a different quantum hardware device. How this works out in practice is shown in \cref{Fig:MPS_tEvo_trotter-d}.
From each one-dimensional potential propagator ($\mathcal{V}^{[j]}_\beta$), we recover a one-dimensional potential and construct Hamiltonians represented as $\left\{ H^{[1]}_{\beta,\gamma}\right\};\left\{H^{[2]}_{\beta,\gamma} \right\}$ in \cref{Fig:MPS_tEvo_trotter-d}, that now represent an array of separate one-dimensional quantum systems. These individual one-dimensional Hamiltonians are directly mapped onto separate ion-trap systems that individually propagate the different components of the entangled quantum system. 

Thus, in higher dimensions, each physical dimension would be independently propagated on distinct ion-trap quantum computers. The victory here comes from the following fact. It is well known that, for correlated systems, as dimensionality increases, the quantum circuit depth of a general unitary goes exponential. For example, for a n-qubit unitary, one has a single $2^n$ by $2^n$ unitary operator, a quantum circuit may, in general, require up to $4^n$ parameters. This exponential scaling poses a prohibitive challenge for computation, even in the quantum domain, let alone for the classical case. The current formalism may provide a family of separate quantum  processes, studied independently and the complexity of such processes grows as $2^{n_1}\times n_{entangle}$, where ${n_1}^{\cal{D}} = n$ for nuclear dimensionality $\cal{D}$. In some sense, the formalism here trades quantum circuit depth for quantum circuit breadth. 
Consequently, while the exponential scaling with nuclear dimensionality may be mitigated, there exists a linear scaling with the extent of entanglement, $n_{entangle}$, that may be present in the system. As chemistry tends to be local, the extent of entanglement is expected to scale as area of the Hilbert space as per the area law of entanglement entropy, rather than the volume of the Hilbert space\cite{Hastings_2007-arealaw,entangle_entop,Anshu2022-bt,tensor_network}.  This results in a reasonable starting point for constructing an approximation for factored quantum dynamics on quantum computing platforms for higher dimensional problems.

\begin{figure}
  \includegraphics[width=0.80\columnwidth]
  {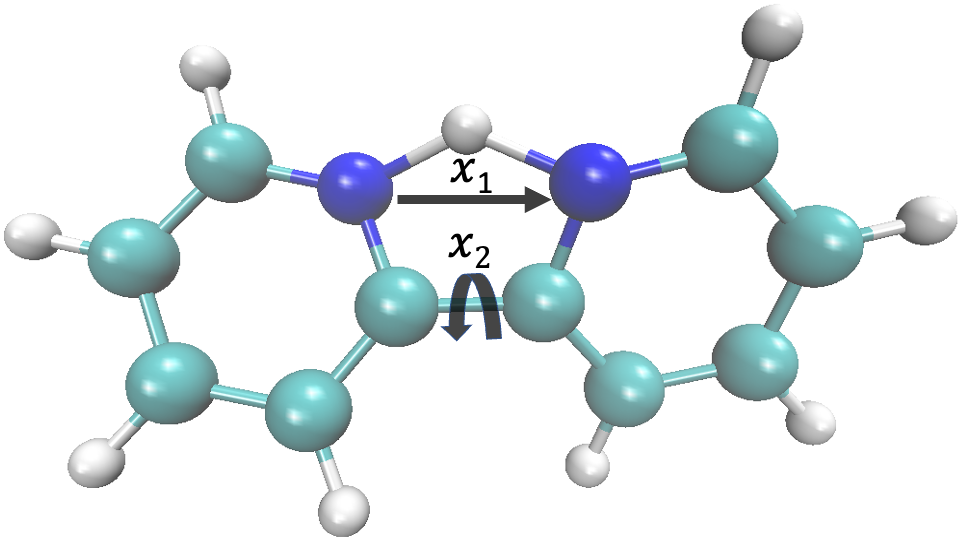}
  \caption{The protonated 2,2'-bipyridine system with the hydrogen bond dimension, $x_1$ and gating dimension, $x_2$, noted. 
  }
  	\label{Fig:Molecule}
\end{figure}
Quantum computations are performed using IonQ's 11-qubit trapped-ion quantum computer, Harmony, and accessed through the cloud via Amazon Braket \cite{wright2019benchmarking, braket}. Each ion encodes an effective quantum bit with long coherence time and near-perfect state preparation and measurement \cite{wright2019benchmarking}. Harmony's laser beam systems can separately address each ion, or pairs of ions, enabling arbitrary single-qubit gate rotations and all-to-all two-qubit gate connectivity. The wavepacket dynamics are simulated and measured on Harmony with 1000 repetitions for each time point. Quantum circuits are executed in parallel such that the maximum number of qubits are in use during a computation.

The system under investigation in this study is a bidentate-chelating agent known as 2,2'-bipyridine. See \cref{Fig:Molecule}. The aromatic nitrogen-containing heterocycles within this molecule holds significance in a variety of applications, particularly in energy storage and catalytic oxidation \cite{Moissette2001-zm,Vitale1999-rr,Graetzel1981-zl}. Notably, the protonated form of 2,2'-bipyridine has garnered widespread attention due to its potential applications as both electron carriers and electron acceptors, while the deprotonated form serves as a potent chelating agent \cite{Howard1996-jw,Steel1990-uj}.

\begin{figure*}[t]
 	\centering
    \subfigure[Quantum (blue) and classical (yellow) time-traces for $\ket{x_1}$]{\includegraphics[width=0.35\textwidth]{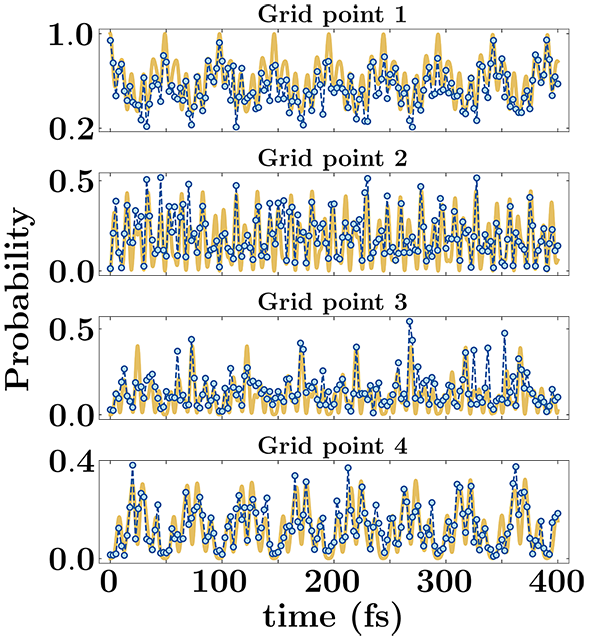}\label{Fig:prob-x_upper_00_2.5_400}}
    \subfigure[Quantum (blue) and classical (yellow) time-traces of $\ket{x_1}$]{\includegraphics[width=0.35\textwidth]{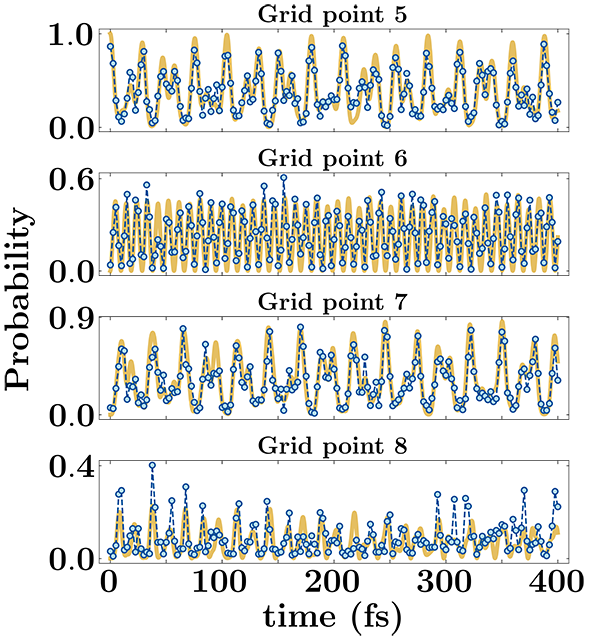}\label{Fig:prob-x_lower_00_2.5_400}}\\
    \subfigure[Quantum (blue) and classical (yellow) time-traces of $\ket{x_2}$]{\includegraphics[width=0.35\textwidth]{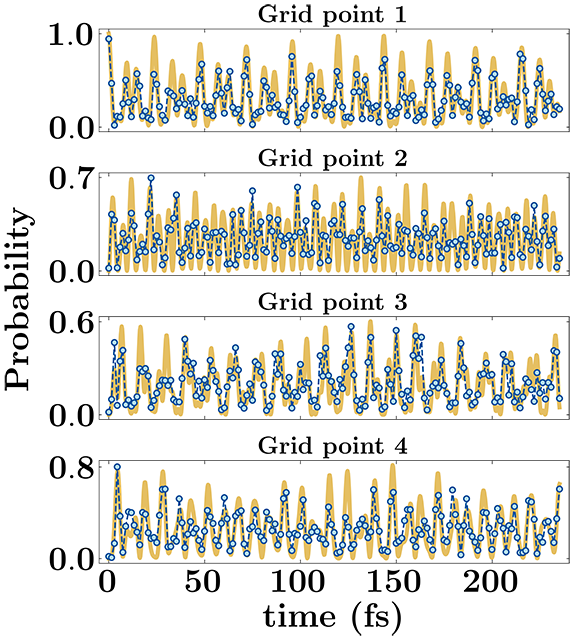}\label{Fig:prob-theta_upper_00_2.5_400}}
    \subfigure[Quantum (blue) and classical (yellow) time-traces of $\ket{x_2}$]{\includegraphics[width=0.35\textwidth]{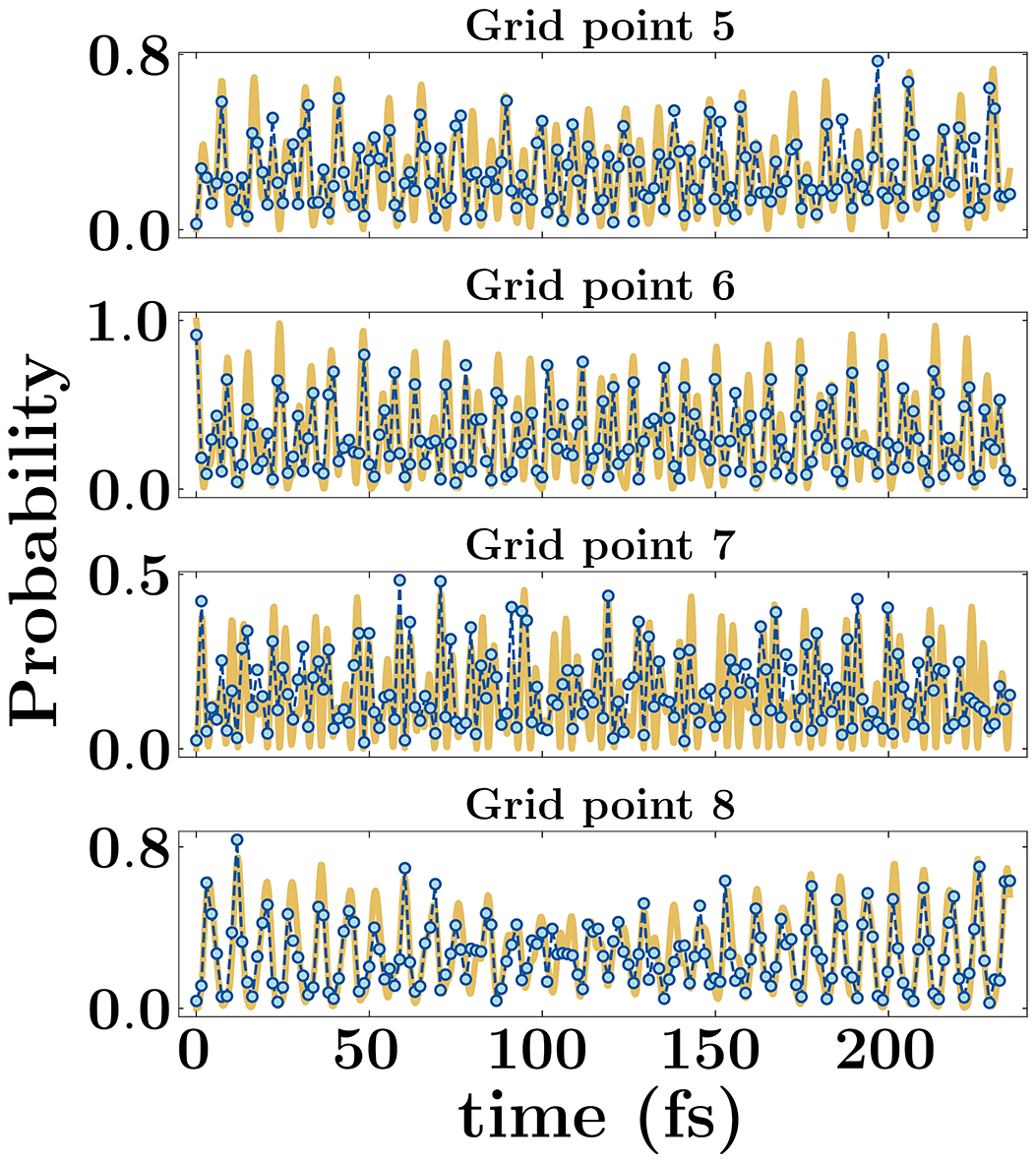}\label{Fig:prob-theta_lower_01_2.5_400}}
    
  	\caption{Illustrating the coherent dynamics of the wavepacket. 
    The time-evolution along specific grid points for the vibrational degree of freedom along the internuclear axis ($x_1$) connecting the two nitrogen atoms in \cref{Fig:Molecule} are shown in Figures (a) and (b). 
   Figures (c) and (d) depict the quantum dynamics corresponding to the torsional degree of freedom ($x_2$) resulting from planar rotations of the pyridyl rings about the C-C bond. The projection of the wavepacket onto each grid point is shown separately. The left and right columns pertain to the two separate blocks of the Hamiltonian. (See SI.) 
   }
\end{figure*}

The system depicted in \cref{Fig:Molecule} is characterized by a shared proton stabilized through a N-H-N hydrogen bond. Furthermore, the ligand exhibits a structural arrangement featuring two planar pyridine rings interconnected by a carbon-carbon single bond, thus introducing a degree of torsional freedom. This torsional rotation of the C-C bond serves as a gating mode, regulating the spatial separation between the donor and acceptor atoms defining the hydrogen bond. Consequently, the torsional rotation of the C-C bond significantly influences the basicity or proton affinity of protonated 2,2'-bipyridine \cite{Howard1996-jw}. In this study, we focus on examining both the torsional angle ($x_2$) of the C-C bond and the corresponding relative proton position along the donor-acceptor axis, denoted as $x_1$, as illustrated in Fig. \ref{Fig:Molecule}.

In our scheme, the potential energy surface is computed through electronic structure methods, details of which can be found in SI. The time evolution of this problem is studied classically and to show that our quantum evaluation process works quite well, in \cref{Fig:prob-x_upper_00_2.5_400,Fig:prob-x_lower_00_2.5_400,Fig:prob-theta_upper_00_2.5_400,Fig:prob-theta_lower_01_2.5_400}, we display the population of the time-evolved wavepackets on each grid basis state as for the classically treated 
and quantum mechanically treated 
problems. Due to the block diagonal nature of the Hamiltonian discussed in Ref. \cite{Debadrita-Mapping-1D-3Qubits}(see SI), each block of the Hamiltonian is simulated separately on two separate two-qubit ion-trap systems. The \cref{Fig:prob-x_upper_00_2.5_400,Fig:prob-x_lower_00_2.5_400} shows the time-traces of the upper and lower block diagonal Hamiltonian corresponding to the vibrational degree of freedom of the molecule ($x_1$) along the internuclear-axis connecting two nitrogen nuclei, while \cref{Fig:prob-theta_upper_00_2.5_400,Fig:prob-theta_lower_01_2.5_400} depicts the time-traces of the upper and lower block diagonal Hamiltonian corresponding to the torsional degree of freedom ($x_2$) resulting from planar rotations of the pyridyl rings about the C-C bond respectively. 
The agreement between the quantum and classical calculations appears to be good, but is gauged in further detail by probing the Fourier transforms of these time traces and comparing those with the respective eigenstates of the nuclear Hamiltonian. 
\begin{figure*}
  \includegraphics[width=\textwidth]{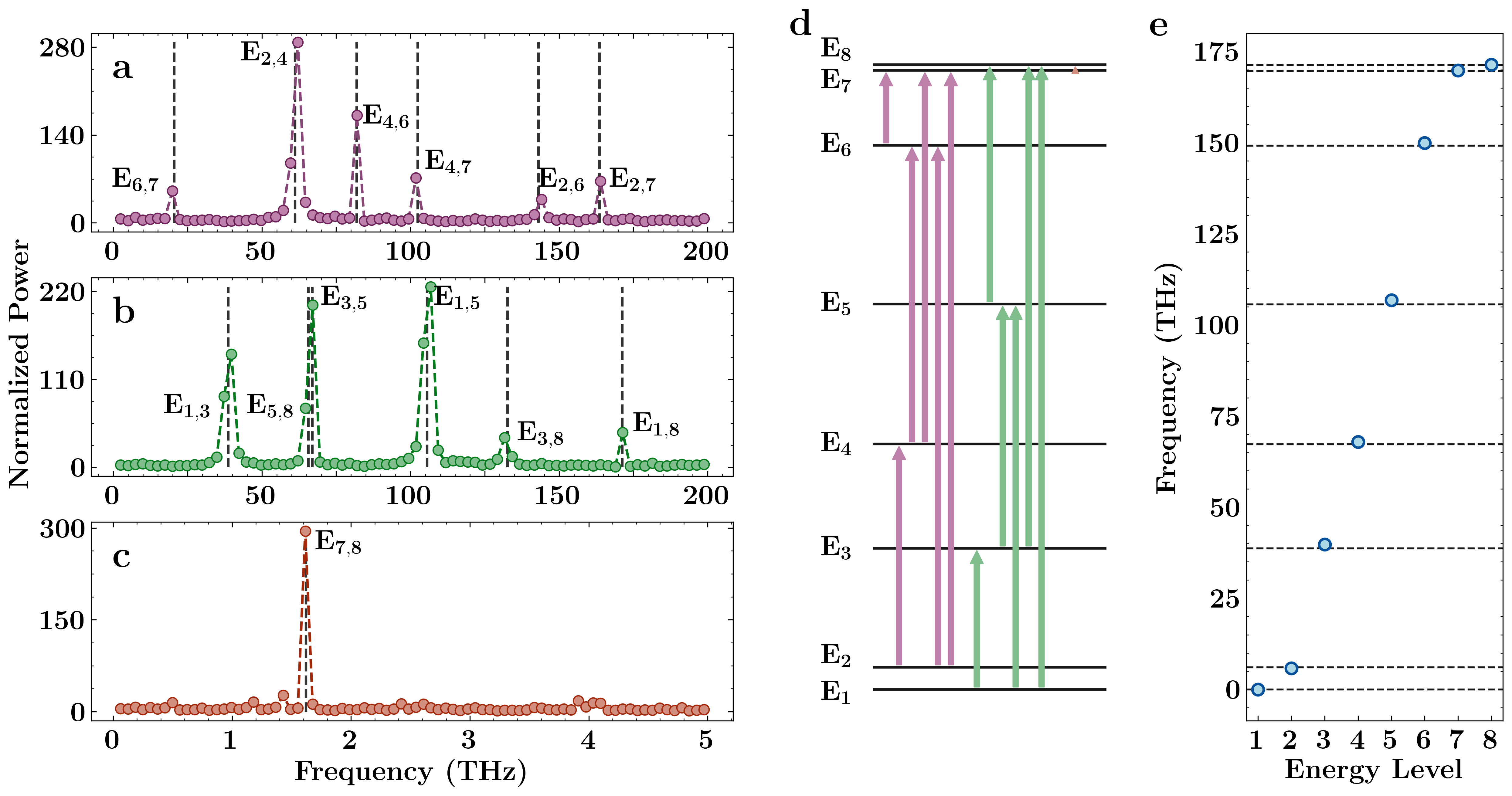}
  \caption{The figure presents the experimentally determined frequency and energy spectra of protonated 2,2'-bipyridine corresponding to the vibrational degree of freedom of transferring proton along the internuclear axis ($x_1$) joining the two nitrogen atoms. Due to the block diagonal nature of the Hamiltonian (see Supplementary Information), each block of the Hamiltonian is simulated separately on two separate two-qubit ion-trap systems. The Fourier transform of the time evolution of the full-Hamiltonian(see SI for theoretical details) allows for the determination of relative energy separations between the eigenstates corresponding to the respective blocks. The panel `a' displays the cumulative Fourier transform obtained from the quantum propagation of the upper block with different initial wavepacket (more details in SI) , while panel `b' illustrates the cumulative Fourier transform derived from the quantum propagation of the lower block of the Hamiltonian along the $x_1$-direction initialized with different wavepackets. The panel `c' shows the cumulative Fourier transform derived from the quantum dynamics of the full-Hamiltonian with two distinct initial wavepackets, providing the relative separation between the two sets of eigenstates corresponding to the lower and upper blocks. The \cref{Fig:ft_x}d presents the extracted frequencies from panels `a' to `c', color-coded according to their parent spectrum. This presentation allows for the experimental determination of the relative energies of all eigenstates. Additionally, \cref{Fig:ft_x}e compares the quantum-computed energy eigenstates of the nuclear Hamiltonian (depicted as blue dots) with the results of exact diagonalization (shown as dashed gray lines). }
  \label{Fig:ft_x}
\end{figure*}

\begin{figure*}
  \includegraphics[width=\textwidth]{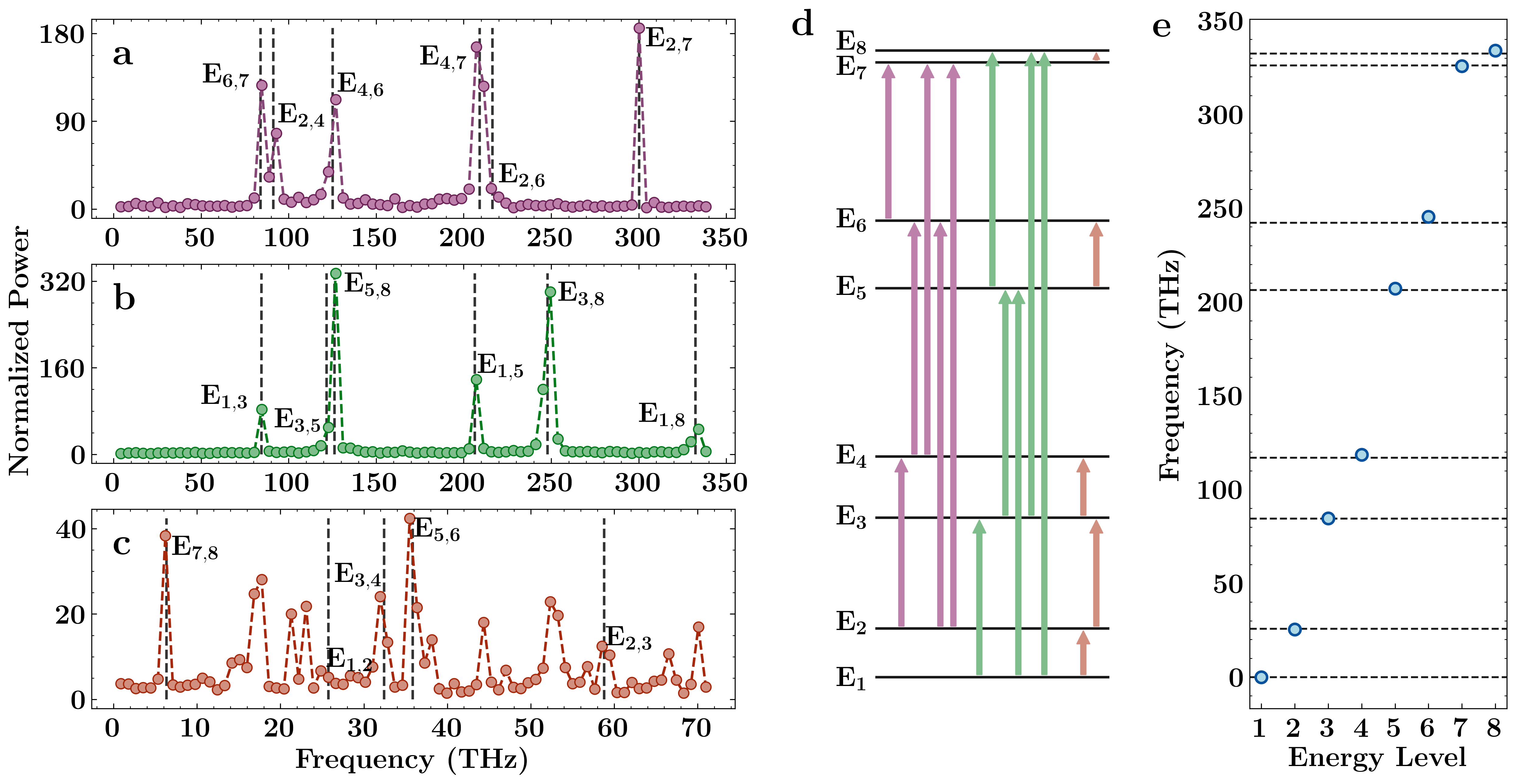}
 \caption{The figure presents the experimentally derived frequency and energy spectra of protonated 2,2'-bipyridine, focusing on the torsional degree of freedom (represented as $x_2$) associated with planar rotations of the pyridyl rings about the C-C bond. Leveraging the block diagonal structure of the Hamiltonian, each block is individually simulated using separate two-qubit ion-trap systems. The panel `a' illustrates the cumulative Fourier transform resulting from quantum propagation of the upper block with 4 distict initial wavepackets (further details in SI), while panel `b' depicts the cumulative Fourier transform obtained from the quantum propagation of the lower block along the $x_2$-direction starting with different initial wavepackets. Panel `c' exhibits the Fourier transform derived from the quantum dynamics of the full-Hamiltonian, elucidating the relative separation between the eigenstates of the lower and upper blocks. Furthermore, in \cref{Fig:ft_theta}d, frequencies extracted from panels `a' to `c' are color-coded by their respective spectra, facilitating the experimental determination of relative energies for all eigenstates. Additionally, \cref{Fig:ft_theta}e presents a comparison between the quantum-computed energy eigenstates of the nuclear Hamiltonian (represented by blue dots) and exact diagonalization results (depicted by dashed gray lines). }
   \label{Fig:ft_theta}
\end{figure*}

In quantum mechanics, time-correlations and their Fourier counterparts are generally used to probe expectation values, But as we discussed in Ref. \cite{Sandia-1D-3Qubits}, and also in SI here, on quantum platforms, the individual terms within a correlation function become accessible to measurement. This provides us with another approach to compute these vibrational properties, which is done here, as was done in Ref. \cite{Sandia-1D-3Qubits}. Specifically, vibrational properties are normally obtained from the Fourier transform of the density-density auto-correlation function. As can be seen in SI, inherently there is a trace in all such observations as required by the expectation  value, that the fundamentals of quantum mechanics requires us to perform. However, on a quantum computer, it is possible to measure the value of density at each basis point directly. Thus, in this publication, as in Ref. \cite{Sandia-1D-3Qubits}, we have utilized this alternate formalism of computing vibrational properties that is not directly available through the fourier transform of density-density auto-correlation function. 

Along the same lines, in \cref{Fig:ft_x,Fig:ft_theta}, we present the experimentally derived frequency and energy spectra of protonated 2,2'-bipyridine, showcasing two distinct vibrational degrees of freedom. The first corresponds to the transfer of a proton along the internuclear axis ($x_1$) joining the two nitrogen atoms, while the second pertains to the torsional motion ($x_2$) associated with planar rotations of the pyridyl rings about the C-C bond. As mentioned earlier, leveraging the block-diagonal nature of the Hamiltonian (see Supplementary Information), each block is simulated separately on two distinct two-qubit ion-trap systems. The Fourier transformation corresponding to the upper and lower blocks of the Hamiltonian is depicted in panels `a' and `b' of \cref{Fig:ft_x,Fig:ft_theta}, respectively, corresponding to their respective degrees of freedom. Additionally, we compute the Fourier transform of the time evolution of the full Hamiltonian (see Supplementary Information for theoretical details) to determine relative energy separations between the eigenstates of the respective blocks. In \cref{Fig:ft_x,Fig:ft_theta}, panel `c' illustrates the Fourier transform derived from the quantum dynamics of the full Hamiltonians. The meaning of y-label in \cref{Fig:ft_x,Fig:ft_theta}(a-c) are described in detail in the Supplementary Information. Additionally, in panel `d', frequencies extracted from panels `a' to `c' are color-coded according to their respective spectra, facilitating the experimental determination of relative energies for all eigenstates. Furthermore, panel `e' in \cref{Fig:ft_x,Fig:ft_theta} presents a comparison between the energy eigenvalues from the quantum (represented by blue dots) and classical (depicted by dashed gray lines) simulations  of the nuclear Hamiltonian corresponding to the respective degree of freedom. 
\begin{figure}[t]
 	\centering
    \includegraphics[width=\columnwidth]{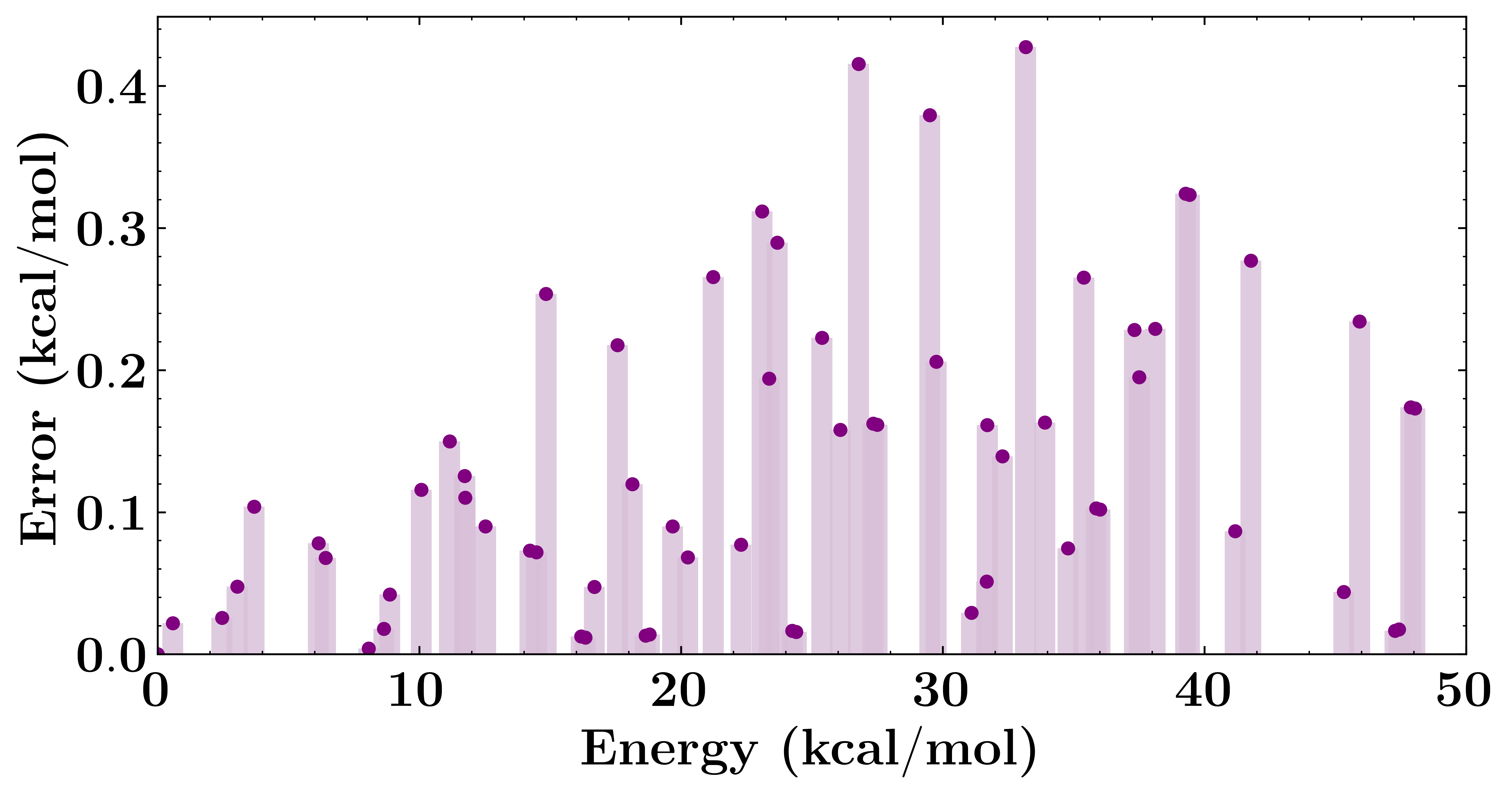}
  	\caption{Mean absolute eigenenergy differences between the quantum computed and classically computed vibrational eigenenergies corresponding to the two chosen modes of protonated 2,2'-bipyridine. Information from Figures \ref{Fig:ft_x}e and \ref{Fig:ft_theta}e are combined to form the vibrational energy spectrum. 
    The mean absolute error between the quantum-computed energies and the exact-diagonalization results is 0.14kcal/mol for first sixty-four energy levels.
   }
  	\label{Fig:Energy2D}
\end{figure}
At the end, from the Fourier transforms in Fig. \ref{Fig:ft_x} and Fig. \ref{Fig:ft_theta}, we find all energy differences (Fig. \ref{Fig:Energy2D}) between eigenvalues that populate a given initial states through our quantum computation on an ion-trap system. The accuracy of these eigenvalues are compared with those obtained from  exact diagonalization. The absolute energy differences between the quantum-computed and those obtained from the exact-diagonalization are shown in 
Figure \ref{Fig:Energy2D}. 
As can be seen the mean absolute error is less than 0.2kcal/mol, thus suggesting chemical accuracy. 
In a real chemical experiment these eigenstates would be thermally populated, thus reducing the number of such eigenvalue differences found in accordance to the experiment but here we find all eigenvalue differences within the initial wavepacket state. Finally, we utilize this information and find  specific eigenstates from these eigenvalue differences, but already the agreement between eigenvalue differences classically computed and quantum computed is  high. 

In conjunction with the findings of Ref. \cite{Sandia-1D-3Qubits}, we have demonstrated that quantum wavepacket dynamics studies can be constructed on ion-trap systems to good accuracy. While in Ref. \cite{Sandia-1D-3Qubits}, the study is shown for a one-dimensional system, here, we have shown that this can be done for a hydrogen transfer mode that is coupled to a gating dimension. We also present a theory that readily generalizes to higher dimensional quantum dynamics problems. 
The complexity of the problem as given by the associated set of quantum circuits and grows with the entanglement in the system. But we do not have an exponential scaling with dimension and the circuit depth only grows in a proportional manner to the number of basis points used per dimension which  may, in general, be fewer than the total number of dimensions in the problem. We believe that this approach offers a robust alternative to the challenges associated with quantum circuit depth and associated complexities in quantum nuclear dynamics.

\section{Acknowledgements}
This work is supported by the U.S. National Science Foundation under award OMA-1936353 and CHE-2311165 for SSI and PR. The quantum computations were performed on IonQ's Harmony quantum computer through the IonQ Academic Research Credits Program to PR and SSI. AD acknowledges support from the John R. and Wendy L. Kindig Foundation. 

\section{Author Contributions}
The theoretical formalism is due to AD and SSI whereas the quantum computations were performed by AJR and PR. 

%


\newpage
\begin{widetext}
\begin{center}
    {\bf For Table of content use only} \\ 
\vspace{1cm}
\includegraphics[width=8.5cm]{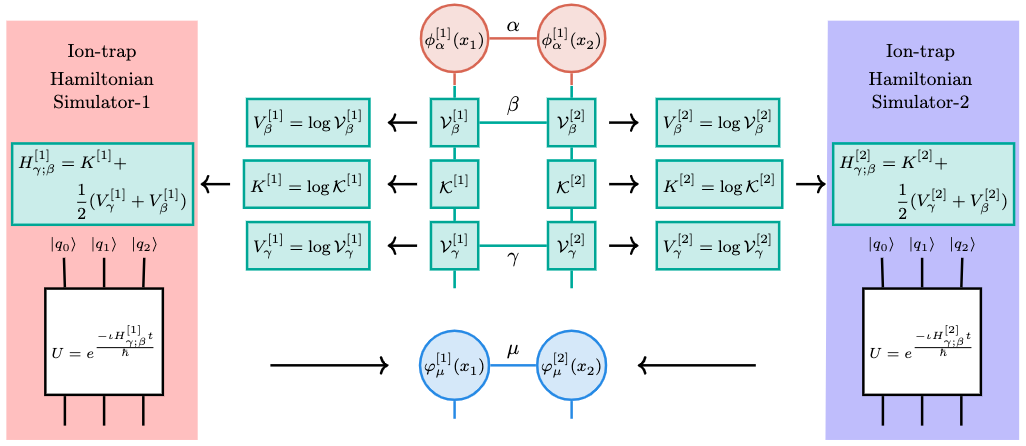}
\end{center}
\end{widetext}

\end{document}


\begin{center}
{\bf Supporting Information}
\end{center}
\title{Quantum nuclear dynamics on a distributed set of ion-trap quantum computing systems}

\author{Anurag Dwivedi}
\affiliation{Department of Chemistry, Indiana University, Bloomington, Indiana 47405, USA}
\affiliation{Indiana University Quantum Science and Engineering Center, Bloomington, Indiana 47405, USA}
\author{A. J. Rasmusson}
\affiliation{Department of Physics, Indiana University, Bloomington, Indiana 47405, USA}
\affiliation{Indiana University Quantum Science and Engineering Center, Bloomington, Indiana 47405, USA}
\author{Philip Richerme}
\affiliation{Department of Physics, Indiana University, Bloomington, Indiana 47405, USA}
\affiliation{Indiana University Quantum Science and Engineering Center, Bloomington, Indiana 47405, USA}
\author{Srinivasan S. Iyengar}
\affiliation{Indiana University Quantum Science and Engineering Center, Bloomington, Indiana 47405, USA}
\affiliation{Department of Chemistry, Indiana University, Bloomington, Indiana 47405, USA}
\date{\today}

\maketitle 

This Supporting Information document is organized as follows. In \cref{multiD-theory} we present our theory for multi-dimensional quantum nuclear wavepacket dynamics that naturally affords itself to implementation on a family of distributed quantum systems. This approach is based on tensor networks, and a specialized version of this general theory is implemented on the IonQ's 11-qubit trapped-ion quantum computer, Harmony, and accessed through the cloud via Amazon Braket \cite{wright2019benchmarking, braket}. A key aspect of this methodology is the development of a set of entangled low-dimensional quantum systems that are implemented here through a quantum-classical interface. The potential energy surfaces used for this purpose are described in \cref{Sec:PES}, and experimental details are provided in \cref{Sec:Experiment}. To compute vibrational properties we use an approach that was described in Ref. \onlinecite{Debadrita-Mapping-1D-3Qubits} and summarized in \cref{Sec:Vibrational}. The range of initial conditions considered for wavepacket dynamics is discussed in \cref{Sec:Initial}. The time-evolution of classical and quantum simulations are compared in \cref{Sec:dynamics}, and the spectral features derived from these are compared in \cref{Sec:spectrum}.

\section{Multi-dimensional wavepacket propagation on a distributed stream of quantum computing systems}
\label{multiD-theory}

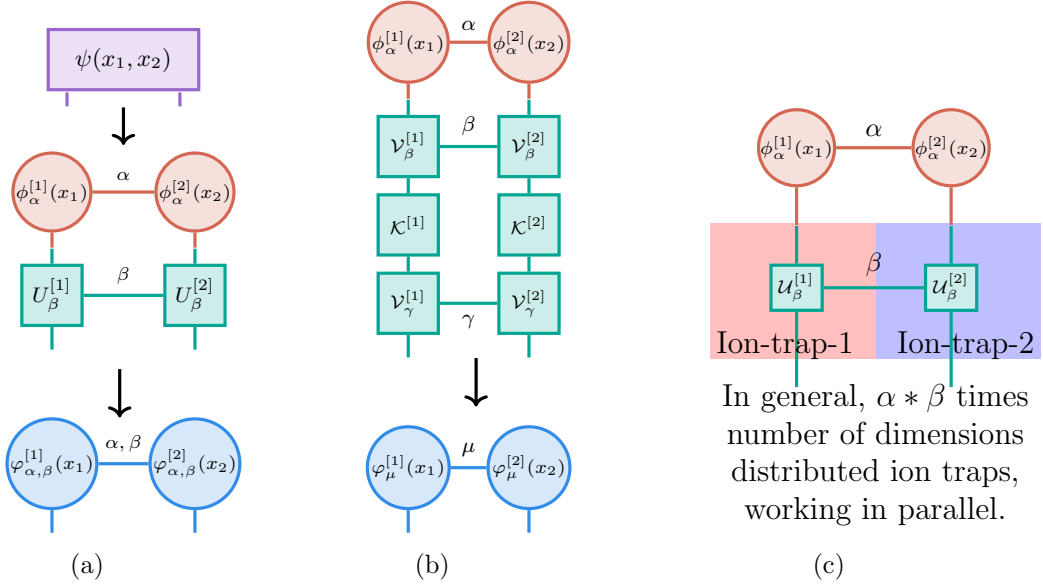
\begin{figure}[tbp]
    \subfigure[]{
\definecolor{bblue}{rgb}{0.19, 0.55, 0.91}
\definecolor{green}{rgb}{0.0, 0.65, 0.58}
\definecolor{purple}{rgb}{0.6, 0.4, 0.8}
\definecolor{orange}{rgb}{0.83, 0.4, 0.32}

\tikzstyle{psi} = [rectangle, very thick, minimum width=2cm, minimum height = 0.8cm, inner sep=0.1cm,draw=purple, fill=purple!20]

\tikzstyle{phi} = [circle, very thick, minimum size=0.8cm, inner sep=0.1mm, draw=orange, fill=orange!20]
    
\tikzstyle{phinew} = [circle, very thick, minimum size=0.7cm, inner sep=0.1mm, draw=bblue, fill=bblue!20]
    
\tikzstyle{U} = [rectangle, very thick, minimum size=0.8cm, inner sep=0.1cm,draw=green, fill=green!20]
    
\tikzset{arw/.style={-Stealth, thick}}
    \begin{tikzpicture}[>=latex, node distance=1cm,
        every edge quote/.style={font=\fontsize{7}{1}},scale=0.5]    
        \node (psi1) [psi, font=\fontsize{9}{1}]{$\psi(x_1,x_2)$};
      
        \node (psix01) [below=0mm of psi1, xshift = -7.5mm, minimum size=0mm, inner sep=0mm]{};
        \node (psix02) [below=0mm of psi1, xshift = +7.5mm, minimum size=0mm, inner sep=0mm]{};

        \node (psix00) [below=2mm of psi1, minimum size=0mm, inner sep=0mm]{};
        \node (ppx01) [below=5mm of psix00, minimum size=0mm, inner sep=0mm]{};
        
        \node (psix1) [below=2mm of psi1, xshift = -7.5mm, minimum size=0mm, inner sep=0mm]{};
        \node (psix2) [below=2mm of psi1, xshift = +7.5mm, minimum size=0mm, inner sep=0mm]{};

        \node (pphi1) [phi,below=8mm of psi1,xshift = -9.5mm, font=\fontsize{7}{1}]{$\phi^{[1]}_\alpha(x_1)$};
        \node (pphi2) [phi, below=8mm of psi1,xshift = +9.5mm, font=\fontsize{7}{1}]{$\phi^{[2]}_\alpha(x_2)$};
               
        \node (ppx1) [below=2mm of pphi1, minimum size=0mm, inner sep=0mm]{};
        \node (ppx2) [below=2mm of pphi2, minimum size=0mm, inner sep=0mm]{};
        
        \node (u1) [U, below=2mm of ppx1, font=\fontsize{8}{1}]{$U^{[1]}_\beta$};
        \node (u2) [U, below=2mm of ppx2, font=\fontsize{8}{1}]{$U^{[2]}_\beta$};
              
        \node (ux1) [below=3mm of u1, minimum size=0mm, inner sep=0mm]{};
        \node (ux2) [below=3mm of u2, minimum size=0mm, inner sep=0mm]{};
        
        \node (eql0sgn) [below=1mm of ux1, xshift=9mm]{};
        \node (eqlsgn) [below=5mm of eql0sgn]{};  
        
        \node (phit1) [phinew, below=12mm of u1, font=\fontsize{7}{1}]{$\varphi^{[1]}_{\alpha,\beta}(x_1)$};
        \node (phit2) [phinew, below=12mm of u2, font=\fontsize{7}{1}]{$\varphi^{[2]}_{\alpha,\beta}(x_2)$};
             
        \node (phitx1) [below=3mm of phit1, minimum size=0mm, inner sep=0mm]{};
        \node (phitx2) [below=3mm of phit2, minimum size=0mm, inner sep=0mm]{};

        \draw[-To, very thick, black, to path={-| (\tikztotarget)}] (psix00) -- (ppx01);
        \draw[-To, very thick, black, to path={-| (\tikztotarget)}] (eql0sgn) edge (eqlsgn);
        
        \path[-]
            
           
            
            (psix01) edge[very thick, purple] node[left,black, yshift=-1mm, font=\fontsize{7}{1}] {} (psix1)
            (psix02) edge[very thick, purple] 
            node[left,black, yshift=-1mm,font=\fontsize{10}{1}] {} (psix2)

            (pphi1) edge[very thick, orange] node[above,black,font=\fontsize{7}{1}] {$\alpha$} (pphi2)

            (pphi1) edge[very thick, orange] node[left,black,yshift=-1mm,font=\fontsize{8}{1}] {} (ppx1)
            (pphi2) edge[very thick, orange] node[left,black,yshift=-1mm,font=\fontsize{8}{1}] {} (ppx2)

            (u1) edge[very thick, green] node[above,black,font=\fontsize{7}{1}] {$\beta$}  node[below,black,font=\fontsize{7}{1}] {} (u2)
            
            (u1) edge[very thick, green] (ppx1)
            (u1) edge[very thick, green] node[left,black,yshift=-0.5mm,font=\fontsize{8}{1}] {}(ux1)
            (u2) edge[very thick, green] (ppx2)
            (u2) edge[very thick, green] node[left,black,yshift=-0.5mm,font=\fontsize{8}{1}] {}(ux2)
            
            (phit1) edge[very thick, bblue] node[above,black,font=\fontsize{7}{1}] {${\alpha,\beta}$} node[below,black] {} 
            (phit2)
            
            (phit1) edge[very thick, bblue] node[left,black,yshift=-0.5mm,font=\fontsize{8}{1}] {}(phitx1)
            (phit2) edge[very thick, bblue] node[left,black,yshift=-0.5mm,font=\fontsize{8}{1}] {}(phitx2)
            ;

    \end{tikzpicture}
        \label{Fig:MPS_tEvo_trotter-a}}
    \subfigure[]
    {
\definecolor{bblue}{rgb}{0.19, 0.55, 0.91}
\definecolor{green}{rgb}{0.0, 0.65, 0.58}
\definecolor{purple}{rgb}{0.6, 0.4, 0.8}
\definecolor{orange}{rgb}{0.83, 0.4, 0.32}

    \tikzstyle{psi} = [rectangle, very thick, minimum width=2cm, minimum height = 0.8cm, inner sep=0.1cm,draw=purple, fill=purple!20]
    
    \tikzstyle{phi} = [circle, very thick, minimum size=0.8cm, inner sep=0.2mm, draw=orange, fill=orange!20]
    
    \tikzstyle{phinew} = [circle, very thick, minimum size=0.8cm, inner sep=0.2mm, draw=bblue, fill=bblue!20]
    
    \tikzstyle{U} = [rectangle, very thick, minimum size=0.8cm, inner sep=0.1cm,draw=green, fill=green!20]
    
    \tikzset{arw/.style={-Stealth, thick}}

    \begin{tikzpicture}[>=latex,node distance=3cm,
        every edge quote/.style={font=\fontsize{7}{1}},
        ]
      

        
        
		\node (pphi1) [phi, xshift = -8mm, font=\fontsize{7}{1}]{$\phi^{[1]}_\alpha(x_1)$};
        \node (pphi2) [phi, right=5mm of pphi1, font=\fontsize{7}{1}]{$\phi^{[2]}_\alpha(x_2)$};

        \node (ppx01) [below=5mm of psix00, minimum size=0mm, inner sep=0mm]{}; 
        
        \node (ppx1) [below=2mm of pphi1, minimum size=0mm, inner sep=0mm]{};
        \node (ppx2) [below=2mm of pphi2, minimum size=0mm, inner sep=0mm]{};
        
        \node (V1) [U, below=2mm of ppx1, font=\fontsize{7}{1}]{$\mathcal{V}^{[1]}_\beta$};
        \node (V2) [U, below=2mm of ppx2, font=\fontsize{7}{1}]{$\mathcal{V}^{[2]}_\beta$};
                
        \node (K1) [U, below=2mm of V1, font=\fontsize{7}{1}]{$\mathcal{K}^{[1]}$};
        \node (K2) [U, below=2mm of V2, font=\fontsize{7}{1}]{$\mathcal{K}^{[2]}$};

        \node (u1) [U, below=2mm of K1, font=\fontsize{7}{1}]{$\mathcal{V}^{[1]}_\gamma$};
        \node (u2) [U, below=2mm of K2, font=\fontsize{7}{1}]{$\mathcal{V}^{[2]}_\gamma$};
             
        \node (ux1) [below=3mm of u1, minimum size=0mm, inner sep=0mm]{};
        \node (ux2) [below=3mm of u2, minimum size=0mm, inner sep=0mm]{};
        
        \node (eql0sgn) [below=10mm of ux1, right=7.5mm of ux1]{};
        \node (eqlsgn) [below=5mm of eql0sgn]{};  
        
        \node (phit1) [phinew, below=12mm of u1, font=\fontsize{7}{1}]{$\varphi^{[1]}_{\mu}(x_1)$};
        \node (phit2) [phinew, below=12mm of u2, font=\fontsize{7}{1}]{$\varphi^{[2]}_{\mu}(x_2)$};
        
        \node (phitx1) [below=3mm of phit1, minimum size=0mm, inner sep=0mm]{};
        \node (phitx2) [below=3mm of phit2, minimum size=0mm, inner sep=0mm]{};

        \draw[-To, very thick, black, to path={-| (\tikztotarget)}] (eql0sgn) edge (eqlsgn);
        
        \path[-]
            
           

            
            (pphi1) edge[very thick, orange] node[above,black,font=\fontsize{8}{1}] {$\alpha$} (pphi2)
           
            (pphi1) edge[very thick, orange] node[left,black,yshift=-1mm,font=\fontsize{8}{1}] {} (ppx1)
            (pphi2) edge[very thick, orange] node[left,black,yshift=-1mm,font=\fontsize{8}{1}] {} (ppx2)
            
            (V1) edge[very thick, green] node[above,black,font=\fontsize{8}{1}] {$\beta$} (V2)
            
            (u1) edge[very thick, green] node[below,black,font=\fontsize{8}{1}] {$\gamma$} (u2)

            (V1) edge[very thick, green] (ppx1)
            (K1) edge[very thick, green] node[left,black,yshift=-0.5mm,font=\fontsize{8}{1}] {}(u1)
            (V2) edge[very thick, green] (ppx2)
            (K2) edge[very thick, green] node[left,black,yshift=-0.5mm,font=\fontsize{8}{1}] {}(u2)
            
            (V1) edge[very thick, green] (K1)
            (V2) edge[very thick, green] (K2)

            (u1) edge[very thick, green] (ux1)
            (u2) edge[very thick, green] (ux2)

            (phit1) edge[very thick , bblue] node[above,black,font=\fontsize{8}{1}] {${\mu}$} node[below,black] {} 
            (phit2)
           
            (phit1) edge[very thick, bblue] node[left,black,yshift=-0.5mm,font=\fontsize{8}{1}] {}(phitx1)
            (phit2) edge[very thick, bblue] node[left,black,yshift=-0.5mm,font=\fontsize{8}{1}] {}(phitx2)
            ;

    \end{tikzpicture} \label{Fig:MPS_tEvo_trotter-c}}
    \subfigure[]{
\definecolor{bblue}{rgb}{0.19, 0.55, 0.91}
\definecolor{green}{rgb}{0.0, 0.65, 0.58}
\definecolor{purple}{rgb}{0.6, 0.4, 0.8}
\definecolor{orange}{rgb}{0.83, 0.4, 0.32}

    \tikzstyle{psi} = [rectangle, very thick, minimum width=2cm, minimum height = 0.6cm, inner sep=0.0cm,draw=purple, fill=purple!20]
    
    \tikzstyle{phi} = [circle, very thick, minimum size=0.6cm, inner sep=0.0mm, draw=orange, fill=orange!20]
    
    \tikzstyle{phinew} = [circle, very thick, minimum size=0.6cm, inner sep=0.0mm, draw=bblue, fill=bblue!20]
    
    \tikzstyle{phiempty} = [circle, very thick, minimum size=0.cm, inner sep=0.0mm, draw=bblue, fill=bblue!20]

    \tikzstyle{U} = [rectangle, very thick, minimum size=0.6cm, inner sep=0.1cm,draw=green, fill=green!20]
    
    \tikzset{arw/.style={-Stealth, thick}}

    \begin{tikzpicture}[>=latex, node distance=0.7cm,
        every edge quote/.style={font=\fontsize{7}{1}}]
      

        
        
        \fill[fill=blue!100!white!25](0,-2.8)--(0,-1)--(2.2,-1)--(2.2,-2.8)--(0,-2.8);
        \fill[fill=red!100!white!25](0,-2.8)--(0,-1)--(-2.2,-1)--(-2.2,-2.8)--(0,-2.8);

        \node (pphiempty) [phiempty, font=\fontsize{7}{1}]{};

        \node (pphi1) [phi, left=5mm of pphiempty, font=\fontsize{7}{1}]{$\phi^{[1]}_\alpha(x_1)$};
        \node (pphi2) [phi, right=10mm of pphi1, font=\fontsize{7}{1}]{$\phi^{[2]}_\alpha(x_2)$};

        \node (ppx01) [below=5mm of psix00, minimum size=0mm, inner sep=0mm]{}; 
        
        \node (ppx1) [below=5mm of pphi1, minimum size=0mm, inner sep=0mm]{};
        \node (ppx2) [below=5mm of pphi2, minimum size=0mm, inner sep=0mm]{};
        
        \node (V1) [U, below=5mm of ppx1, font=\fontsize{7}{1}]{$\mathcal{U}^{[1]}_\beta$};
        \node (V2) [U, below=5mm of ppx2, font=\fontsize{7}{1}]{$\mathcal{U}^{[2]}_\beta$};         

             
        \node (ux1) [below=10mm of V1, minimum size=0mm, inner sep=0mm]{};
        \node (ux2) [below=10mm of V2, minimum size=0mm, inner sep=0mm]{};

        
        \path[-]
            
           

            
            (pphi1) edge[very thick, orange] node[above,black,font=\fontsize{10}{1}] {$\alpha$} (pphi2)
           
            (pphi1) edge[very thick, orange] node[left,black,yshift=-1mm,font=\fontsize{7}{1}] {} (ppx1)
            (pphi2) edge[very thick, orange] node[left,black,yshift=-1mm,font=\fontsize{7}{1}] {} (ppx2)
            
            (V1) edge[very thick, green] node[above,black,font=\fontsize{10}{1}] {$\beta$} (V2)
            

            (V1) edge[very thick, green] (ppx1)
            (V1) edge[very thick, green] node[left,black,yshift=-0.5mm,font=\fontsize{7}{1}] {}(ux1)
            (V2) edge[very thick, green] (ppx2)
            (V2) edge[very thick, green] node[left,black,yshift=-0.5mm,font=\fontsize{7}{1}] {}(ux2)
            
           
            ;
        \filldraw[black] (1.2,-2.3) circle (0pt)node[below]{Ion-trap-2};
        \filldraw[black] (-1.2,-2.3) circle (0pt)node[below]{Ion-trap-1};
        \filldraw[black] (0.0,-3) circle (0pt)node[below]{In general, $\alpha*\beta$ times};
        \filldraw[black] (0.0,-3.5) circle (0pt)node[below]{number of dimensions};
        \filldraw[black] (0.0,-4) circle (0pt)node[below]{distributed ion traps,};        
        \filldraw[black] (0.0,-4.5) circle (0pt)node[below]{working in parallel.};        
           
    \end{tikzpicture}

\label{Fig:MPS_tEvo_trotter-b}}
\caption{\label{Fig:MPS_tEvo_trotter} Figure \ref{Fig:MPS_tEvo_trotter-a} depicts  the time evolution of the tensor network representation of a wavepacket. Figure \ref{Fig:MPS_tEvo_trotter-c} enumerates the action of \ref{Fig:MPS_tEvo_trotter-a} when the time-evolution is written in Trotterized form. Here, the initial vector is given by entanglement variables $\alpha$ while the Trotterized propagator has two entangled variables, $\beta$ and $\gamma$, which combine to determine the index $\mu \equiv \left( \alpha, \beta, \gamma\right)$ for the propagated system. Figure \ref{Fig:MPS_tEvo_trotter-b} is derived from Figures \ref{Fig:MPS_tEvo_trotter-a} and \ref{Fig:MPS_tEvo_trotter-c} and concisely depicts the simultaneous study of quantum dynamics for individual one-dimensional wavefunctions on distinct ion-trap quantum computers.}
\end{figure}

We present the theory here to construct quantum nuclear wavepacket dynamics calculations on a stream of quantum processors. At the outset, the initial quantum nuclear state is expressed as the matrix product state (MPS)\cite{tensor_network}, where the $N$-dimensional quantum nuclear state is written as a family of one-dimensional functions, 
\begin{align}
\psi(\vb{\bar{x}}) = \sum_{\bar{\boldsymbol\alpha}}^{\bar{\boldsymbol\eta}} \tensor*{\phi}{^{[1]}_{}^{}_{\alpha_1}}(x_1) \qty[\prod_{j=2}^{N\text{-} 1} \tensor*{\phi}{^{[j]}_{}^{}_{\alpha_{j\text{-} 1}}_{\alpha_j}}(x_j)] \tensor*{\phi}{^{[N]}_{}_{\alpha_{N\text{-} 1}}}(x_N),
	\label{Eq:MPS_WF} 
\end{align}
Here, each term $\phi^{[j]}$ denotes a one-dimensional function, with the summation in \cref{Eq:MPS_WF} extending over the so-called entanglement variables, bond dimensions, or Schmidt rank\cite{tensor_network} $\bar{\boldsymbol\alpha} \equiv \qty{ \alpha_1, \alpha_2, \ldots, \alpha_{N-1} }$. The respective limits of this summation are denoted by $\bar{\boldsymbol\eta} \equiv \qty{\eta_1,\eta_2,\ldots,\eta_{N-1}}$. 
In the current publication, we have two quantum nuclear degrees of freedom, that is,  $N=2$. Thus, the above expression is a Schmidt decomposition\cite{Schmidt_SVD}:
\begin{align}
	\psi(x_1,x_2) = \sum_{{\alpha}_1}^{{\eta}_1} \tensor*{\phi}{^{[1]}_{}^{}_{\alpha_1}}(x_1)  \tensor*{\phi}{^{[2]}_{}_{\alpha_{1}}}(x_2),
	\label{Eq:Schmidt} 
\end{align}
For the two dimensional case treated here, the bond dimension is $\alpha_1$, and in the diagramatic represenatation of Eq. (\ref{Eq:Schmidt}) in  \cref{Fig:MPS_tEvo_trotter}(a), 
$\alpha_1$ is replaced by the symbol $\alpha$. 
Thus, in the two dimensional case, the system is defined by a set of one-dimensional functions: $\left\{ \tensor*{\phi}{^{[1]}_{}^{}_{\alpha}}(x_1) \right\};\left\{ \tensor*{\phi}{^{[2]}_{}_{\alpha}}(x_2) \right\}$ as illustrated using orange circles in \cref{Fig:MPS_tEvo_trotter}(a). 

It is noteworthy that writing the initial wavefunction into MPS stands as a pivotal step in this approach. Not only does it reduce the computational complexity in representing a quantum state, but it also furnishes us with a distributed quantum computing protocol. This paper essentially develops a parallel stream of one-dimensional propagators mapped onto independent ion-trap systems as depicted in \cref{Fig:MPS_tEvo_trotter}(c).

\subsection{Action of time-evolution operator on initial MPS wavefunction}\label{Sec:TN-time-evolution}
We represent the Hamiltonian for the wavepacket dynamics in the coordinate representation, where the kinetic energy operator for each nuclear dimension is separately represented using ``Distributed Approximating Functionals'' (DAF)\cite{DAFprop,discreteDAF,qwaimd-TCAreview},
\begin{align}
K(x_i,x_i^{\prime}) =& K(\left\vert x_i-x_i^{\prime}\right\vert) \nonumber \\ =&
\frac{-\hbar^2}{4m\sigma^3\sqrt{2\pi}}
\exp \left\{ -\frac{ {\left( x_i - x_i^\prime \right)}^2}
{2 {\sigma}^2} \right\} 
\sum_{n=0}^{M_{DAF}/2} {\left( \frac{-1}{4} \right)}^n \frac{1}{n!} 
H_{2n+2} \left( \frac{ x_i - x_i^\prime }{ \sqrt{2} \sigma} \right).
\label{DAFderivative}
\end{align}
The DAF provides an analytic banded Toeplitz representation for the
the kinetic energy operator. The banded-Toeplitz representation of the DAF approximation for the kinetic energy operator, 
where the property of its matrix elements, $K_{ij} \equiv K{(\vert} i-j {\vert )}$, 
has a critical role in reducing the nuclear Hamiltonian to the form of the ion-trap Hamiltonian in Ref. \onlinecite{Debadrita-Mapping-1D-3Qubits}, and also here in obtaining circuit representations in \cref{Sec:Block-diagonal,Sec:Experiment}. The quantities $\left\{ H_{2n+2}\left( \frac{ x_i - x_i^\prime }{ \sqrt{2} \sigma} \right) \right\}$ are the even order Hermite polynomials that only depend on the spread separating the grid basis vectors, $\ket{x_i}$ and $\ket{x_i^{\prime}}$, and $M_{DAF}$ and $\sigma$ are parameters that together determine the accuracy and efficiency of the resultant approximate kinetic energy operator. Appendix D in Ref. \onlinecite{Debadrita-Mapping-1D-3Qubits} provides a brief summary of the DAF approach for approximating functions, their derivatives and propagated forms. Thus,  the Hamiltonian describing the system under investigation is 
\begin{align}
    \hat{H}\left(\vb{\bar{x}},\vb{\bar{x}}^\prime\right) = \sum_i {K}\left({{x}_i},{x}_i^\prime\right) + \delta \left( \vb{\bar{x}}-\vb{\bar{x}}^\prime \right) \hat{V}\left(\vb{\bar{x}},\vb{\bar{x}}^\prime\right),
\end{align} 
where the potential energy operator, $\hat{V}\left(\vb{\bar{x}},\vb{\bar{x}}^\prime\right)$ is diagonal in the coordinate representation and is obtained from electronic structure calculations. Furthermore, the potential energy operator is not naturally separable into dimensions like in the case of the kinetic energy operator. While this does present a challenge, we present a solution in \cref{Sec:1DHam}. An additional challenge in quantum dynamics arises from the need to approximate the action of the time-evolution operator on a given state vector $\Psi$, that is, $e^{-i\hat{H}\Delta t/\hbar} \ket{\Psi}$. In this study, we employ the widely used Trotter-Suzuki factorization \cite{Campbell1897-tj,Baker1901-cj,Trotter,Suzuki1976-us,Nelson-Trotter,qwaimd}. This factorization method offers a straightforward approximation of the operator, where the accuracy depends on the size of the time-step $\Delta t$. Specifically, the Trotter symmetric split operator \cite{Trotter,Nelson-Trotter} expansion of $e^{-i\hat{H}\Delta t/\hbar}$, at second order in $\Delta t$, is
\begin{align}
	e^{-i\hat{H}\Delta t/\hbar} &=
 e^{-i\hat{V}\left(\vb{\bar{x}}\right)\Delta t/2\hbar} \left\{ \prod_i e^{-i\hat{K}\left({{x}_i},{x}_i^\prime\right)\Delta t/\hbar}\right\} e^{-i\hat{V}\left(\vb{\bar{x}^\prime}\right)\Delta t/2\hbar} + O(\Delta t^3) \nonumber \\ &=
 e^{-i\hat{V}\left(\vb{\bar{x}}\right)\Delta t/2\hbar} \left\{ \prod_i\tensor*{\mathcal{K}}{^{[i]}}(x_i,x_i^\prime) \right\} e^{-i\hat{V}\left(\vb{\bar{x}^\prime}\right)\Delta t/2\hbar} + O(\Delta t^3) 
	\label{Eq:Trotter}
\end{align}
that is, the time-evolution operator is now expressed in terms of the potential propagator $e^{-i\hat{V}\Delta t/2\hbar}$ and the individual kinetic propagators $e^{-i\hat{K}\left({{x}_i},{x}_i^\prime\right)\Delta t/\hbar} \equiv \tensor*{\mathcal{K}}{^{[i]}}(x_i,x_i^\prime)$ using expression given in Eq. (\ref{DAFderivative}). 

Additionally, as noted above, the non-separable nature of the potential arising from electronic structure does present a problem and here, as in Eq. (\ref{Eq:Schmidt}), we represent $e^{-i\hat{V}\left(\vb{\bar{x}}\right)\Delta t/2\hbar}$ as a tensor network, which for two-dimensions is given by
\begin{align}
	e^{-i\hat{V}\left(\vb{\bar{x}}\right)\Delta t/2\hbar} = \sum_{\beta} \tensor*{\mathcal{V}}{^{[1]}_{\beta}^{}}(x_1) \tensor*{\mathcal{V}}{^{[2]}_{\beta}}(x_2).
 	\label{Eq:UV-TN-1}
 \end{align}
Here, $\left\{ \tensor*{\mathcal{V}}{^{[1]}_{}^{}_{\beta}} (x_1)\right\};\left\{ \tensor*{\mathcal{V}}{^{[2]}_{}^{}_{\beta}}(x_2) \right\}$ are the one-dimensional functions obtained from the Schmidt decomposition\cite{Schmidt_SVD} of the potential propagator $e^{-i\hat{V}\left(\vb{\bar{x}}\right)\Delta t/2\hbar}$ and the sum runs over the entanglement variable $\beta$. 


Finally, the time-evolution of the MPS nuclear wavefunction described in Equation \ref{Eq:Schmidt} is achieved through Equations \ref{Eq:Trotter} and \ref{Eq:UV-TN-1} as  
%
\begin{align}
    \nonumber
    \psi(\bar{\vb{x}};t+\Delta t) &=  \sum_{\alpha,\beta,\gamma}\int \dd{x'_1}\dd{x'_2}\\
    &\quad\times \qty[\tensor*{\mathcal{V}}{^{[1]}_{\gamma}}(x_1) \tensor*{\mathcal{K}}{^{[1]}}(x'_1,x_1) \tensor*{\mathcal{V}}{^{[1]}_{\beta}}(x'_1)]\phi^{[1]}_{\alpha}(x'_1)\nonumber\\
    &\quad\times\qty[\tensor*{\mathcal{V}}{^{[2]}_{\gamma}}(x_2) \tensor*{\mathcal{K}}{^{[2]}}(x'_2,x_2) \tensor*{\mathcal{V}}{^{[2]}_{\beta}}(x'_2)]\phi^{[2]}_{\alpha}(x'_{2})
    \label{Eq:MPS_tEvo}
\end{align}
and this process is described by \cref{Fig:MPS_tEvo_trotter}(b).
As a result, the action of the time-evolution operator on the initial MPS, is transformed into a set of parallel streams of one-dimensional effective quantum propagations and this is concisely represented in \cref{Fig:MPS_tEvo_trotter}(c).
\subsection{Computing effective 1D Hamiltonians from 1D potential propagators}\label{Sec:1DHam}
\begin{figure}[tbp]
    \input{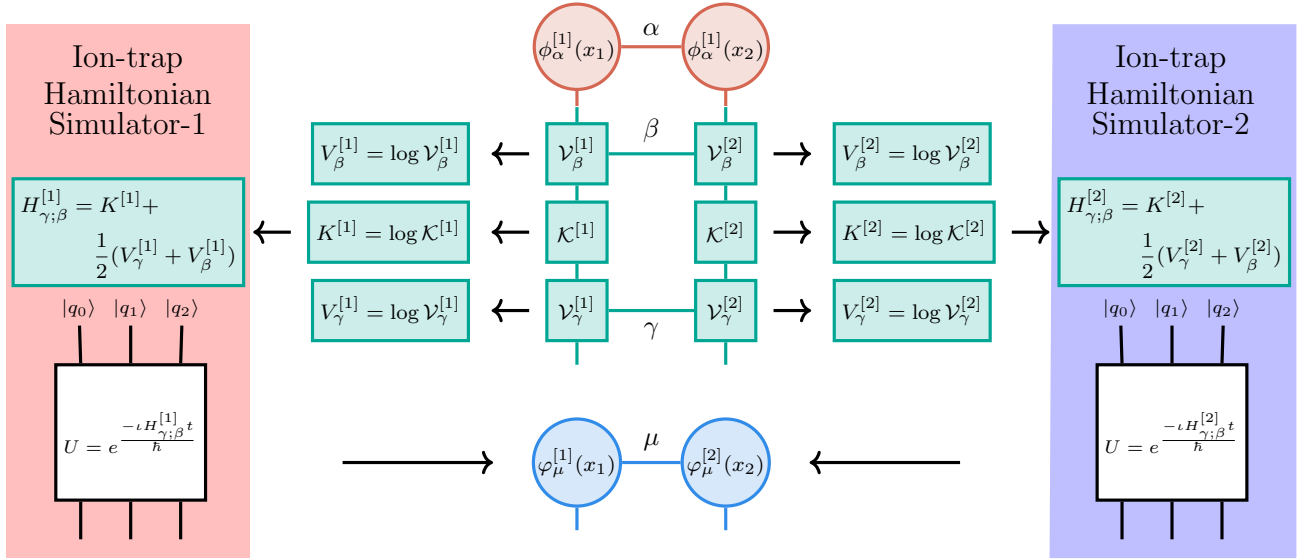}
    \caption{A detailed exposition of \cref{Fig:MPS_tEvo_trotter-a} that precisely outlines how the distributed simulations are carried out. The individual one-dimensional potentials are derived from each one-dimensional propagator as shown, yielding a family of effective one-dimensional Hamiltonians for the system. These individual one-dimensional Hamiltonians are then be independently and concurrently simulated on separate ion-trap quantum computers.}
    \label{Fig:MPS_tEvo_trotter-d}
\end{figure}
We first note that the direct action of each Trotterized propagator block in \cref{Eq:MPS_tEvo} and in \cref{Fig:MPS_tEvo_trotter}(b) depicts the action of one-dimensional propagators, obtained from tensor-network decomposition of the {\em potential-propagator},
on the MPS components of the initial wavepacket. 
Towards simplifying this action and making direct analogues to ion-trap Ising Hamiltonains, as discussed in Ref. \onlinecite{Debadrita-Mapping-1D-3Qubits}, we write each of the potential-propagator terms, $\left\{\mathcal{V}^{[j]}_{\beta_{1}}(x_{j})\right\};\left\{ \mathcal{V}^{[j]}_{\gamma_{1}}(x_{j}) \right\}$, encoding information about the one dimensional potentials as,
\begin{align}
    \nonumber\mathcal{V}^{[j]}_{\beta}(x_{j})
    ={\cal A}^{[j]}_{\beta}(x_{j}) \times \exp{-\imath {V}^{[j]}_{\beta}(x_{j}) \Delta t/2\hbar}\\
    \mathcal{V}^{[j]}_{\gamma}(x_{\textit{\textit{\textit{\textit{}}}}j})
    ={\cal A}^{[j]}_{\gamma}(x_{j}) \times \exp{-\imath {V}^{[j]}_{\gamma}(x_{j}) \Delta t/2\hbar}
    \label{Eq:V-A-S}
\end{align}
where, $\left\{{V}^{[j]}_{\beta}(x_{j})\right\};\left\{ {V}^{[j]}_{\gamma}(x_{j}) \right\}$
are effective one dimensional potentials obtained by taking the logarithmic of the potential propagators, that is, 
\begin{align}
{\log \cal A}^{[j]}_{\beta}(x_{j}) &= Re \left[ \log \mathcal{V}^{[j]}_{\beta}(x_{j}) \right] \label{Eq:log_A}\\ 
{V}^{[j]}_{\beta}(x_{j}) &= -\frac{2\hbar}{\Delta t} Im \left[ \log \mathcal{V}^{[j]}_{\beta}(x_{j}) \right]
\label{Eq:log_V}
\end{align}
to obtain a family of reduced dimensional Hamiltonians
\begin{align}
H^{[j]}_{\gamma;\beta}(x_j, x'_j) =K^{[j]}\left({{x}_j},{x}_j^\prime\right)+\frac{1}{2}({V}^{[j]}_{\gamma}(x_{i})+{V}^{[j]}_{\beta}(x'_{j}))
\label{Eq:MPS_Ham}
\end{align}
as depicted within the pink and blue blocks in \cref{Fig:MPS_tEvo_trotter-d}. 
In this publication, ${\cal A}^{[j]}_{\beta}(x_{j}) = 1$, and hence \cref{Eq:log_A} does not contribute to the trajectories. 
This feature enables the parallel computation of nuclear dynamics across distinct ion-trap quantum hardware. Consequently, the TN representation establishes effective one-dimensional subsystems, thereby defining corresponding `one-dimensional' Hamiltonians for each of these subsystems, 
\begin{flalign}
    \noindent
	\nonumber
	\psi(\bar{\vb{x}};t+\Delta t) &= \sum_{\alpha,\beta,\gamma}\int \dd{x'_1}\dd{x'_2}\\
    &\quad\times\qty[\exp{-\imath H_{\gamma;\beta}^{[1]}(x_1, x'_1)\Delta t/\hbar}]\phi^{[1]}_{\alpha}(x'_{1})\nonumber\\
	&\quad\times\qty[\exp{-\imath H^{[2]}_{\gamma;\beta}(x_2, x'_2)\Delta t/\hbar}]\phi^{[2]}_{\alpha}(x'_2).
	\label{Eq:MPS_tEvo-LogHam3}
\end{flalign}

where, the effective `one-dimensional' Hamiltonians obey,
\begin{align}
    \exp{-\imath H^{[j]}_{\gamma;\beta}(x_j, x'_j)\Delta t/\hbar} \equiv
    \exp{{-i\left[K^{[j]}(x_j, x'_j)+
    \frac{1}{2}\left({V}^{[j]}_{\gamma}(x_{j})+{V}^{[j]}_{\beta}(x'_{j})\right)\right] 
    \Delta t/\hbar}}
    \label{Eq:Ham1}
\end{align}
The equations above produce a stream of one-dimensional unitary propagations that may be directly mapped to quantum simulators as depicted in \cref{Fig:MPS_tEvo_trotter-d}. 
The one dimensional Hamiltonians, $\left\{ H_{\gamma;\beta}^{[1]}\right\};\left\{H_{\gamma;\beta}^{[2]} \right\}$ represent independent and parallel, quantum computing steps, to be performed on independent quantum hardware systems (see \cref{Fig:MPS_tEvo_trotter-d}) and each of these propagate the representative one-dimensional systems. 

\begin{figure}[t]
 	\centering
    \subfigure[]{\includegraphics[width=0.35\textwidth]{Figures_ionq/Bipyridine_Molecule.png}\label{Fig:Molecule}}
    \subfigure[]{\includegraphics[width=0.35\textwidth]{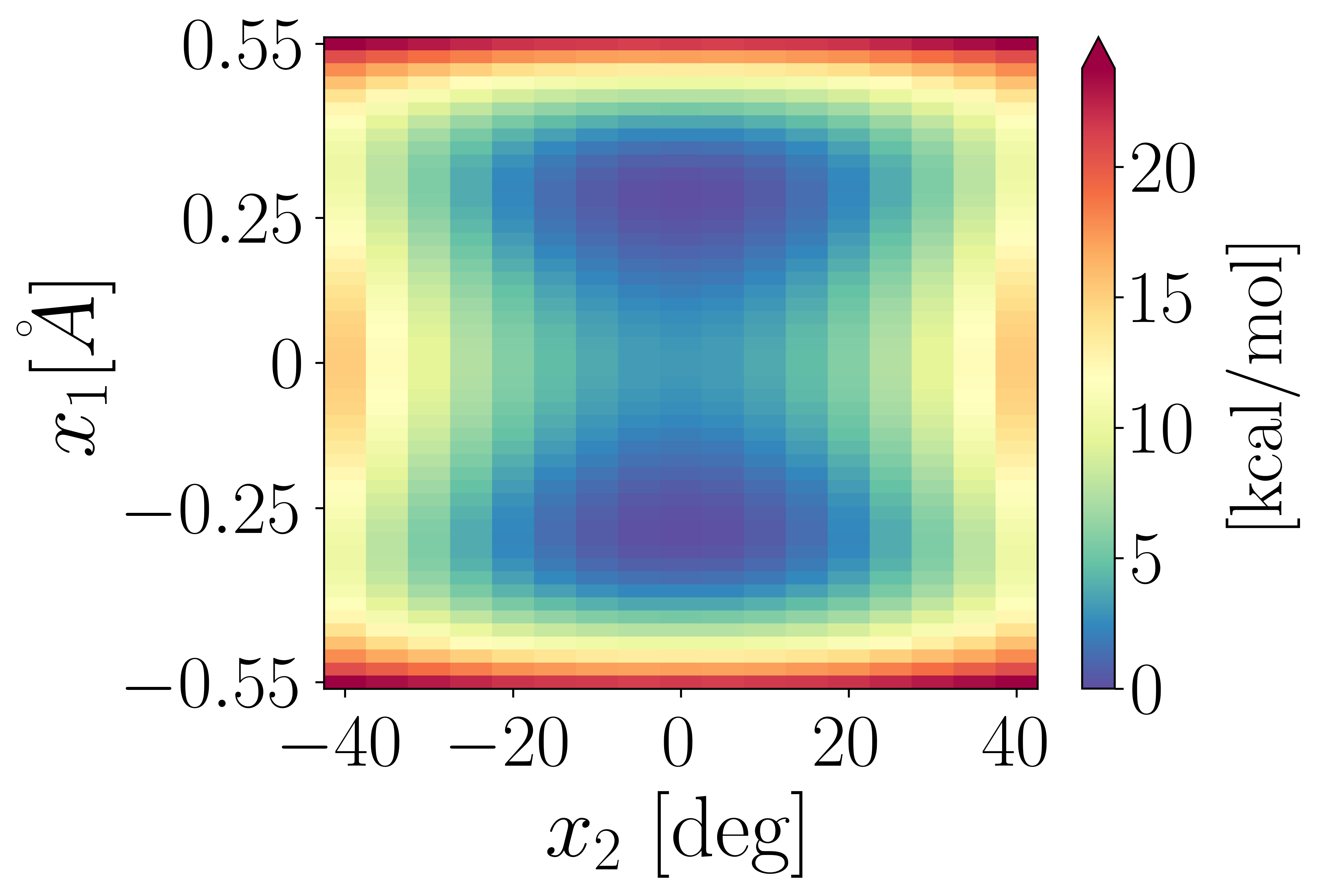}\label{Fig:pes}}
    \caption{Figure (a) shows the protonated 2,2'-bipyridine system with the two dimensions, $x_1$ and $x_2$, correspond to the vibrational degree of freedom of transferring proton along the internuclear axis joining the two nitrogen atoms, and the torsional degree of freedom due to planar rotations of the pyridyl rings about the C-C bond. Figure (b) shows the colormap plot of the potential energy surface for the molecule.}
    \label{Fig:molecule_pes}
\end{figure}
\section{Potential energy surfaces (PES) for quantum dynamics}\label{Sec:PES}
In this section, we provide a concise overview of the construction process for the Potential Energy Surface (PES) (see \cref{Fig:molecule_pes}(b)) tailored to the protonated 2,2’-bipyridine system (shown in \cref{Fig:molecule_pes}(a)), crucial for elucidating the TN propagation scheme. To capture the system's relevant degrees of freedom accurately, we employ cylindrical coordinates $(x_1,x_2)$, where $x_1$ signifies the relative position of the shared hydrogen nucleus along the NN axis, and $x_2$ denotes the dihedral angle between the two pyridyl rings surrounding the C-C bond.

The PES is constructed by identifying stationary points for each fixed dihedral angle $x_2\in[\ang{-35},\ang{35}]$, positioning the shared proton at the midpoint of the internuclear axis between the nitrogen nuclei. Notably, each of these torsional geometries yields a transition state with one imaginary frequency upon diagonalizing the Hessian matrix. This imaginary frequency corresponds to the vibrational mode associated with proton transfer along the internuclear axis connecting the nitrogen nuclei.

\begin{table*}[t]
    \centering
    \begin{tabular*}{\textwidth}{@{\extracolsep{\fill}} cc}
    \hline \hline
    \vspace*{2mm}
    No. of grid points along the $x_1$-dimension  & 8 \\ \vspace*{2mm}
    No. of grid points along the $x_2$-dimension & 8\\ \vspace*{2mm}
    Grid size along the $x_1$-dimension & 1.1\r{A} \\ \vspace*{2mm}
    Range of angles along the $x_2$-dimension & $[-35^{\circ}$,$35^{\circ}]$ \\ \vspace*{2mm}
    Level of theory & B3LYP/6-311++G(d,p)//B3LYP/6-311++G(d,p)\\
    \hline \hline
    \end{tabular*}
\caption{Characteristics of the grid over which 2D potential surface is created.}
\label{Tab:grid}
\end{table*}

Subsequently, maintaining a fixed torsional geometry (that is, a constant value of $x_2$), we compute a series of one-dimensional effective potentials. These potentials, contingent upon the proton's position along the internuclear axis, are computed via a density functional theory (DFT) methodology. Specifically, we employ a hybrid generalized gradient approximation (GGA) functional B3LYP in conjunction with an atom-centered Gaussian basis set 6-311++G(d,p). To compute the Potential Energy Surface (PES), we employ a grid consisting of $2^3=8$ points along the internuclear axis connecting the two nitrogen atoms for each of the 8 specific dihedral angles within the range of [-35°, 35°]. We provide a summary of the primary features utilized in the calculation of the PES in \cref{Tab:grid}.
The resultant rectangular $8\times 8$ grid along $(x_1,x_2)$ has potential values as shown in \cref{Fig:molecule_pes}(b). This potential has inversion symmetry, $V(x_1,x_2) = V(-x_1,-x_2)$, and is a result of the symmetry in the molecule shown in \cref{Fig:molecule_pes}(a). This symmetry of the potential is exploited here, using the methods discussed in Ref. \onlinecite{Debadrita-Mapping-1D-3Qubits}, to block-diagonalize the molecular Hamiltonian an obtain an efficient quantum computing algorithm. The process is discussed below. 

\subsection{Block diagonalization of one-dimensional nuclear Hamiltonians, $\left\{ H_{\gamma;\beta}^{[1]}\right\};\left\{ H_{\gamma;\beta}^{[2]} \right\}$ owing to symmetric nature of PES in \cref{Fig:molecule_pes}(b)}\label{Sec:Block-diagonal}
As discussed in Ref. \onlinecite{Debadrita-Mapping-1D-3Qubits}, the choice of kinetic energy operator in Eq. (\ref{DAFderivative}) and symmetric nature of potential energy operator highlighted above, allows for the one-dimensional nuclear Hamiltonians, $\left\{ H_{\gamma;\beta}^{[1]}\right\};\left\{ H_{\gamma;\beta}^{[2]} \right\}$, to be represented in block-diagonal form. 
This process aims to facilitate the independent propagation of each block across an additional parallel set of ion-trap systems. 
While the algorithm presented above (see Figure \ref{Fig:MPS_tEvo_trotter}(c)) already yields as many multiple channels as entanglement dimensions times the number of physical dimensions, the block-diagonalization discussed here yields an additional factor of two in terms of number of distributed set of quantum processes. 
The kinetic energy operator, when expressed in the Distributed Approximating Functional (DAF) representation \cite{DAFprop-PRL,discreteDAF}, takes the form of a banded Toeplitz structure where each sub-diagonal element has a constant value, as mentioned in  \cref{Sec:TN-time-evolution}. Additionally, the Schmidt decomposed potentials, $\left\{ {V}^{[j]}_{\gamma}(x_{i})\right\}$ derived from \cref{Eq:V-A-S,Eq:log_V} are symmetric as we note in \cref{Sec:1DHam}.
Matrices with these two key properties (the Toeplitz form of the kinetic energy operator and the symmetric nature of the one-dimensional effective potentials) can be transformed into a block-diagonal form by applying a Givens unitary transformation\cite{Debadrita-Mapping-1D-3Qubits}. Specifically, the $(i,l)$-the element of $H_{\gamma;\beta}^{[1]}$ and $H_{\gamma;\beta}^{[2]}$ may be transformed as per
\begin{align}
\tilde{H}{^{[j]}_{{\gamma;\beta},i,l}}= \frac{1}{2}(H^{[j]}_{{\gamma;\beta},i,l} + \alpha_{l}H^{[j]}_{{\gamma;\beta},i,n_j-l} + \alpha_{i}H^{[j]}_{{\gamma;\beta},n_j-i,l} + \alpha_{i}\alpha_{l}H^{[j]}_{{\gamma;\beta},n_j-i,n_j-l})
\end{align}
Here, $\alpha_i = \text{sgn}\left[i-\frac{n_j}{2}\right]$, and the grid point indices range from 0 to $n_j=2^{D_j}-1$, where $D_j$ represents the number of qubits used to simulate the respective one-dimensional Hamiltonian. The diagonal blocks of the unitary transformed Hamiltonian $\tilde{H}_{\gamma;\beta}^{[j]}$ can then be expressed as:
\begin{align}
    \tilde{H}^{[j]}_{\gamma;\beta,i,l} = \left[K^{[j]}_{i,l} + \alpha_{i}K^{[j]}_{i,n_j-l}\right] + \frac{1}{4}\left[(V^{[j]}_{\gamma,i}+V^{[j]}_{\beta,i})+(V^{[j]}_{\gamma,n_j-l}+V^{[j]}_{\beta,n_j-l})\right]\delta_{i,l}
    \label{Eq:on-diagonal}
\end{align}
Meanwhile, the elements of the off-diagonal blocks are given by:
\begin{align}
    \tilde{H}^{[j]}_{\gamma;\beta,i,l} = \frac{1}{4}\left[(V^{[j]}_{\gamma,i})-V^{[j]}_{\gamma,n_j-l})+(V^{[j]}_{\beta,i}-V^{[j]}_{\beta,n_j-l})\right]\delta_{i,l}
    \label{Eq:off-diagonal}
\end{align}
Due to the inherent symmetry of the potential energy surface, as discussed in \cref{Sec:PES}, and observed in other hydrogen-bonded systems\cite{qwaimd-wavelet,Meyer-MCTDH-Zundel,diken2005h3o2zundel}, the off-diagonal blocks of the transformed Hamiltonian (\cref{Eq:off-diagonal}) are uniformly zero. In addition to the strategies outlined in \cref{Sec:1DHam}, the symmetry of the potential here, enables further parallelization of simulation of these effective one-dimensional Hamiltonians. 
\section{Initial Conditions for wavepacket dynamics}\label{Sec:Initial}
\begin{table}[]
    \centering
\begin{tabular*}{\textwidth}{@{\extracolsep{\fill}} c|cccc}
        \hline \hline
        \vspace*{2mm}
            Simulation type & \multicolumn{4}{c}{Initial Wavepackets\footnote{Four different initial wavepackets were considered as shown on the four columns below. Notation: $\delta(x_1-x_1^1)$ implies the initial wavepacket has a non-zero value at grid point ``1'' (the superscript) for dimension ``1'' (the subscript)}}\\
            \hline
            \vspace*{2mm}
            $x_1$-upper block\footnote{Using the block diagonalization scheme discussed in Section \ref{Sec:Block-diagonal}.} & $\delta(x_1-x_1^1)$ & $\delta(x_1-x_1^2)$ & $\delta(x_1-x_1^3)$ & $\delta(x_1-x_1^4)$\\
            \vspace*{2mm}
            $x_1$-lower block\footnotemark[2] & $\delta(x_1-x_1^5)$ & $\delta(x_1-x_1^6)$ & $\delta(x_1-x_1^7)$ & $\delta(x_1-x_1^8)$\\
            \vspace*{2mm}
            $x_2$-upper block\footnotemark[2] & $\delta(x_2-x_2^1)$ & $\delta(x_2-x_2^2)$ & $-$ & $-$\\
            \vspace*{2mm}
            $x_2$-lower block\footnotemark[2] & $\delta(x_2-x_2^6)$ & $\delta(x_2-x_2^7)$ & $\delta(x_2-x_2^8)$ & $-$\\
            \hline
            \vspace*{2mm}
            $x_1$-full\footnote{Without the block diagonalization scheme discussed in Section \ref{Sec:Block-diagonal}. That is using the full Hamiltonian without block diagonalization.} & $\delta(x_1-x_1^1)$ & $\delta(x_1-x_1^2)$ & $-$ & $-$\\
            \vspace*{2mm}
            $x_2$-full\footnotemark[3] & $\delta(x_2-x_2^1)$ & $-$ & $-$ & $-$\\
            \hline \hline
\end{tabular*}
    \caption{Initial wavepacket  details. The $x_1$ and $x_2$ simulations are conducted on a distributed set of two ion-traps}
    \label{Tab:Psi_0}
\end{table}
The dynamics of many initial wavepackets were studied and these initial wavepackets are shown in Table \ref{Tab:Psi_0}. In all cases, the initial wavepackets were chosen as Dirac delta functions localized at specific grid points as shown in Table \ref{Tab:Psi_0}. The reason for this is as follows. Since our goal was to obtain spectral properties from the dynamics, it is needed that all eigenstates be captured within a specific initial wavepacket state. Delta functions by their choice have this property. Secondly, delta functions typically have a high energy expectation value and hence in some sense these choices provide the hardest test cases for our simulation. As we see from the agreement in vibrational properties, the quantum simulations are highly robust. 

\begin{table*}[t!]
        \centering
        \begin{tabular*}{\textwidth}{@{\extracolsep{\fill}} c|c|cccc|cccc}
        \hline\hline
        \vspace*{2mm}
            Simulation type & Size & \multicolumn{4}{c|}{Quantum Simulation} & \multicolumn{4}{c}{Classical Simulation} \\
            \hline
            \vspace*{2mm}
            & & \begin{tabular}{@{}c@{}} $\Delta t$\\$(\unit{\femto\second})$ \end{tabular} & \begin{tabular}{@{}c@{}} $T$\\$(\unit{\femto\second})$ \end{tabular} & \begin{tabular}{@{}c@{}} $\Delta \omega$\footnote{Nyquist theorem}\\$(\unit{\tera\hertz})$ \end{tabular} & \begin{tabular}{@{}c@{}} $\omega_{\text{max}}$\footnote{$\omega_{\text{max}} = \frac{1}{2\Delta t}$}\\$(\unit{\tera\hertz})$ \end{tabular} & \begin{tabular}{@{}c@{}} $\Delta t$\\$(\unit{\femto\second})$ \end{tabular} & \begin{tabular}{@{}c@{}} $T$\\$(\unit{\femto\second})$ \end{tabular} & \begin{tabular}{@{}c@{}} $\Delta \omega$\footnotemark[1]\\$(\unit{\tera\hertz})$ \end{tabular} & \begin{tabular}{@{}c@{}} $\omega_{\text{max}}$\footnotemark[2]\\$(\unit{\tera\hertz})$ \end{tabular} \\
            \hline
            \vspace*{2mm}
            $x_1$-upper/lower block & $4\times 4$ & 2.50 & 400 & 1.250 & 200.0 & 0.25 & 400 & 1.250 & 2000 \\
            \vspace*{2mm}
            $x_2$-upper/lower block & $4\times 4$ & 1.47 & 235 & 2.125 & 340.4 & 0.25 & 235 & 2.125 & 2000 \\
            \vspace*{2mm}
            $x_1$-full Hamiltonian & $8\times 8$ & 100.00 & 16000 & 0.031 & 5.0 & 0.25 & 16000 & 0.031 & 2000 \\
            \vspace*{2mm}
            $x_2$-full Hamiltonian & $8\times 8$ & 7.00 & 1120 & 0.445 & 71.4 & 0.25 & 1120 & 0.445 & 2000 \\
            \hline\hline
        \end{tabular*}
        \caption{Simulation details.}
        \label{Tab:Simulation}
\end{table*}
Furthermore, given the block-diagonalization scheme discussed above, two separate sets of simulations were conducted as indicated by the top half and bottom half of Table \ref{Tab:Psi_0}. In essence, the symmetry of the potential allows us to further reduce complexity. Each of these initial wavepackets, depending on which Hamiltonian blocks they are associated with, undergo simulation on IonQ's ion-trap quantum hardware. Details regarding the total propagation time and simulation time-steps are provided in \cref{Tab:Simulation}. 
Additionally, the time evolution of the initial wavepacket on classical hardware is computed as follows:
\begin{align}
	\nonumber
    \ket{\psi_{\gamma;\beta}(x_{j};t)} &= e^{-iH^{[j]}_{\gamma;\beta}t/\hbar}\ket{\psi_{\gamma;\beta}(x_{j};0)} \\ &= \sum_i e^{-iE^{[j]}_{\gamma;\beta,i}t/\hbar}\ket{\chi^{[j]}_{\gamma;\beta,i}}\braket{\chi^{[j]}_{\gamma;\beta,i}}{\psi_{\gamma;\beta}(x_{j};0)},
	\label{Eq:Classical}
\end{align}
Here, $H_{\gamma;\beta}^{[j]}$ represents the effective one-dimensional Hamiltonians (as derived in \cref{Sec:1DHam}). $\ket{\chi^{[j]}_{\gamma;\beta,i}}$ represents the eigenfunctions that satisfy the eigenvalue equation $H^{[j]}_{\gamma;\beta}\ket{\chi_{\gamma;\beta,i}} = E^{[j]}_{\gamma;\beta,i}\ket{\chi^{[j]}_{\gamma;\beta,i}}$ with eigenvalue $E^{[j]}_{\gamma;\beta,i}$. The ket $\ket{\psi_{\gamma;\beta}(x_{j};0)}$ represents the initial wavepackets as detailed in \cref{Tab:Psi_0} for each simulation.
On the contrary for the quantum computation, multiple quantum circuits ${\cal U}(n\Delta t)$ are created and these act on the initial wavepacket to arrive at the wavepacket at a certain time step. In principle this last step can also be parallelized but is not considered in this publication. 

\section{Vibrational Spectral behavior from time-evolution of any initial state}\label{Sec:Vibrational}

The computation of the shared-proton wavepacket dynamics provides a means to accurately determine its vibrational frequencies and is given by the fourier transform of the density-density time auto-correlation function, ${\text{Tr}}[\rho(0)\rho(t)]$ as
\begin{align}
    \int_{-\infty}^{+\infty} dt\, 
    e^{{\imath \omega t}} \; {\text{Tr}}[\rho(0)\rho(t)] &= \int_{-\infty}^{+\infty} dt\, e^{{\imath \omega t}}
    \; {\text{Tr}}\left[\ket{\chi(0)}\bra{\chi(0)}\sum_{i,j}c_{i}(0)c_{j}^{*}(0) 
    e^{\imath (E_{i}-E_{j}) t/{\hbar}}
    \ket{\phi_{i}}\bra{\phi_{j}}\right] \nonumber \\ 
&=\sum_{i,j}|c_{i}(0)|^{2}|c_{j}(0)|^{2} \; \delta \left(\omega-(E_{i}-E_{j})/\hbar \right),
    \label{Eq:Density-timecorrelation-FT}
\end{align}
where time-dependent density matrix is $\rho(t) = \ket{\chi(t)}\bra{\chi(t)}$. Then, one may rewrite the right side of \cref{Eq:Density-timecorrelation-FT} using convolution theorem\cite{Numerical-Recipes} as described in Ref. \onlinecite{Sandia-1D-3Qubits}, to obtain
\begin{align}
    & \int dx  {
    \left\{ 
    \left\vert 
    \int_{-\infty}^{+\infty} dt  e^{{\imath \omega t}} \;
    \sum_{i,j}c_{i}(0)c_{j}^{*}(0) 
    e^{\imath (E_{i}-E_{j}) t/{\hbar}}
    {\phi_{i}(x)}
    {\phi_{j}(x)} 
    \right\vert 
    \right\}
    }^2 +
    \nonumber \\ & 
    \int_{x\neq x^\prime} dx \,dx^\prime \left\vert \int_{-\infty}^{+\infty} dt  e^{{\imath \omega t}} 
    \sum_{i,j}c_{i}(0)c_{j}^{*}(0) 
    e^{\imath (E_{i}-E_{j}) t/{\hbar}}
    {\phi_{i}(x)}
    {\phi_{j}(x^\prime)} \right\vert^2.
    \label{Density-timecorrelation-FT-2} 
\end{align}
or
\begin{align}
     \int_{-\infty}^{+\infty} dt\, 
    e^{{\imath \omega t}} \; {\text{Tr}}[\rho(0)\rho(t)]  &= \int dx  {
    \left\{ 
    \left\vert 
    \int_{-\infty}^{+\infty} dt  e^{{\imath \omega t}} \; \rho(x,x;t)
    \right\vert 
    \right\}
    }^2 \nonumber \\ &+
    \int_{x\neq x^\prime} dx \,dx^\prime \left\vert \int_{-\infty}^{+\infty} dt  e^{{\imath \omega t}} \; \rho(x,x^\prime;t) \right\vert^2.
    \label{Eq:Density-timecorrelation-FT-2} 
\end{align}
While standard approaches may compute the results from Eq. (\ref{Density-timecorrelation-FT-2}), a quantum computer, such as the ones utilized in Ref. \onlinecite{Sandia-1D-3Qubits} and in this publication, have the ability to obtain wavepacket density information (that is the diagonal elements of the wavepacket density, $\rho(x,x;t)$) and compute the individual terms inside the first term of the integral above as, 
\begin{align}  
    {\cal I} (\omega; x) & = \int_{-\infty}^{+\infty} dt\,  e^{{\imath \omega t}} \; \rho(x,x; t) \\
    & = \int_{-\infty}^{+\infty} dt\,  e^{{\imath \omega t}} \;
    \sum_{i,j}c_{i}(0)c_{j}^{*}(0) 
    e^{\imath (E_{i}-E_{j}) t/{\hbar}}
    {\phi_{i}(x)}
    {\phi_{j}(x)} \nonumber \\
    & = \sum_{i,j} \delta(\omega-(E_{i}-E_{j})) c_{i}(0)c_{j}^{*}(0) {\phi_{i}(x)}
    {\phi_{j}(x)},
    \label{Eq:Density-timecorrelation-FT-3} 
\end{align}
which makes the first term in \cref{Eq:Density-timecorrelation-FT-2},
\begin{align}  
{\cal P} (\omega) = \int dx \; {\left\vert {\cal I} (\omega; x) \right\vert}^2
    \label{Eq:integ-Density-timecorrelation-FT-3} 
\end{align}
By measuring \cref{Eq:Density-timecorrelation-FT-3} on an ion-trap quantum computer in Ref. \onlinecite{Sandia-1D-3Qubits}, we have provided a new approach to determine spectroscopic features in complex systems using quantum computing platforms. 

However, as can be seen from \cref{Tab:Psi_0}, there are a variety of initial wavepackets used here, and hence each resultant trajectory would independently lead to the quantities in \cref{Eq:Density-timecorrelation-FT-3,Eq:integ-Density-timecorrelation-FT-3}. Thus, in the results section, we have first individual computed the quantity in \cref{Eq:Density-timecorrelation-FT-3,Eq:integ-Density-timecorrelation-FT-3} to obtain ${\cal I} (\omega; x) \rightarrow {\cal I}_{\chi_0} (\omega; x)$ and ${\cal P} (\omega) \rightarrow {\cal P}_{\chi_0} (\omega)$, where the respective quanities are now designated using the choice of initial wavepacket. We then compute the cumulative spectrum from all of these wavepackets and denote these as
\begin{align}
\sum_{\forall \chi_0} {{\cal P}_{\chi_0} (\omega)} \rightarrow {\cal P} (\omega)
\label{Eq:integ-Density-timecorrelation-FT-4} 
\end{align}
 to compute vibrational properties. Thermal weights could also be introduced at this stage but it is not done in this paper since these would only affect the peak heights and not the frequencies.

\section{Experimental implementation on a distributed set of ion-trap systems}\label{Sec:Experiment}

Experiments are performed on IonQ's 11-qubit trapped-ion quantum computer, Harmony, and accessed through the cloud via Amazon Braket \cite{wright2019benchmarking, braket}. Harmony's laser beam systems can separately address each ion, or pairs of ions, enabling arbitrary single-qubit gate rotations and all-to-all two-qubit gate connectivity. The wavepacket dynamics are simulated and measured on Harmony with 1000 repetitions for each time point. Quantum circuits are executed in parallel such that the maximum number of qubits are in use. Finally, a classical Fourier transform of the data is used to reveal the frequency components of the vibrational motion using the formalism mentioned in \cref{Sec:Vibrational}.

Quantum circuits are compiled directly from the target unitary operators. This approach reduces error accumulation due to continuous analog simulation or due to Trotterization errors which may degrade the spectral signal in the frequency domain. Quantum circuits have the same gate layout for a given Hamiltonian, but vary the single-qubit rotation angles to simulate evolution for a given time. The largest contribution to errors comes from two-qubit gate operations (largest by an order of magnitude) across all time points. Variations in error due to single-qubit gate rotations are negligible. Since we aim to measure vibrational spectra, time-varying errors in the dynamics are more impactful than constant-amplitude errors. We observe that in our compilation strategy, two-qubit gate errors lead to a notable loss in oscillation amplitudes, but the frequency domain spectra remain robust.

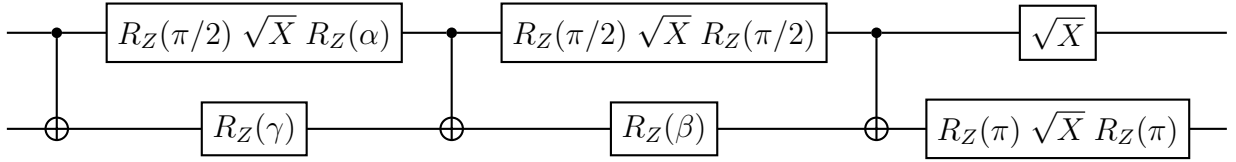
\begin{figure}
    \centering
    $
    \begin{quantikz}
    \lstick{} & \ctrl{1} & \gate{R_Z(\pi/2) \; \sqrt{X} \; R_Z(\alpha)} & \ctrl{1} & \gate{R_Z(\pi/2) \; \sqrt{X} \; R_Z(\pi/2)} & \ctrl{1} & \gate{\sqrt{X}} & \qw \\
    \lstick{} & \targ{} & \gate{R_Z(\gamma)} & \targ{} & \gate{R_Z(\beta)} & \targ{} & \gate{R_Z(\pi) \; \sqrt{X} \; R_Z(\pi)} & \qw\\
    \end{quantikz}
    $
    \caption{General quantum circuit diagram for simulating the dynamics of the two-qubit block Hamiltonians ($x_1$-upper/lower and $x_2$-upper/lower). The parameters $\gamma$, $\beta$, and $\alpha$ are varied per time point.}
    \label{fig:two-qubit-circuit}
\end{figure}
The two-qubit unitary time evolution blocks of the $x_1$-upper/lower and $x_2$-upper/lower Hamiltonians are compiled into quantum circuits of CNOT gates and single-qubit rotations using Cartan's $KAK$ decomposition, which guarantees a maximum of three two-qubit gates given our system size \cite{tucci2005introduction, earnest2021pulse}. As mentioned previously, the single-qubit rotations encode the time point values. A generalized quantum circuit diagram is shown in Fig. \ref{fig:two-qubit-circuit}. Lastly, the circuit is compiled into the native gate basis supported by Harmony as described in Ref. \onlinecite{wright2019benchmarking}. A two-qubit evolution for any two-qubit block may look like the following circuit where single-qubit gate parameters $\gamma$, $\beta$, and $\alpha$ depend on the evolution time $t$\cite{earnest2021pulse}.

\begin{figure}
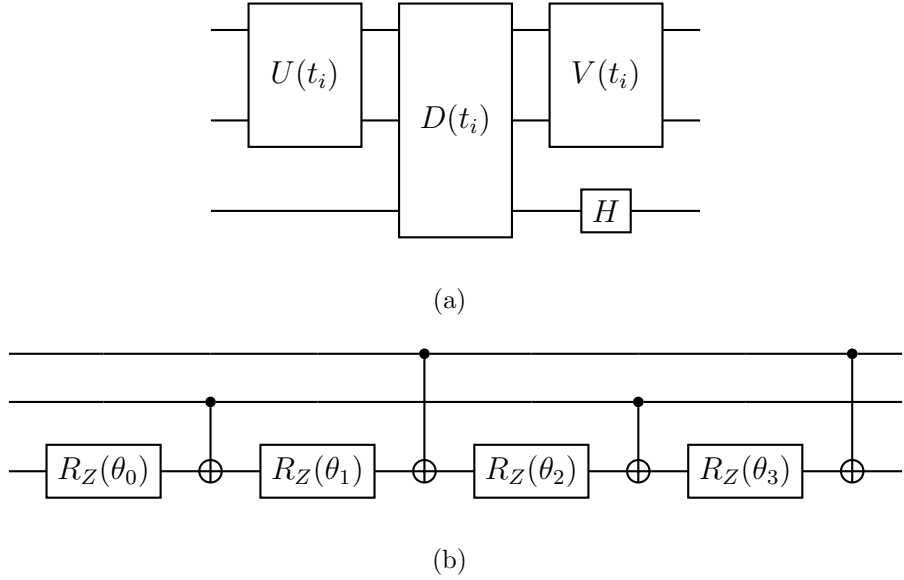

    \centering
    \subfigure[]{\input{Fig_UDU}}
    \subfigure[]{\input{Fig_RzCXRzCXRzCXRzCX}}
    \caption{(a) The general quantum circuit diagram for simulating the full three-qubit block Hamiltonian dynamics ($x_1$-full and $x_2$-full). State preparation and measurement are not shown. The additional Hadamard gate maps the basis back to the grid basis as described in the text. (b) A generalized quantum circuit diagram for the diagonal unitary $D(t_i)$.}
    \label{fig:three-qubit-circuit}
\end{figure}

The three-qubit unitary time evolution of the $x_1$-full and $x_2$-full Hamiltonians is compiled as follows, with a general quantum circuit diagram shown in \cref{fig:three-qubit-circuit}. First, the Single Value Decomposition (SVD) separates the unitary operator of the given Hamiltonian at a specified time point $t_i$ into three parts $U(t_i)$, $D(t_i)$, and $V(t_i)$ as given by 
\begin{equation}
    \begin{pmatrix}U_0 & 0 \\ 0 & U_1\end{pmatrix} = \begin{pmatrix} U & 0 \\ 0 & U \end{pmatrix} \begin{pmatrix} D & 0 \\ 0 & D^{\dagger} \end{pmatrix} \begin{pmatrix} V & 0 \\ 0 & V \end{pmatrix}
\end{equation}
where $U_0$ and $U_1$ are the known two-qubit upper and lower blocks of the unitary evolution and $0$'s are $4 \times 4$ blocks. Since $U_0 = UDV$ and $U_1 =  UD^{\dagger}V$,
\begin{equation}
    U_0 U_1^{\dagger} = U D^2 U^{\dagger} \;.
\end{equation}
From the eigenvalue decomposition of $U_0 U_1^{\dagger}$, we can solve for $U$ and $D$. We find $V$ using the relation $V = D^{-1} U^{-1} U_0$ defined earlier. Next, these matrices are complied as elements of the quantum circuit. Cartan's $KAK$ decomposition is used for the two-qubit $U(t_i)$ and $V(t_i)$ unitaries where each contributes at most three two-qubit gates and some single-qubit rotations. The remaining unitary $D(t_i)$ is diagonal which is decomposed into four two-qubit gates and four single-qubit $R_Z(\theta)$ rotations as shown in Fig. \ref{fig:three-qubit-circuit}(b) \cite{shende2005synthesis}.

\begin{table}
    \centering
    \begin{tabular*}{\textwidth}{@{\extracolsep{\fill}} ccc } 
     \hline\hline
     \; Computation \; & \; Givens-transformed grid \; & Shuffled \\
     \hline
     \vspace*{2mm}
     $\ket{000}$ & $(\vert x^0\rangle + \ket{x^7})/\sqrt{2}$ & \; $(\vert x^0\rangle + \ket{x^7})/\sqrt{2}$ \; \\ \vspace*{2mm}
     $\ket{001}$ & $(\vert x^1\rangle + \ket{x^6})/\sqrt{2}$ & \; $(\vert x^1\rangle + \ket{x^6})/\sqrt{2}$ \; \\ \vspace*{2mm}
     $\ket{010}$ & $(\vert x^2\rangle + \ket{x^5})/\sqrt{2}$ & \; $(\vert x^2\rangle + \ket{x^5})/\sqrt{2}$ \; \\ \vspace*{2mm}
     $\ket{011}$ & $(\vert x^3\rangle + \ket{x^4})/\sqrt{2}$ & \; $(\vert x^3\rangle + \ket{x^4})/\sqrt{2}$ \; \\ \vspace*{2mm}
     $\ket{100}$ & $(\vert x^4\rangle - \ket{x^3})/\sqrt{2}$ & \; $(\vert x^7\rangle - \ket{x^0})/\sqrt{2}$ \; \\ \vspace*{2mm}
     $\ket{101}$ & $(\vert x^5\rangle - \ket{x^2})/\sqrt{2}$ & \; $(\vert x^6\rangle - \ket{x^1})/\sqrt{2}$ \; \\ \vspace*{2mm}
     $\ket{110}$ & $(\vert x^6\rangle - \ket{x^1})/\sqrt{2}$ & \; $(\vert x^5\rangle - \ket{x^2})/\sqrt{2}$ \; \\ \vspace*{2mm}
     $\ket{111}$ & $(\vert x^7\rangle - \ket{x^0})/\sqrt{2}$ & \; $(\vert x^4\rangle - \ket{x^3})/\sqrt{2}$ \; \\
     \hline\hline
    \end{tabular*}
    \caption{Enumeration of the different bases. The computation basis are the bitstrings measured in the experiment. The Givens-transformed grid basis is in terms of the grid basis states $\ket{x^i}$. The shuffled basis is used experimentally to reduce the quantum circuit. It is similar to the Givens-transformed grid basis with the eighth and fifth states swapped and the seventh and sixth states swapped.}
    \label{tab:experiment-transform-bases}
\end{table}

With this procedure of generating a quantum circuit given one of the three-qubit Hamiltonians ($x_1$-full or $x_2$-full), the dynamics can be simulated in the grid basis on the trapped-ion quantum computer. The final step is to append a series of gates which map the wavepacket onto the grid basis. We hand-optimize this circuit by adding just a single Hadamard gate by shuffling the lower block of states: the eighth and fifth states are switched, as are the seventh and sixth states, as shown in the third column of Table \ref{tab:experiment-transform-bases}.
For example, a Hadamard acting on the second qubit of the $\ket{000}$ state, in the shuffled basis, gives a one-to-one mapping between the computation basis and the grid basis $\ket{x^i}$:
\begin{align}
    H \ket{000} &= (\vert 000\rangle + \ket{100})/\sqrt{2} \\
     & = (\vert x^0\rangle + \ket{x^7} + \ket{x^7} - \ket{x^0})) / 2 \\
     & = \ket{x^7}
\end{align}
This generalizes to all states. Working in the shuffled basis has an additional simplification for state preparation. The states of interest in the transformed basis are entangled states which require two-qubit gates. However, in the shuffled basis, the initial states are no longer entangled states and single-qubit rotations can prepare them.

\section{Comparison of Wavepacket time-evolution for all initial wavepackets from quantum and classical simulations}\label{Sec:dynamics}
\begin{figure}[t]
    \centering
    \subfigure[Quantum (blue) and classical (yellow) time-traces for $\ket{x_1}$ for initial wavepacket $\Psi_0(x_1) = \delta(x_1-x^1_1)$]{\label{Fig:prob-x-upper-00}
        \includegraphics[width=0.35\textwidth]{Figures_ionq/Propagation_plots/4x4_x_upper/x_upper_00_2.5_400-lower.png}}
    \subfigure[Quantum (blue) and classical (yellow) time-traces for $\ket{x_1}$ for initial wavepacket $\Psi_0(x_1) = \delta(x_1-x^2_1)$]{\label{Fig:prob-x-upper-01}
        \includegraphics[width=0.35\textwidth]{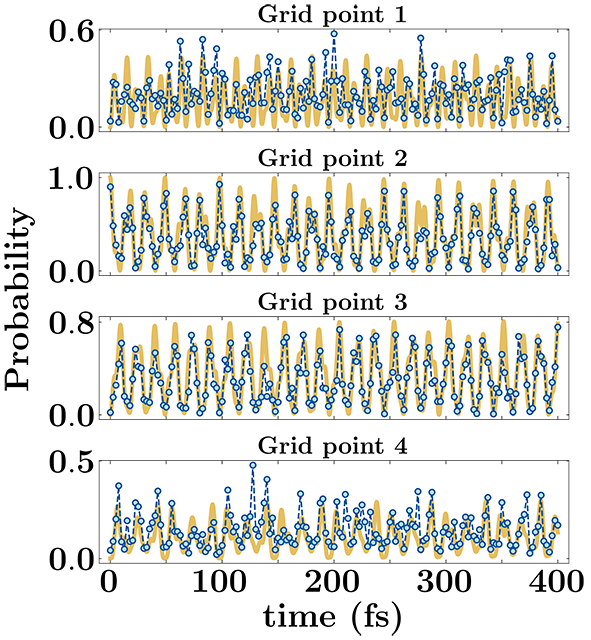}}\\
    \subfigure[Quantum (blue) and classical (yellow) time-traces for $\ket{x_1}$ for initial wavepacket $\Psi_0(x_1) = \delta(x_1-x^3_1)$]{\label{Fig:prob-x-upper-10}
        \includegraphics[width=0.35\textwidth]{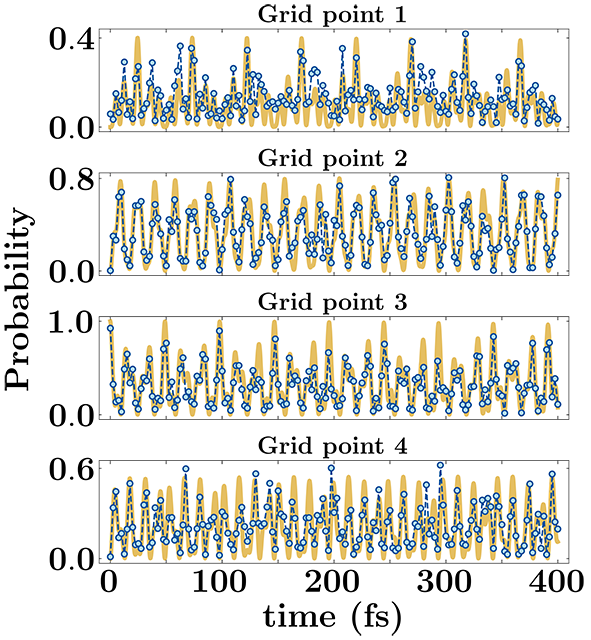}}
    \subfigure[Quantum (blue) and classical (yellow) time-traces for $\ket{x_1}$ for initial wavepacket $\Psi_0(x_1) = \delta(x_1-x^4_1)$]{\label{Fig:prob-x-upper-11}
        \includegraphics[width=0.35\textwidth]{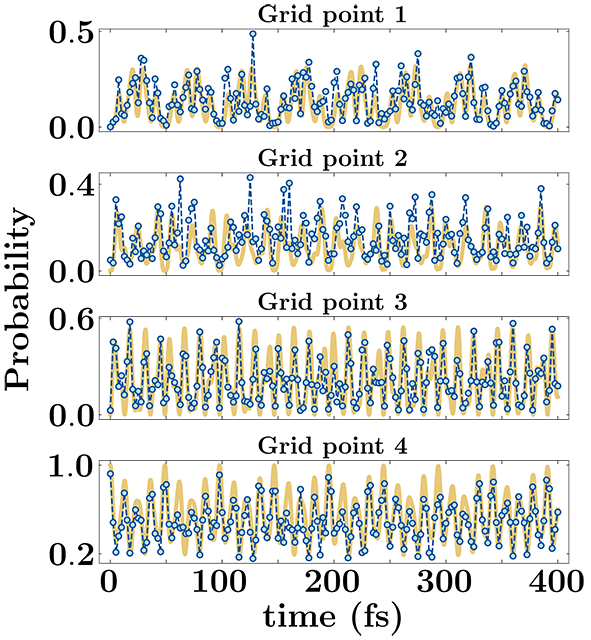}}    \caption{Illustrating the coherent evolution of the wavepacket pertaining to the upper block of the Hamiltonian along the $x_1$ direction, encapsulating the vibrational dynamics along the internuclear axis connecting the two nitrogen atoms within the molecular structure depicted in \cref{Fig:Molecule}. The temporal evolutions depicted in each subfigure is initialized at distinct grid points as delineated in the subfigure captions, with each time-traces projected onto specific grid points denoted in the panel titles. Solid yellow lines denote the exact numerical solution to the Schrödinger equation, calculated classically, while the blue curve represents outcomes derived from ion-trap dynamics.}
    \label{Fig:x-upper}
\end{figure}
This section compares the quantum dynamics trajectories obtained using the classical and quantum hardware for  
the initial wavepackets provided in \cref{Tab:Psi_0}. As shown in Figs. \ref{Fig:MPS_tEvo_trotter}(c) and \ref{Fig:MPS_tEvo_trotter-d}, the quantum dynamics simulation for each of the effective one-dimensional Hamiltonians, $H^{[j]}_{\gamma,\beta}$, has been conducted independently. Since these one-dimensional Hamiltonians can be further expressed in a block-diagonal form, as discussed in \cref{Sec:Block-diagonal}, each of these diagonal sub-blocks is simulated independently on both quantum and classical computers. The information regarding simulation time-steps and total propagation time is delineated in \cref{Tab:Simulation}.  

\begin{figure}[t]
 	\centering
    \subfigure[Quantum (blue) and classical (yellow) time-traces for $\ket{x_1}$ for initial wavepacket $\Psi_0(x_1) = \delta(x_1-x^5_1)$]{\label{Fig:prob-x-lower-00}
    \includegraphics[width=0.35\textwidth]{Figures_ionq/Propagation_plots/4x4_x_lower/x_lower_00_2.5_400-lower.png}}
    \subfigure[Quantum (blue) and classical (yellow) time-traces for $\ket{x_1}$ for initial wavepacket $\Psi_0(x_1) = \delta(x_1-x^6_1)$]{\label{Fig:prob-x-lower-01}
    \includegraphics[width=0.35\textwidth]{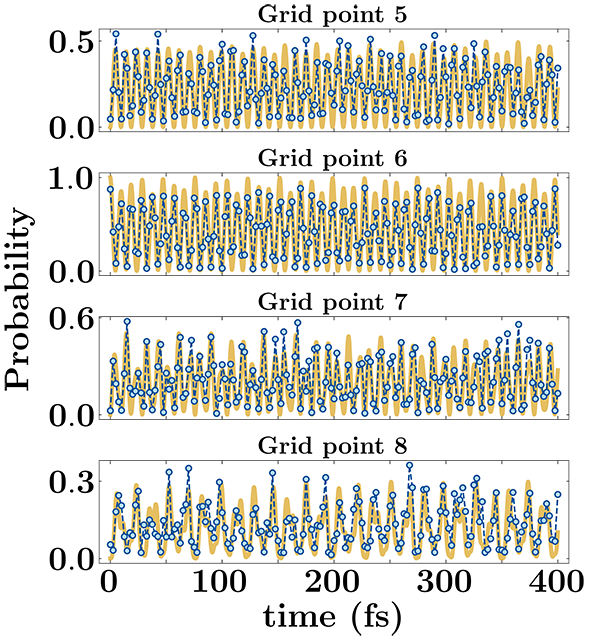}}\\
    \subfigure[Quantum (blue) and classical (yellow) time-traces for $\ket{x_1}$ for initial wavepacket $\Psi_0(x_1) = \delta(x_1-x^7_1)$]{\label{Fig:prob-x-lower-10}
    \includegraphics[width=0.35\textwidth]{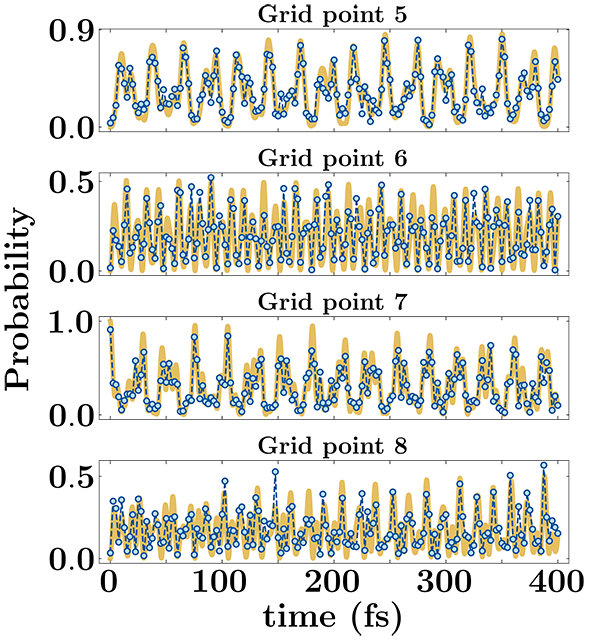}}
    \subfigure[Quantum (blue) and classical (yellow) time-traces for $\ket{x_1}$ for initial wavepacket $\Psi_0(x_1) = \delta(x_1-x^8_1)$]{\label{Fig:prob-x-lower-11}
    \includegraphics[width=0.35\textwidth]{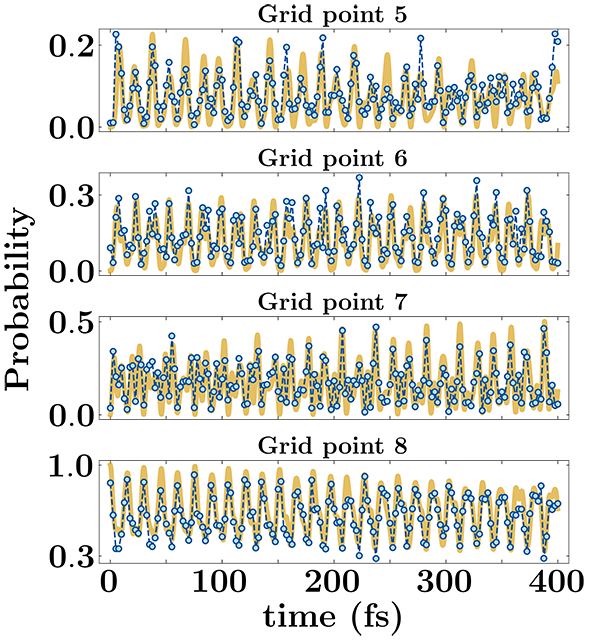}}
  	\caption{Showcasing the coherent evolution of the wavepacket associated with the lower block of the Hamiltonian along the $x_1$ direction, capturing the vibrational dynamics along the internuclear axis linking the two nitrogen atoms within the molecular structure portrayed in \cref{Fig:Molecule}. The temporal evolution depicted in each subfigure is initiated from distinct grid points, as specified in the subfigure captions, with each time-trace projected onto specific grid points indicated in the panel titles.}
   \label{Fig:x-lower}
\end{figure}

\begin{figure}[t]
 	\centering
    \subfigure[Quantum (blue) and classical (yellow) time-traces for $\ket{x_1}$ for initial wavepacket $\Psi_0(x_1) = \delta(x_1-x^1_1)$]{\label{Fig:prob-x-full-000-111}
    \includegraphics[width=0.70\textwidth]{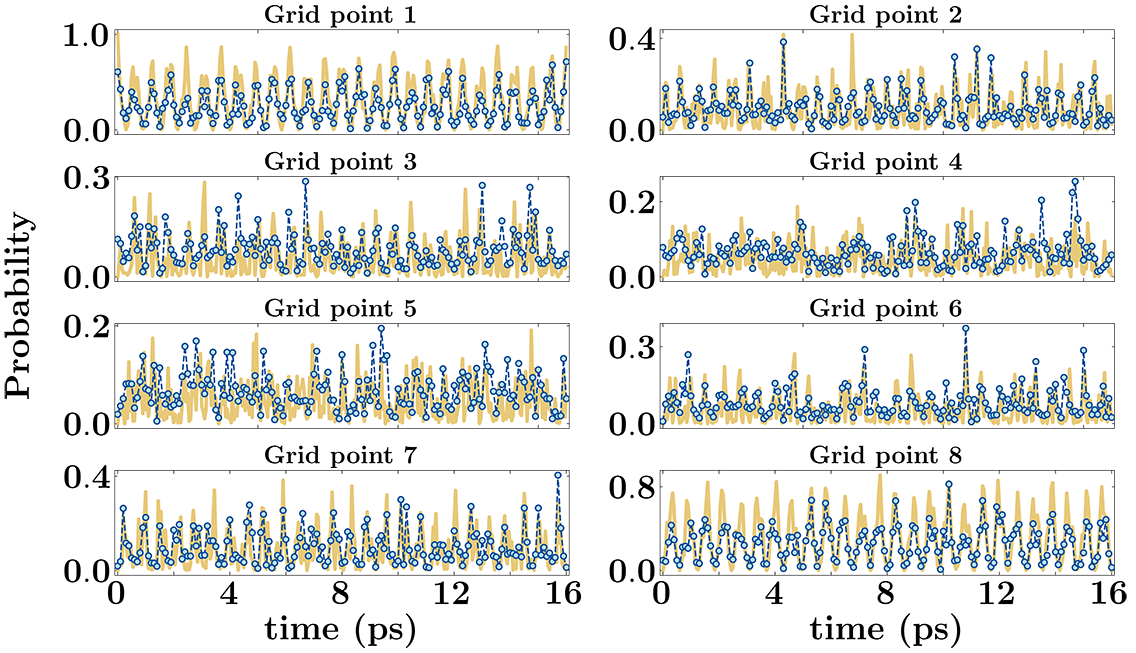}}\\
    \subfigure[Quantum (blue) and classical (yellow) time-traces for $\ket{x_1}$ for initial wavepacket $\Psi_0(x_1) = \delta(x_1-x^2_1)$]{\label{Fig:prob-x-full-001-110}
    \includegraphics[width=0.70\textwidth]{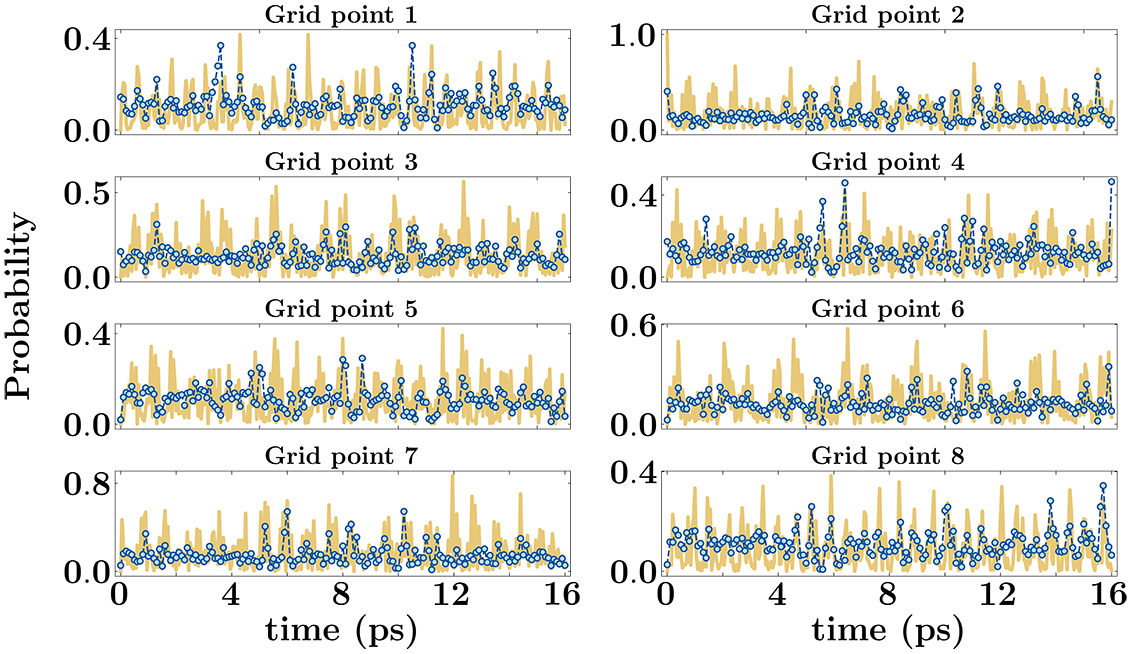}}
    \caption{The coherent evolution of the wavepacket associated with the full $8\times8$ Hamiltonian along the $x_1$ direction as shown in the molecular structure in \cref{Fig:Molecule}. The temporal evolution depicted in Figure (a) is initialized at the first grid point while the wavepacket in Figure (b) is initialized at the second grid point. The time-traces in each panel correspond to the time-evolution of the wavepacket projected onto specific grid points indicated in the panel titles.}
    \label{Fig:x-full}
\end{figure}

\begin{figure}[t]
 	\centering
    \subfigure[Quantum (blue) and classical (yellow) time-traces for $\ket{x_2}$ for initial wavepacket $\Psi_0(x_2) = \delta(x_2-x^1_2)$]{\label{Fig:prob-theta-upper-00}
    \includegraphics[width=0.35\textwidth]{Figures_ionq/Propagation_plots/4x4_theta_upper/theta_upper_00_1.47fs_235fs-lower.png}}
    \subfigure[Quantum (blue) and classical (yellow) time-traces for $\ket{x_2}$ for initial wavepacket $\Psi_0(x_2) = \delta(x_2-x^2_2)$]{\label{Fig:prob-theta-upper-01}
    \includegraphics[width=0.35\textwidth]{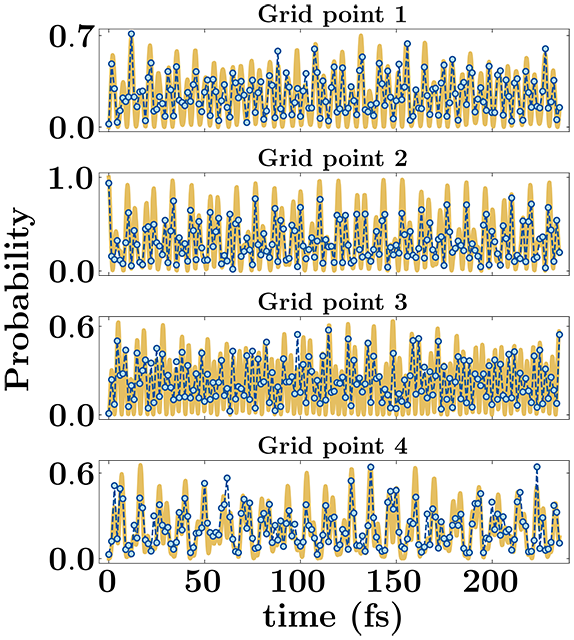}}
    \caption{The figures illustrate the temporal evolution of the wavepacket associated with the upper block of the Hamiltonian, focusing on the $x_2$ direction, capturing the torsional degree of freedom arising from planar rotations of the pyridyl rings about the C-C bond within the molecular architecture (refer to \cref{Fig:Molecule}). The temporal evolutions depicted in each subfigure is initialized at distinct grid points as delineated in the subfigure captions, with each time-traces projected onto specific grid points indicated in the panel titles.}
    \label{Fig:theta-upper}
\end{figure}

\begin{figure}[t]
 	\centering
    \subfigure[Quantum (blue) and classical (yellow) time-traces for $\ket{x_2}$ for initial wavepacket $\Psi_0(x_2) = \delta(x_2-x^6_2)$]{\label{Fig:prob-theta-lower-01}
    \includegraphics[width=0.30\textwidth]{Figures_ionq/Propagation_plots/4x4_theta_lower/theta_lower_01_1.47fs_235fs-lower.png}}
    \subfigure[Quantum (blue) and classical (yellow) time-traces for $\ket{x_2}$ for initial wavepacket $\Psi_0(x_2) = \delta(x_2-x^7_2)$]{\label{Fig:prob-theta-lower-10}
    \includegraphics[width=0.30\textwidth]{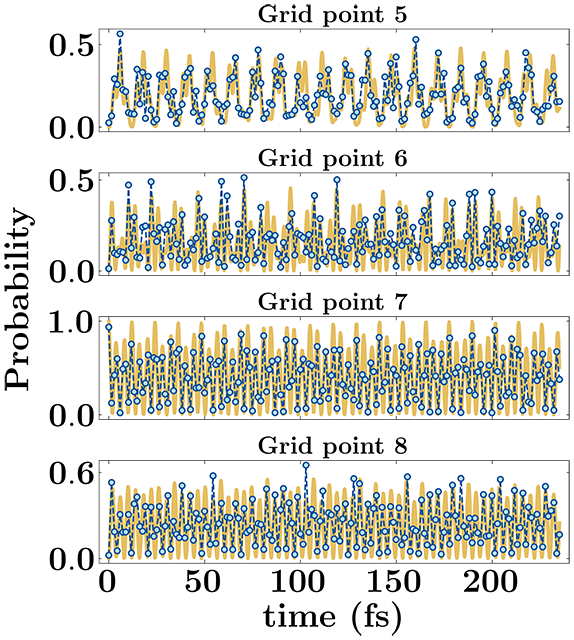}}
    \subfigure[Quantum (blue) and classical (yellow) time-traces for $\ket{x_2}$ for initial wavepacket $\Psi_0(x_2) = \delta(x_2-x^8_2)$]{\label{Fig:prob-theta-lower-11}
    \includegraphics[width=0.30\textwidth]{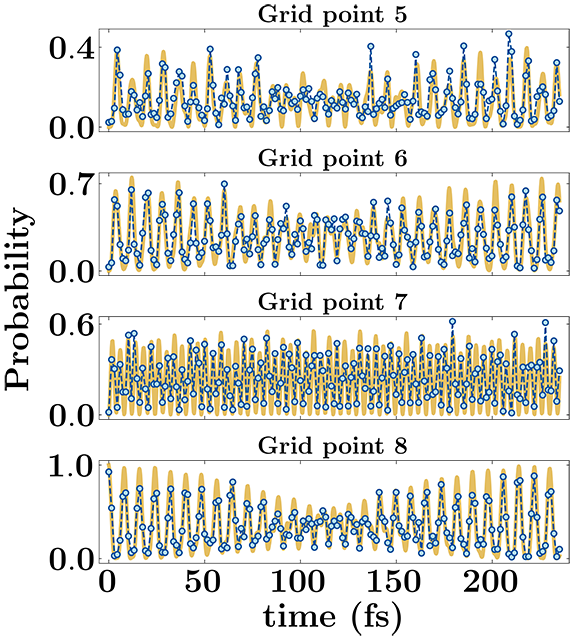}}  	\caption{The time-evolution of the wavepacket linked with the lower block of the Hamiltonian, emphasizing the torsional degree of freedom along the $x_2$ direction within the molecular configuration illustrated in \cref{Fig:Molecule}. The temporal evolution depicted in each subfigure is initiated from distinct grid points, as specified in the subfigure captions, Additionally, each time-trace is projected onto specific grid points as indicated in the panel titles.}
    \label{Fig:theta-lower}
\end{figure}

\cref{Fig:x-upper,Fig:x-lower,Fig:x-full,Fig:theta-upper,Fig:theta-lower} complement figure 5 in the paper and provide a complete picture of our study. The \cref{Fig:x-upper} shows the dynamics for the upper-block of the block diagonal Hamiltonian, $\tilde{H}^{[1]}_{\gamma,\beta}$, for the $x_1$ dimension, encoding the proton vibration along the donor-acceptor axis connecting the two nitrogen atoms in the structure shown in \cref{Fig:molecule_pes}(a). Specifically, \cref{Fig:x-upper}(a) presents the projection of the time-traces at distinct grid points ($x^1_1, x^2_1, x^3_1, x^4_1$) in 4 distinct panels. Note on notation: while the subscripts here represents dimensions, the superscripts refers to grid points. Hence, the four time traces in \cref{Fig:x-upper} refer to the time evolution of various nuclear wavepacket on each of the four grid points. The blue time dots in \cref{Fig:x-upper} are obtained from measurements on the ion-trap quantum computer and the yellow traces are from classical computation. Here, the initial wavepacket is localized at grid point `1' and has a zero value at all other grid points, that is, $\Psi_0(x_1) = \delta(x_1-x^1_1)$. Similarly, \cref{Fig:x-upper}(b-d) shows the time-traces when the initial wavepackets are $\Psi_0(x_1) = \delta(x_1-x^2_1)$, $\Psi_0(x_1) = \delta(x_1-x^3_1)$ and $\Psi_0(x_1) = \delta(x_1-x^4_1)$ respectively. 

The dynamics of the lower block of the block diagonal Hamiltonian $\tilde{H}^{[1]}_{\gamma,\beta}$, for the $x_1$ dimension, are illustrated in fig. \ref{Fig:x-lower}. Specifically, \cref{Fig:x-lower}(a) presents the projection of the time-traces at distinct grid points ($x^5_1, x^6_1, x^7_1, x^8_1$) in four distinct panels. Note: Grid points 5-8 appear on the lower block of the Hamiltonian. Here, the initial wavepacket is localized at grid point '5' and has a zero value at all other grid points, that is, $\Psi_0(x_1) = \delta(x_1-x^5_1)$. Similarly, \cref{Fig:x-lower}(b-d) show the time-traces when the initial wavepackets are $\Psi_0(x_1) = \delta(x_1-x^6_1)$, $\Psi_0(x_1) = \delta(x_1-x^7_1)$ and $\Psi_0(x_1) = \delta(x_1-x^8_1)$ respectively. Once again, solid yellow lines denote the exact numerical solution to the Schrödinger equation, calculated classically, while the blue dots represents outcomes derived from ion-trap quantum computation. These $4\times4$ upper and lower block of the one-dimensional Hamiltonian $\tilde{H}^{[1]}_{\gamma,\beta}$ are simulated on a 2-qubit ion-trap systems. Detailed information regarding the time-steps and total propagation time for each of the time-traces can be found in \cref{Tab:Simulation}. 

As noted above, the symmetry of the potential allows the block diagonalization of the effective one-dimensional Hamiltonians. Time-evolution of the separate blocks along with the their respective Fourier transforms (discussed in the next section), yield information regarding the vibrational states for the system being studied. However, one of the critical features of Fourier analysis provided in \cref{Sec:Vibrational} is that only eigenvalue differences are obtained from the Fourier transform. This in general does not present a problem since most chemical spectroscopy experiments also yield eigenvalue differences. However, here, obtaining eigenvalue differences for each of the Hamiltonian blocks separately implies that the overall spectrum of the system is not accessible. Hence, to obtain the relative placements of the eigenvalues of the separate blocks, we also compute here the time-evolution of the system arising from the non-block-diagonalized Hamiltonian. The resultant quantum dynamics simulation of the $8\times8$ full-Hamiltonian, $\tilde{H}^{[1]}_{\gamma,\beta}$, has been conducted on a 3-qubit ion-trap system. As noted, this simulation aims to determine the relative energy differences between the eigenstates obtained from the Fourier transform of the dynamics simulation of the upper and lower blocks of the block-diagonal Hamiltonian. In \cref{Fig:x-full}, we illustrate the quantum dynamics of the Hamiltonian, $\tilde{H}^{[1]}_{\gamma,\beta}$, for two different initial wavepackets: $\Psi_0(x_1) = \delta(x_1-x^1_1)$ and $\Psi_0(x_1) = \delta(x_1-x^2_1)$,
as shown in \cref{Fig:x-full}(a-b) respectively. Each of these figures presents the projection of the time-traces at distinct grid points ($x^1_1, x^2_1, x^3_1, x^4_1, x^5_1, x^6_1, x^7_1, x^8_1$) in eight distinct panels. 

\cref{Fig:theta-upper} illustrates the dynamics for the upper block of the block-diagonal Hamiltonian $\tilde{H}^{[2]}_{\gamma,\beta}$, along the $x_2$ direction. This encodes the torsional degree of freedom that acts here as a gating mode and regulates proton transfer (larger angles have larger donor-acceptor distances). The gating mode originates here from planar rotations of the pyridyl rings about the C-C bond within the molecular structure depicted in \cref{Fig:molecule_pes}(a). Specifically, \cref{Fig:theta-upper}(a) displays the projection of the time-traces at distinct grid points ($x^1_2, x^2_2, x^3_2, x^4_2$) in four distinct panels. In these panels, the initial wavepacket is localized at grid point '1' and has a zero value at all other grid points, defined as $\Psi_0(x_2) = \delta(x_2-x^1_2)$. Similarly, \cref{Fig:theta-upper}(b) presents the time-traces when the initial wavepacket is $\Psi_0(x_2) = \delta(x_2-x^1_2)$. In both figures, solid yellow lines represent the dynamics simulated classically through the exact numerical solution to the Schrödinger equation, while the blue dots indicate outcomes derived from ion-trap dynamics.

The dynamics of the $4\times4$ lower block of the block diagonal Hamiltonian $\tilde{H}^{[2]}_{\gamma,\beta}$ is simulated on a two-qubit ion-trap system. 
Specifically, \cref{Fig:theta-lower}(a) presents the projection of the time-traces at distinct grid points ($x^5_1, x^6_1, x^7_1, x^8_1$) in four distinct panels. Here, the initial wavepacket is localized at grid point '6' and has a zero value at all other grid points, that is, $\Psi_0(x_2) = \delta(x_2-x^6_2)$. Similarly, \cref{Fig:theta-lower}(b-c) depict the time-traces when the initial wavepackets are $\Psi_0(x_2) = \delta(x_2-x^7_2)$ and $\Psi_0(x_2) = \delta(x_2-x^8_2)$ respectively.

\begin{table}[]
    \centering
    \begin{tabular*}{\textwidth}{@{\extracolsep{\fill}}ccc}
        \hline \hline
        \vspace*{2mm}
        Simulation type & Initial Wavepacket & Wavepacket Error \\
        \hline
        \vspace*{2mm}
        \multirow{4}{*}{$x_1$-upper block} & $\delta(x_1-x_1^1)$ & 0.278\\
        \vspace*{2mm}
        & $\delta(x_1-x_1^2)$ & 0.200\\
        \vspace*{2mm}
        & $\delta(x_1-x_1^3)$ & 0.192\\
        \vspace*{2mm}
        & $\delta(x_1-x_1^4)$ & 0.264\\
        \hline
        \vspace*{2mm}
        \multirow{4}{*}{$x_1$-lower block} & $\delta(x_1-x_1^5)$ & 0.221\\
        \vspace*{2mm}
        & $\delta(x_1-x_1^6)$ & 0.162\\
        \vspace*{2mm}
        & $\delta(x_1-x_1^7)$ & 0.236\\
        \vspace*{2mm}
        & $\delta(x_1-x_1^8)$ & 0.232\\
        \hline
        \vspace*{2mm}
        \multirow{2}{*}{$x_2$-upper block} & $\delta(x_2-x_2^1)$ & 0.199\\
        \vspace*{2mm}
        & $\delta(x_2-x_2^2)$ & 0.233\\
        \hline
        \vspace*{2mm}
        \multirow{3}{*}{$x_2$-lower block} & $\delta(x_2-x_2^6)$ & 0.226\\
        \vspace*{2mm}
        & $\delta(x_2-x_2^7)$ & 0.188\\
        \vspace*{2mm}
        & $\delta(x_2-x_2^8)$ & 0.222\\
        \hline
        \vspace*{2mm}
        \multirow{2}{*}{$x_1$-full} & $\delta(x_1-x_1^1)$ & 0.080\\
        \vspace*{2mm}
        & $\delta(x_1-x_1^2)$ & 0.090\\
        \hline
        \vspace*{2mm}
        $x_2$-full & $\delta(x_2-x_2^1)$ & 0.077\\
        \hline \hline
    \end{tabular*}
    \caption{Wavepacket Error}
    \label{Tab:Wavepacket Error}
\end{table}
\begin{figure}
  \includegraphics[width=0.7\textwidth]{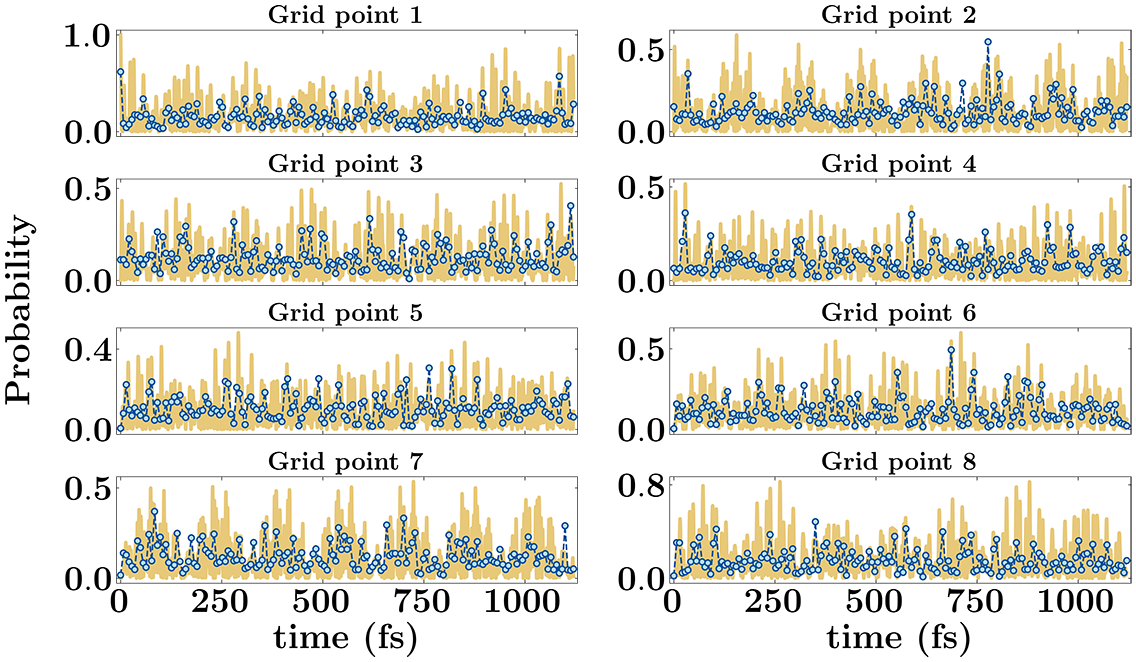}
  \caption{The figure illustrates the quantum dynamics associated with the torsional degree of freedom ($x_2$), arising from planar rotations of the pyridyl rings around the C-C bond. The initial wavepacket is $\Psi_0(x_2) = \delta(x_2-x^1_2)$, and the projection onto individual grid points is presented separately in each panel.}
  \label{Fig:prob-theta-full-000-111}
\end{figure}
The quantum dynamics simulation of the $8\times8$ full Hamiltonian, $\tilde{H}^{[2]}_{\gamma,\beta}$, has been executed on a 3-qubit ion-trap system. This simulation is aimed at determining the relative energy differences between the eigenstates derived from the Fourier transform of the dynamics simulations of the upper and lower blocks of the block-diagonal Hamiltonian. In \cref{Fig:prob-theta-full-000-111}, we present the quantum dynamics of the Hamiltonian $\tilde{H}^{[2]}_{\gamma,\beta}$ for the initial wavepacket $\Psi_0(x_2) = \delta(x_2-x^1_2)$. It showcases the projection of the time-traces at distinct grid points ($x^1_2, x^2_2, x^3_2, x^4_2, x^5_2, x^6_2, x^7_2, x^8_2$) in eight distinct panels. 
While the dynamics' trends are effectively conveyed in the figures presented in \cref{Fig:x-upper,Fig:x-lower,Fig:x-full,Fig:theta-upper,Fig:theta-lower}, we further quantify the numerical difference between the quantum and classical simulations as follows:
\begin{align}
    \Delta\Psi = \sqrt{\frac{1}{NT}\int_0^{T}\dd{t}\int_0^{N} \dd{x_j} \|\psi_{\text{Classical}}(x_j;t)- \psi_{\text{Quantum}}(x_j;t)\|^2},
\label{Eq:T-Ave_Error}
\end{align} 
Here, $\psi_{\text{Quantum}}$ represents the wavepacket at time, $t$, obtained from the ion-trap, while $\psi_{\text{Classical}}$ signifies the corresponding wavepacket derived from exact diagonalization results. The variables $T$ and $N$ denote the total propagation time and the size of the Hamiltonian involved in simulating each Hamiltonian block, respectively. These errors between $\psi_{\text{Quantum}}$ and $\psi_{\text{Classical}}$ for each simulation are summarized in \cref{Tab:Wavepacket Error}.
As noted, the error is of the order of $10-20\%$ loss in probability and this is consistent with that seen in Ref. \onlinecite{Sandia-1D-3Qubits}. However, as will be seen below,  and as already noted in Ref. \onlinecite{Sandia-1D-3Qubits}, the spectral features are in greater agreement since the error in quantum computation appears to not be frequency dependent.
\section{Comparison of spectral features obtained from the Wavepacket time-evolution}\label{Sec:spectrum}
\begin{figure}[t]
 	\centering
    \includegraphics[width=0.75\textwidth]{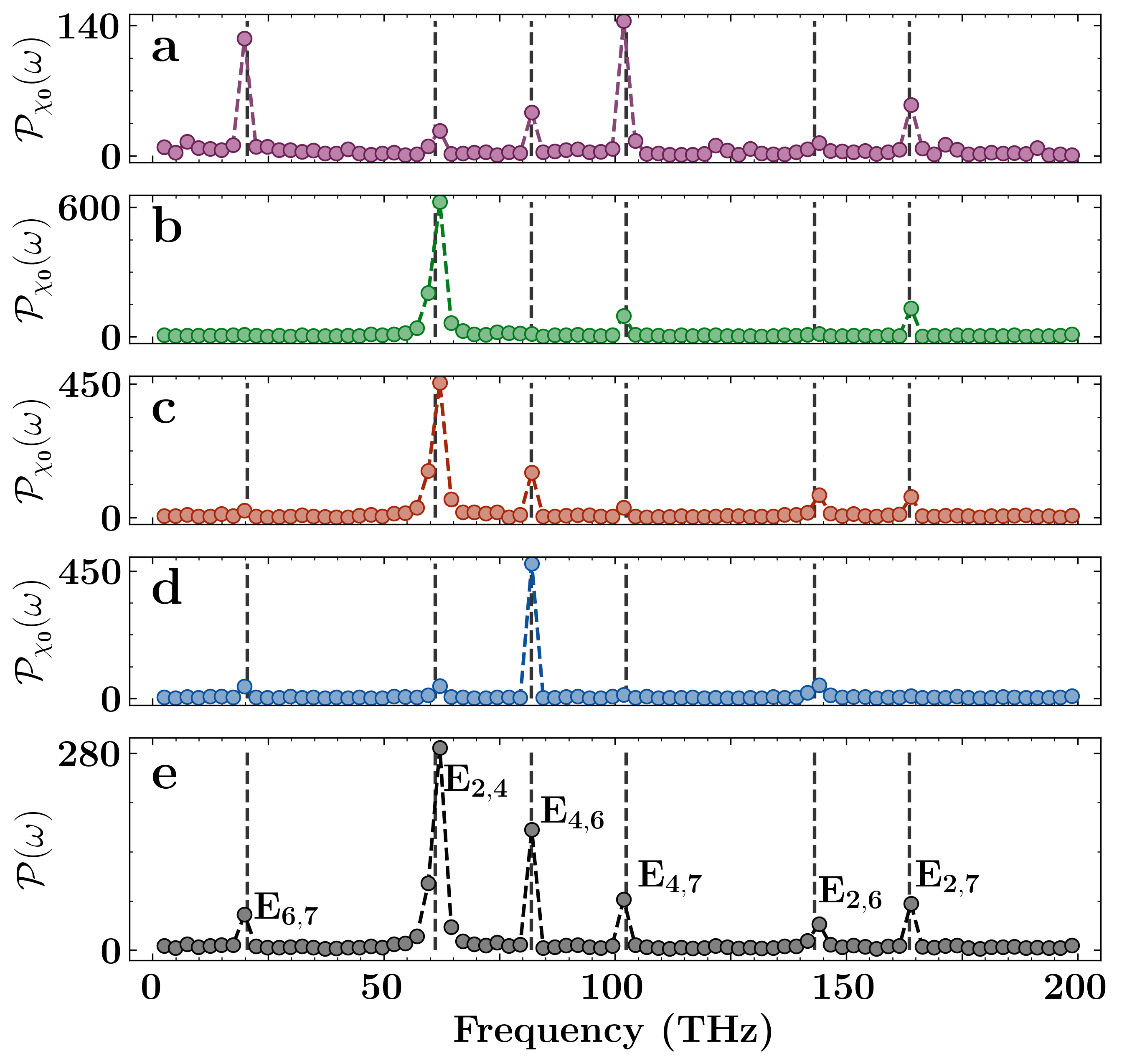}
    \caption{The figure depicts the experimentally determined relative energy separation among the four eigenstates associated with the upper block of the Hamiltonian, which corresponds to the vibrational degree of freedom involving proton transfer along the internuclear axis ($x_1$) connecting the two nitrogen atoms in protonated 2,2'-bipyridine. These separations are obtained through the Fourier transform of the time evolution of the upper block of the Hamiltonian, considering four distinct initial states as detailed in \cref{Tab:Psi_0}. Panel `a' showcases ${\cal P}_{\chi_0}(\omega)$, mentioned in \cref{Eq:integ-Density-timecorrelation-FT-3}, resulting from the quantum propagation of the upper block, as depicted in \cref{Fig:x-upper}(a), with the initial wavepacket $\Psi_0^1=\delta(x_1-x_1^1)$ projected onto four different grid points $\left\{x_1^1,x_1^2,x_1^3,x_1^4\right\}$. Similarly, panels `b', `c', and `d' display the Fourier transforms derived from the quantum propagation (see \cref{Fig:x-upper}(b-d)) of the upper block of the Hamiltonian along the $x_1$-direction, each initialized with wavepackets $\Psi_0^2=\delta(x_1-x_1^2)$, $\Psi_0^3=\delta(x_1-x_1^3)$, $\Psi_0^4=\delta(x_1-x_1^4)$, respectively. 
    Finally, panel 'e' illustrates the ${\cal P}(\omega)$, mentioned in \cref{Eq:integ-Density-timecorrelation-FT-4}, obtained from summing the quantities in panels `a', `b', `c', and `d'. Dashed gray lines and labels show predicted frequencies from exact diagonalization of the nuclear Hamiltonian.}
    \label{Fig:FT_x_upper}
\end{figure}
\begin{figure}[t]
 	\centering
    \includegraphics[width=0.75\textwidth]{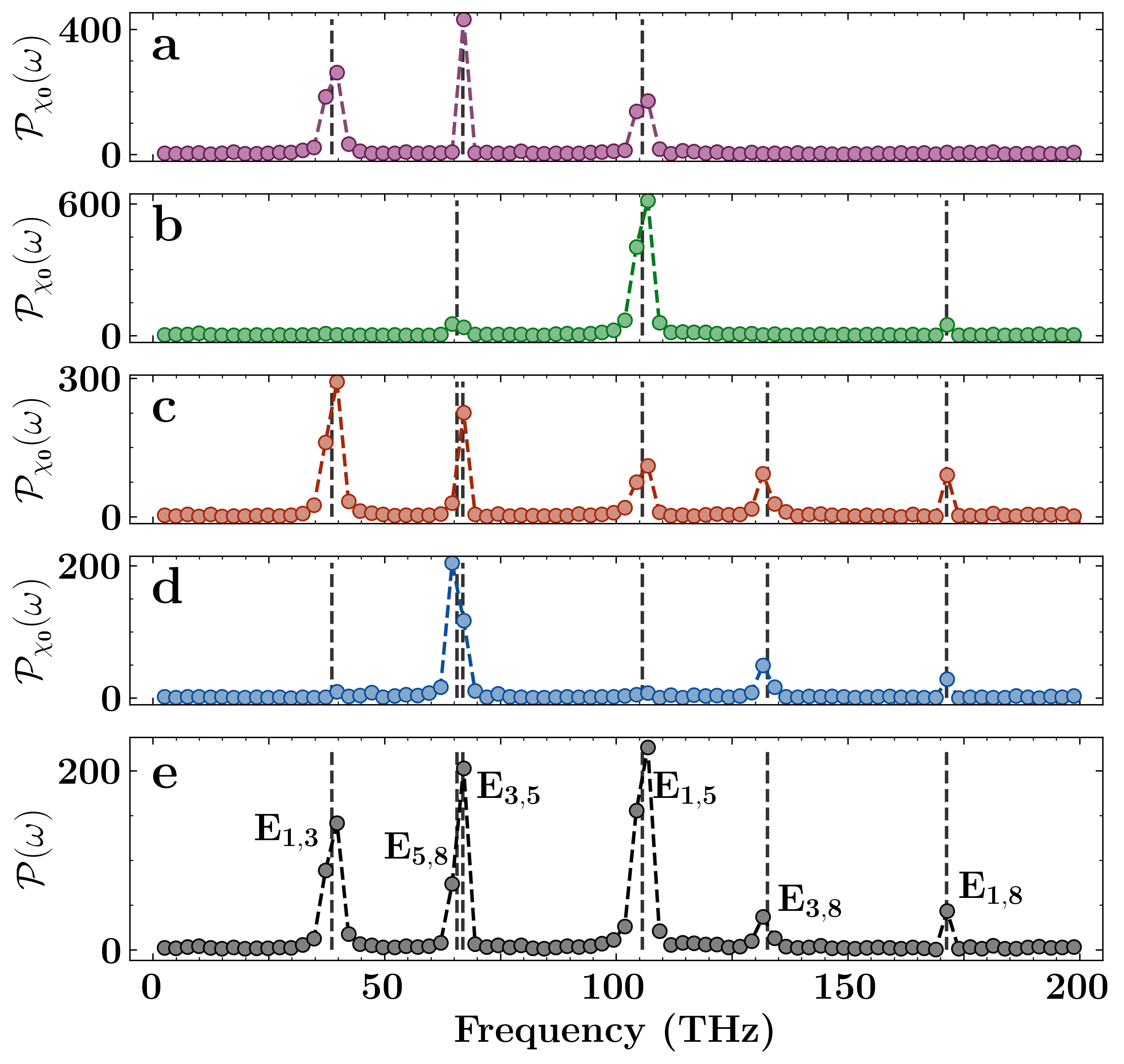}
    \caption{The figure illustrates the experimentally determined relative energy separation among the four eigenstates associated with the lower block of the Hamiltonian, corresponding to the vibrational degree of freedom involving proton transfer along the internuclear axis ($x_1$) connecting the two nitrogen atoms in protonated 2,2'-bipyridine. Panel `a' showcases ${\cal P}_{\chi_{0}} (\omega)$, as defined in \cref{Eq:integ-Density-timecorrelation-FT-3}, derived from the Fourier transforms ${\cal I} (\omega, x_1)$ (refer to \cref{Eq:Density-timecorrelation-FT-3}) of the individual panels in \cref{Fig:x-lower}(a), with the initial wavepacket $\Psi_0^1=\delta(x_1-x_1^5)$. Similarly, panels `b', `c', and `d' display the Fourier transforms obtained from the quantum propagation (see \cref{Fig:x-lower}(b-d)) of the lower block of the Hamiltonian along the $x_1$-direction, each initialized with wavepackets $\Psi_0^2=\delta(x_1-x_1^6)$, $\Psi_0^3=\delta(x_1-x_1^7)$, $\Psi_0^4=\delta(x_1-x_1^8)$, respectively. These panels present  ${\cal P}_{\chi_{0}} (\omega)$ obtained from ${\left\vert {\cal I} (\omega; x_1) \right\vert}^2$ summed over the four grid points $\left\{x_1^5,x_1^6,x_1^7,x_1^8\right\}$ as mentioned in \cref{Eq:integ-Density-timecorrelation-FT-3}. Finally, panel e' illustrates \({\cal P}(\omega)\), as defined in \cref{Eq:integ-Density-timecorrelation-FT-4}, obtained by summing the quantities in panels a', b', c', and `d'. Dashed gray lines and labels indicate the predicted frequencies from the exact diagonalization of the nuclear Hamiltonian.}
    \label{Fig:FT_x_lower}
\end{figure}
\begin{figure}[t]
 	\centering
    \includegraphics[width=0.75\textwidth]{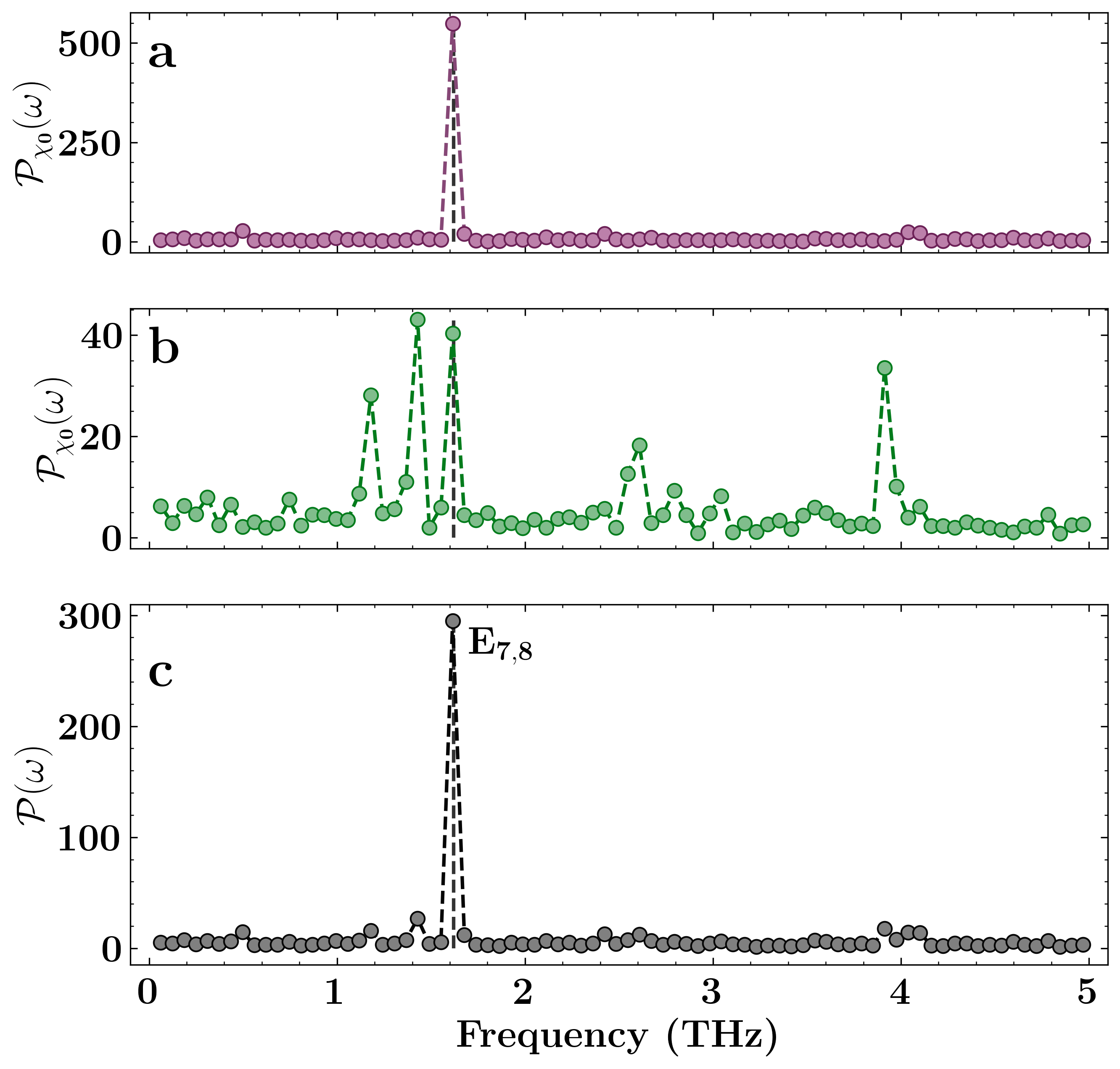}
    \caption{The figure illustrates the experimentally determined relative energy separation among the eight eigenstates of the Hamiltonian, corresponding to the vibrational degree of freedom involving proton transfer along the internuclear axis ($x_1$) connecting the two nitrogen atoms in protonated 2,2'-bipyridine. Each of the time-traces shown in \cref{Fig:x-full}(a) with the initial wavepacket $\Psi_0^1=\delta(x_1-x_1^1)$ projected onto eight different grid points $\left\{x_1^1, x_1^2, x_1^3, x_1^4, x_1^5, x_1^6, x_1^7, x_1^8\right\}$ 
    undergoes Fourier transformation to compute ${\cal I}(\omega, x_1)$. These transformations are then summed to obtain ${\cal P}_{\chi_{0}} (\omega)$ as described in \cref{Eq:integ-Density-timecorrelation-FT-3}, resulting in the energy separation presented in panel 'a' of this figure. Similarly, panel `b' displays the average Fourier transforms derived from the quantum propagation (see \cref{Fig:x-full}(b)) of the Hamiltonian along the $x_1$-direction initialized with wavepacket $\psi_0^2=\delta(x_1-x_1^2)$. Finally, panel `c' illustrates the sum of the Fourier transforms, ${\cal P}(\omega)$, obtained from panels `a' and `b'. Dashed gray lines and labels indicate predicted frequencies from exact diagonalization of the nuclear Hamiltonian.}
    \label{Fig:FT_x_full}
\end{figure}
\begin{figure}[t]
 	\centering
    \includegraphics[width=0.75\textwidth]{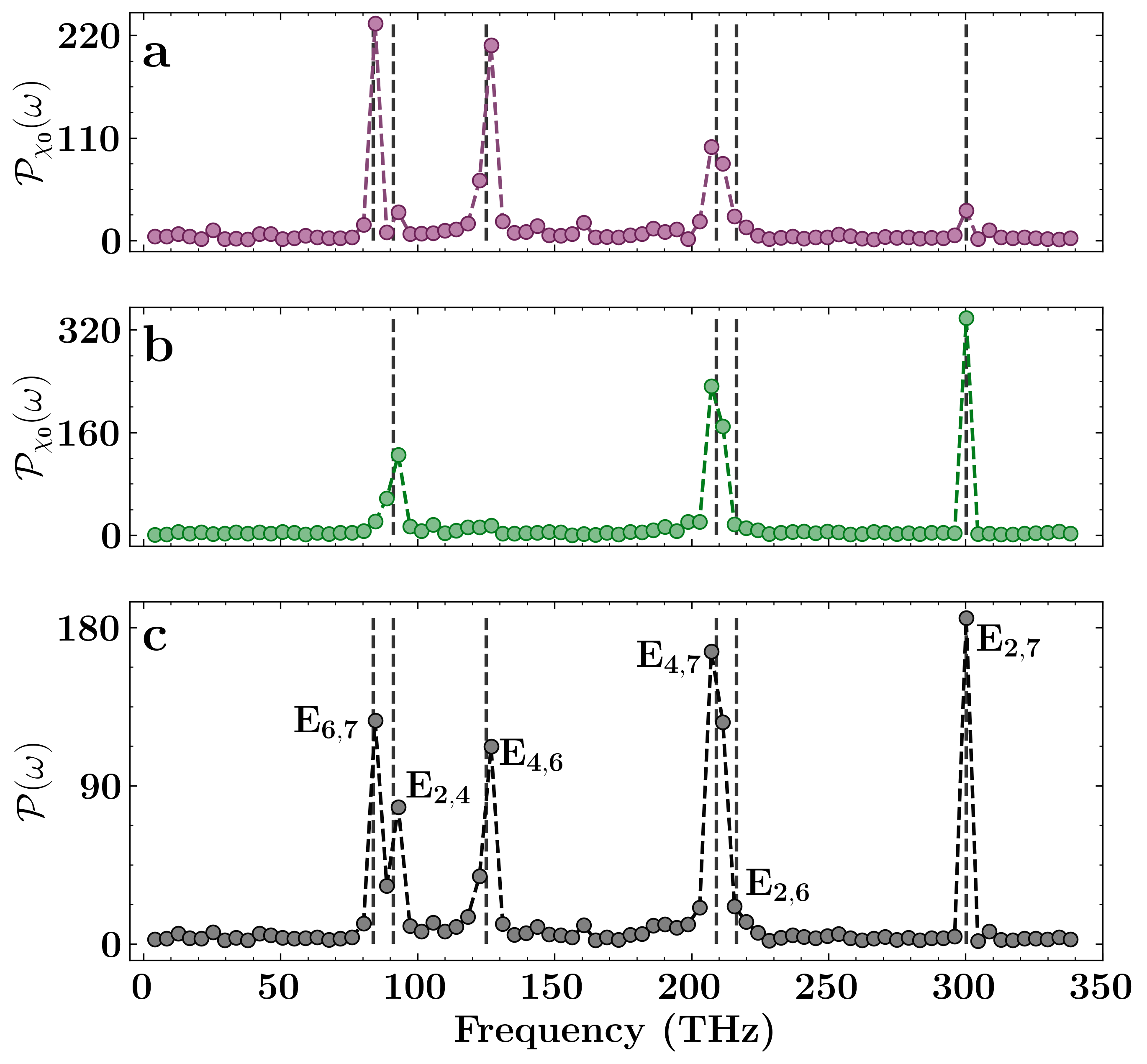}
    \caption{The figure illustrates the experimentally determined relative energy separation among the four eigenstates associated with the upper block of the Hamiltonian, corresponding to the torsional degree of freedom arising from planar rotations of the pyridyl rings about the C-C bond ($x_2$) in protonated 2,2'-bipyridine. These separations are obtained through the Fourier transform of the time evolution of the upper block of the Hamiltonian, considering two distinct initial states as detailed in \cref{Tab:Psi_0}. Specifically, panel `a' showcases the Fourier transform, ${\cal P}_{\chi_{0}} (\omega)$, resulting from the quantum propagation of the upper block, as depicted in \cref{Fig:theta-upper}(a), with the initial wavepacket $\Psi_0^1=\delta(x_2-x_2^1)$ projected onto four different grid points $\left\{x_2^1,x_2^2,x_2^3,x_2^4\right\}$. Panel `b' displays the Fourier transform derived from the quantum propagation (see \cref{Fig:theta-upper}(b)) of the upper block of the Hamiltonian along the $x_2$-direction, initialized with wavepackets $\Psi_0^2=\delta(x_2-x_2^2)$. Finally, panel `c' illustrates the sum of the Fourier transforms obtained from panels `a' and `b'. Dashed gray lines and labels indicate predicted frequencies from the exact diagonalization of the nuclear Hamiltonian.}
    \label{Fig:FT_theta_upper}
\end{figure}
\begin{figure}[t]
 	\centering
    \includegraphics[width=0.75\textwidth]{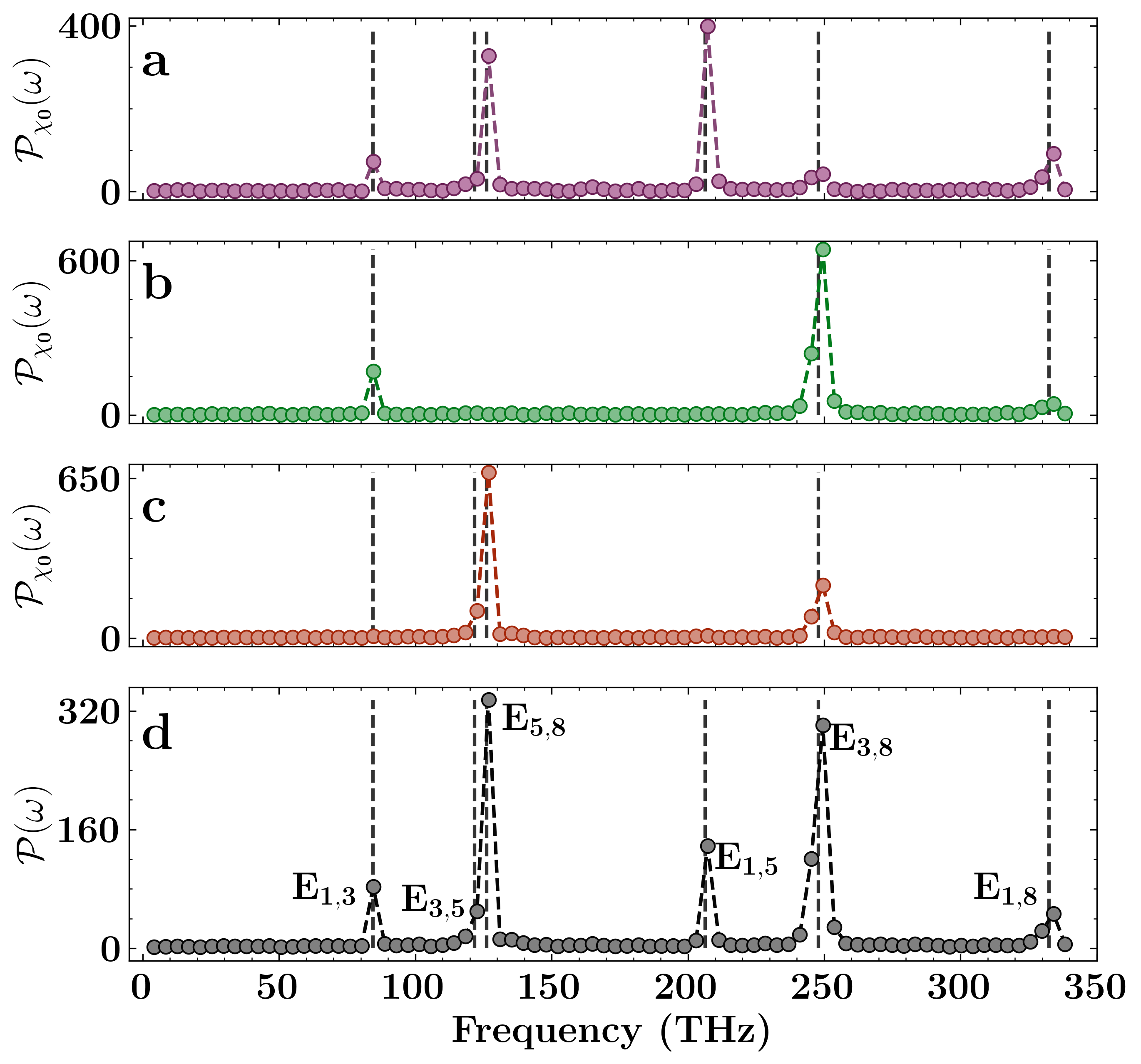}
    \caption{The figure illustrates the experimentally determined relative energy separation among the four eigenstates associated with the lower block of the Hamiltonian, corresponding to the torsional degree of freedom arising from planar rotations of the pyridyl rings about the C-C bond ($x_2$) in protonated 2,2'-bipyridine. Panel `a' showcases the average Fourier transform resulting from the quantum propagation of the lower block, as depicted in \cref{Fig:theta-lower}(a), with the initial wavepacket $\Psi_0^1=\delta(x_2-x_2^6)$ projected onto four different grid points $\left\{x_2^5,x_2^6,x_2^7,x_2^8\right\}$. Similarly, panels `b' and `c' display the Fourier transform derived from the quantum propagation (see \cref{Fig:theta-lower}(b-c)) of the lower block of the Hamiltonian along the $x_2$-direction, initialized with wavepackets $\Psi_0^2=\delta(x_2-x_2^7)$ and $\Psi_0^3=\delta(x_2-x_2^8)$ respectively. Finally, panel `d' illustrates the sum of the Fourier transforms obtained from panels `a', `b', and `c'. Dashed gray lines and labels indicate predicted frequencies from the exact diagonalization of the nuclear Hamiltonian.}
    \label{Fig:FT_theta_lower}
\end{figure}
\begin{figure}[t]
 	\centering
    \includegraphics[width=0.75\textwidth]{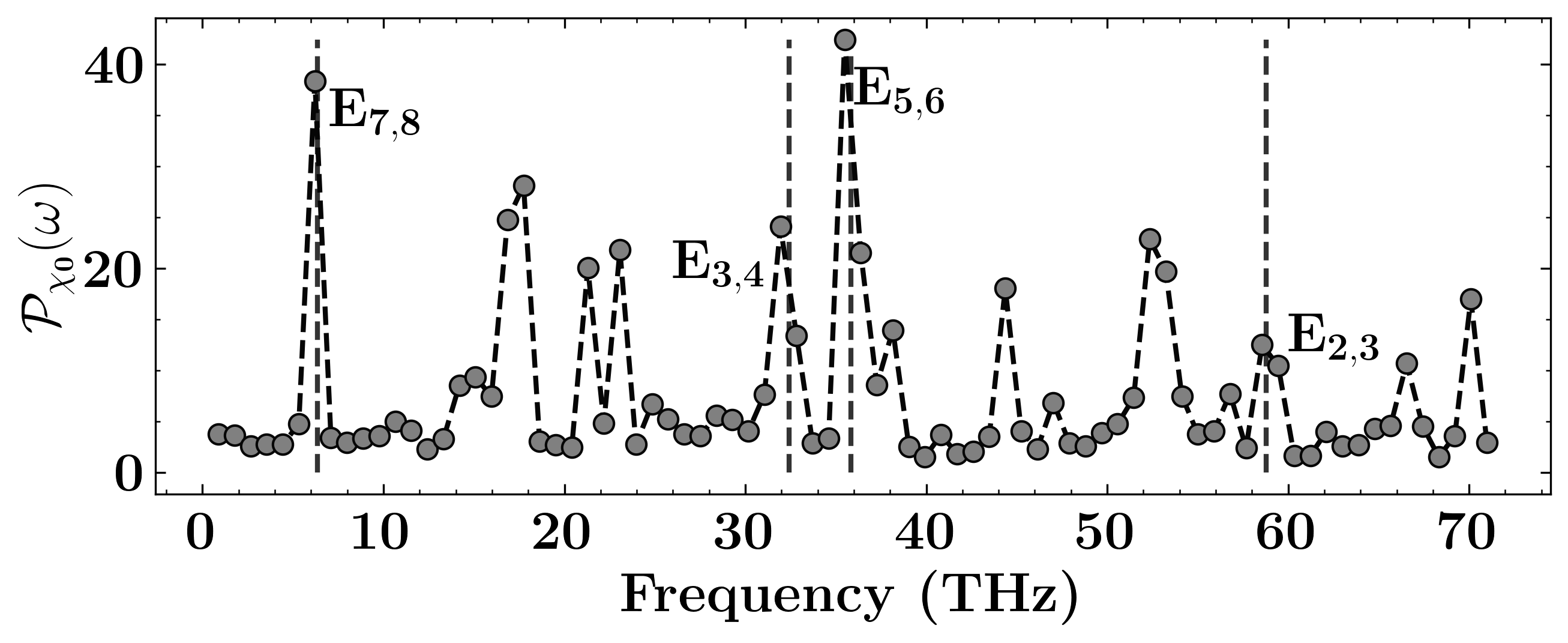}
    \caption{The figure illustrates the experimentally determined relative energy separation among the eight eigenstates of the Hamiltonian, corresponding to the torsional degree of freedom arising from planar rotations of the pyridyl rings about the C-C bond ($x_2$) in protonated 2,2'-bipyridine. Each of the time-traces shown in \cref{Fig:prob-theta-full-000-111} with the initial wavepacket $\Psi_0^1=\delta(x_2-x_2^1)$ projected onto eight different grid points $\left\{x_2^1, x_2^2, x_2^3, x_2^4, x_2^5, x_2^6, x_2^7, x_2^8\right\}$ is Fourier transformed and summed to produce the energy separation in the figure here. Dashed gray lines and labels indicate predicted frequencies from the exact diagonalization of the nuclear Hamiltonian.}
    \label{Fig:FT_theta_full}
\end{figure}

The Fourier transform of the quantum dynamics shown in \cref{Fig:x-upper,Fig:x-lower,Fig:x-full,Fig:theta-upper,Fig:theta-lower,Fig:prob-theta-full-000-111}, is computed using the formalism outlined in \cref{Sec:Vibrational}. Specifically, we calculated ${\cal I}(\omega, x_1)$ from \cref{Eq:Density-timecorrelation-FT-3} for each subfigure (that is each initial wavepacket) in the mentioned figures. Subsequently, the ${\cal P} (\omega)$ values for each of the initial wavepackets is cumulated and the results are presented in \cref{Fig:FT_x_upper,Fig:FT_x_lower,Fig:FT_x_full,Fig:FT_theta_upper,Fig:FT_theta_lower,Fig:FT_theta_full}. 

But before we delve into the details of the analysis, an important factor to consider while inspecting the Fourier transform figures is the range of vertical axes, ${\cal P}_{\chi_0} (\omega)$ and ${\cal P} (\omega)$. To understand this, we use the Cauchy-Schwarz inequality\cite{Riesznagy} in Eq. (\ref{Eq:Density-timecorrelation-FT-3}) to obtain $\|{\cal I}_{\chi_0} (\omega; x)\| < T_{\chi_0}$, where $T_{\chi_0}$ is the total length of simulation (see \cref{Tab:Simulation}). As a result, $\| {{\cal P}_{\chi_0} (\omega)} \| < {T_{\chi_0}}^2$ and these inequalities, as we will see, are consistent with the Fourier transform figures here.

This analysis aims to extract the relative energy differences between the corresponding eigenstates. Subsequently, these energy differences are utilized to construct the energy spectrum of the molecule as illustrated in \cref{Fig:Energy2D} and in Figure 8 of the main paper. The resulting energy spectrum is then compared with the energies obtained through the exact diagonalization of the nuclear Hamiltonian. The frequency spacing, $\Delta\omega$, and the maximum frequency for each computed Fourier transform are presented in \cref{Tab:Simulation}.

The Fourier transform of the quantum dynamics depicted in \cref{Fig:x-upper}, associated with the upper block of the block-diagonal Hamiltonian $\tilde{H}^{[1]}_{\gamma,\beta}$, is computed and presented in \cref{Fig:FT_x_upper}. This analysis aims to extract the relative energy differences between the respective four eigenstates associated with it. Specifically, the Fourier transform of each panel in \cref{Fig:x-upper}(a) is conducted to obtain ${\cal I}(\omega, x_1)$ and integrated over grid space to obtain ${\cal P}_{\chi_{0}} (\omega)$ 
and the resultant Fourier transform shown in panel `a' in \cref{Fig:FT_x_upper}. The resultant Fourier transform corresponds to the upper block of the Hamiltonian $\tilde{H}^{[1]}_{\gamma,\beta}$, with an initial wavepacket $\Psi_0(x_1) = \delta(x_1-x^1_1)$. Similarly, panels `b', `c', and `d' show the resultant Fourier transform obtained from 
each subfigure in \cref{Fig:x-upper}(b-d), respectively 
associated with the corresponding distinct initial wavepackets summarized in \cref{Tab:Psi_0} under the `$x_1$-upper block' row. Finally, panel 'e' illustrates the cumulative 
Fourier transforms, 
obtained from panels `a', `b', `c', and `d'. Dashed gray lines and labels indicate predicted frequencies from the exact diagonalization of the nuclear Hamiltonian and the dots represent the data obtained from the quantum computation. Visually the agreement looks near perfect and quantitative analysis of errors in eigenvalues obtained from this analysis is presented later. 

Similarly, the Fourier transform of the quantum dynamics portrayed in \cref{Fig:x-lower}, linked with the lower block of the block-diagonal Hamiltonian $\tilde{H}^{[1]}_{\gamma,\beta}$, is computed and illustrated in \cref{Fig:FT_x_lower}. This examination aims to discern the relative energy differences between the respective four eigenstates associated with it. Specifically, the Fourier transform of each panel in \cref{Fig:x-lower}(a) is performed to obtain ${\cal I}(\omega, x)$ and then integrated over grid space to obtain ${\cal P}_{\chi_{0}} (\omega)$. 
The resultant Fourier transform shown in panel `a' of \cref{Fig:FT_x_lower}. This resultant Fourier transform corresponds to the lower block of the Hamiltonian $\tilde{H}^{[1]}_{\gamma,\beta}$, with an initial wavepacket $\Psi_0(x_1) = \delta(x_1-x^5_1)$. Similarly, panels `b', `c', and `d' display the resultant Fourier transform obtained by integrating ${\cal I}(\omega, x_1)$ 
for each subfigure in \cref{Fig:x-lower}(b-d), respectively, associated with the corresponding distinct initial wavepackets summarized in \cref{Tab:Psi_0} under the `$x_1$-lower block' row. Finally, panel 'e' showcases the accumulation of Fourier transforms from all initial wavepackets
obtained from panels `a', `b', `c', and `d'. Once again, dashed gray lines and labels indicate predicted frequencies from the exact diagonalization of the nuclear Hamiltonian and the dots represent the data obtained from the quantum computation.

The Fourier transforms obtained from the quantum dynamics illustrated in \cref{Fig:x-full} of the Hamiltonian $\tilde{H}^{[1]}_{\gamma,\beta}$, which represents the vibrational degree of freedom of the proton along the internuclear axis between the two nitrogen nuclei in \cref{Fig:Molecule}, are depicted in \cref{Fig:FT_x_full}. Each panel in \cref{Fig:x-full}(a) undergoes Fourier transformation to compute ${\cal I}(\omega, x_1)$, as mentioned in \cref{Eq:integ-Density-timecorrelation-FT-3}. This is further integrated to derive the resultant Fourier transform ${\cal P}_{\chi_{0}} (\omega)$. 
Panel `a' of \cref{Fig:FT_x_lower} shows the resultant Fourier transform corresponding to the quantum dynamics of the Hamiltonian $\tilde{H}^{[1]}_{\gamma,\beta}$ with an initial wavepacket $\Psi_0(x_1) = \delta(x_1-x^1_1)$. Similarly, panel `b' corresponds to the Fourier transform associated with the initial wavepacket $\Psi_0(x_1) = \delta(x_1-x^2_1)$. Panel `c' illustrates the accumulation of Fourier transforms for all initial wavepackets 
obtained from 
the Fourier transforms in panels a' and `b'. The purpose of the Fourier transform of the full Hamiltonian $\tilde{H}^{[1]}_{\gamma,\beta}$ is to compute the relative energy separations between the eigenstates associated with the upper and lower blocks of the Hamiltonian.

The Fourier transform of the quantum dynamics depicted in \cref{Fig:theta-upper}, corresponding to the upper block of the block-diagonal Hamiltonian $\tilde{H}^{[2]}_{\gamma,\beta}$, is calculated and presented in \cref{Fig:FT_theta_upper}. Each panel in \cref{Fig:theta-upper}(a) undergoes Fourier transformation to compute ${\cal I}(\omega, x_2)$ as described in \cref{Eq:integ-Density-timecorrelation-FT-3}. This is then integrated over the grid space to produce the resultant Fourier transform ${\cal P}_{\chi_{0}} (\omega)$, shown in panel `a' of \cref{Fig:FT_theta_upper}. This transform corresponds to the upper block of the Hamiltonian $\tilde{H}^{[2]}_{\gamma,\beta}$ with the initial wavepacket $\Psi_0(x_2) = \delta(x_2-x^1_2)$. Similarly, panel b' displays the Fourier transform ${\cal P}_{\chi_{0}} (\omega)$, obtained from the integration of Fourier transforms ${\cal I}(\omega, x_2)$ for each panel in \cref{Fig:theta-upper}(b), associated with the initial wavepacket $\Psi_0(x_2) = \delta(x_2-x^2_2)$. Finally, panel c' illustrates the accumulated Fourier transform, ${\cal P}(\omega)$, 
derived from the transforms shown in panels a' and b'.

Similarly, the Fourier transform of the quantum dynamics depicted in \cref{Fig:theta-lower}, associated with the lower block of the block-diagonal Hamiltonian $\tilde{H}^{[2]}_{\gamma,\beta}$, is computed and presented in \cref{Fig:FT_theta_lower}. Specifically, the Fourier transform of each panel in \cref{Fig:theta-lower}(a) is performed to calculate ${\cal I}(\omega, x_2)$ in \cref{Eq:integ-Density-timecorrelation-FT-3} and then integrated to obtain the resultant Fourier transform, ${\cal P}_{\chi_{0}} (\omega)$,  shown in panel `a' of \cref{Fig:FT_theta_lower}. This resultant Fourier transform corresponds to the lower block of the Hamiltonian $\tilde{H}^{[2]}_{\gamma,\beta}$, with an initial wavepacket $\Psi_0(x_2) = \delta(x_2-x^6_1)$. Similarly, panels `b' and `c' display the resultant Fourier transform obtained by integrating the quantity ${\cal I}(\omega, x_2)$ over grid space, computed for each panel in \cref{Fig:theta-lower}(b-c), respectively, associated with the corresponding distinct initial wavepackets summarized in \cref{Tab:Psi_0} under the `$x_2$-lower block' row. Finally, panel `d' showcases the sum of the Fourier transforms obtained from panels `a', `b', and `c'. The peaks are compared with the predicted frequencies from the exact diagonalization of the nuclear Hamiltonian and shown as gray lines and labels in the figure.

The Fourier transforms derived from the quantum dynamics depicted in \cref{Fig:prob-theta-full-000-111} of the Hamiltonian $\tilde{H}^{[2]}_{\gamma,\beta}$, which represents the torsional degree of freedom of the proton arising from the planar rotation of the planes containing the pyridyl rings in \cref{Fig:Molecule}, are showcased in \cref{Fig:FT_theta_full}. Each panel in \cref{Fig:prob-theta-full-000-111} undergoes Fourier transformation to calculate ${\cal I}(\omega, x_2)$ and subsequent integration over grid space to yield the resultant Fourier transform, ${\cal P}_{\chi_{0}} (\omega)$ in \cref{Eq:integ-Density-timecorrelation-FT-3},  as shown in \cref{Fig:FT_theta_full}. This resultant Fourier transform corresponds to the quantum dynamics of the Hamiltonian $\tilde{H}^{[2]}_{\gamma,\beta}$, with an initial wavepacket $\Psi_0(x_2) = \delta(x_2-x^1_2)$. The Fourier transform of the full Hamiltonian $\tilde{H}^{[2]}_{\gamma,\beta}$ aims to compute the relative energy separations between the eigenstates associated with the upper and lower blocks of the Hamiltonian.

\begin{figure}
  \includegraphics[width=\textwidth]{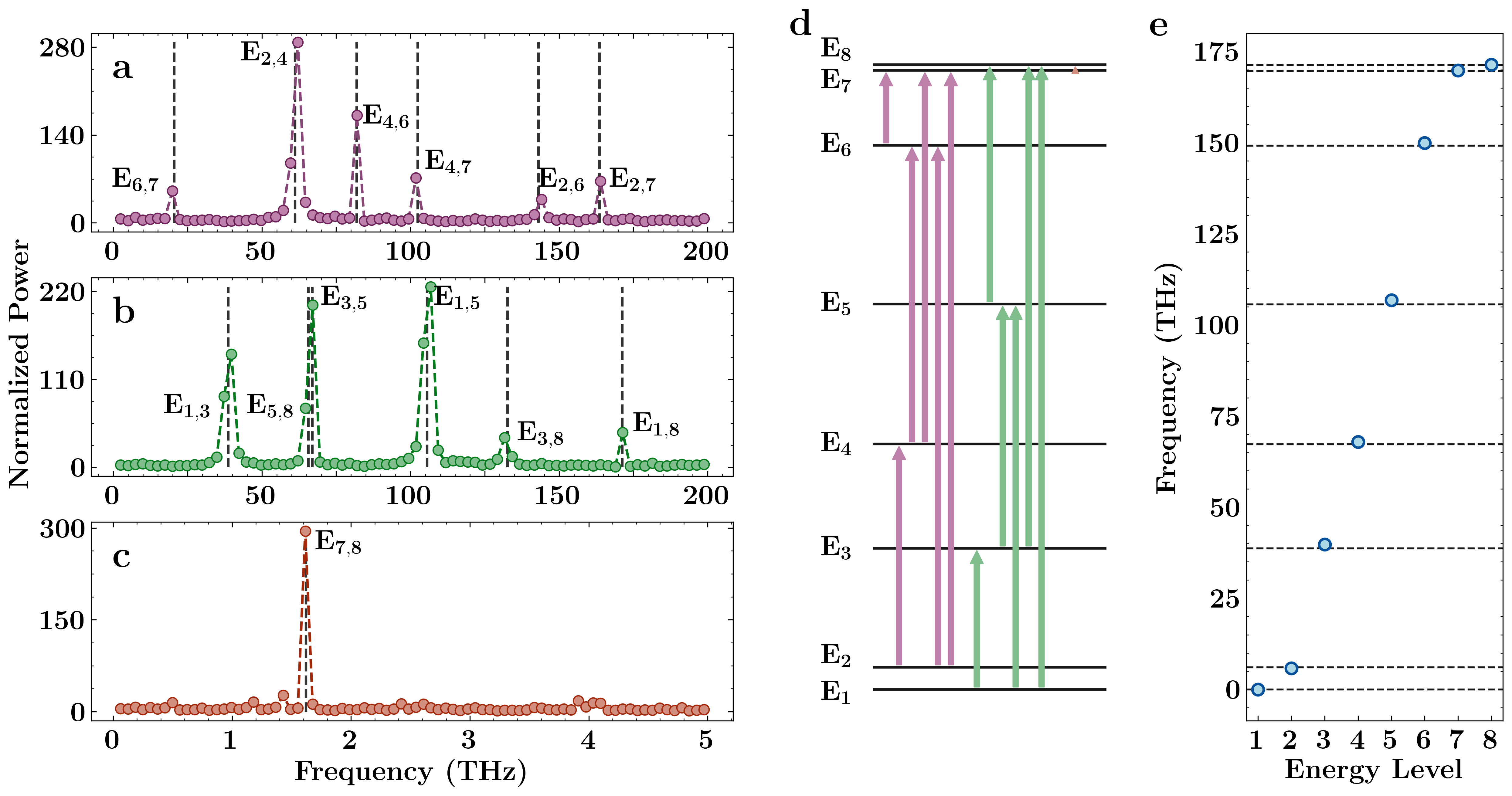}
  \caption{The figure presents the experimentally determined frequency and energy spectra of protonated 2,2'-bipyridine corresponding to the vibrational degree of freedom of transferring proton along the internuclear axis ($x_1$) joining the two nitrogen atoms. Due to the block diagonal nature of the Hamiltonian (see \cref{Sec:Block-diagonal}), each block of the Hamiltonian is simulated separately on two separate two-qubit ion-trap systems. The Fourier transform of the time evolution of the full-Hamiltonian allows for the determination of relative energy separations between the eigenstates corresponding to the respective blocks. The panel `a' displays the average Fourier transform obtained from the quantum propagation of the upper block with different initial wavepacket (see detailed plot in \cref{Fig:x-upper}) , while panel `b' illustrates the average Fourier transform derived from the quantum propagation of the lower block of the Hamiltonian along the $x_1$-direction initialized with different wavepackets (see \cref{Fig:x-lower}). The panel `c' shows the average Fourier transform derived from the quantum dynamics of the full-Hamiltonian with two distinct initial wavepackets, also shown in \cref{Fig:FT_x_full}, providing the relative separation between the two sets of eigenstates corresponding to the lower and upper blocks. The \cref{Fig:ft_x}d presents the extracted frequencies from panels `a' to `c', color-coded according to their parent spectrum. This presentation allows for the experimental determination of the relative energies of all eigenstates. Additionally, \cref{Fig:ft_x}e compares the quantum-computed energy eigenstates of the nuclear Hamiltonian (depicted as blue dots) with the results of exact diagonalization (shown as dashed gray lines).}
  \label{Fig:ft_x}
\end{figure}
Finally, the Fourier transforms 
obtained from the \cref{Fig:FT_x_upper,Fig:FT_x_lower,Fig:FT_x_full} are utilized to get the energy spectrum of the molecule, as depicted in \cref{Fig:ft_x}, corresponding to the vibrational degree of freedom associated with the transferring hydrogen along the internuclear axis connecting two nitrogen atoms. The panel `a' in \cref{Fig:ft_x}, which is imported from \cref{Fig:FT_x_upper}, displays the cumulative Fourier transform 
obtained from the quantum propagation of the upper block with different initial wavepacket, while panel `b' (see \cref{Fig:x-lower} for detailed plot) illustrates the cumulative Fourier transform derived from the quantum propagation of the lower block of the Hamiltonian along the $x_1$-direction initialized with different wavepackets. The panel `c', also shown in \cref{Fig:FT_x_full}, shows the cumulative Fourier transform derived from the quantum dynamics of the full-Hamiltonian with two distinct initial wavepackets providing the relative separation between the two sets of eigenstates corresponding to the lower and upper blocks. In \cref{Fig:ft_x}d, the extracted frequencies from panels `a' to `c' are presented, color-coded according to their parent spectrum, facilitating the experimental determination of the relative energies of all eigenstates. Additionally, \cref{Fig:ft_x}e compares the quantum hardware-computed energy eigenstates of the nuclear Hamiltonian (depicted as blue dots) with the results of exact diagonalization (shown as dashed gray lines).

\begin{figure}
  \includegraphics[width=\textwidth]{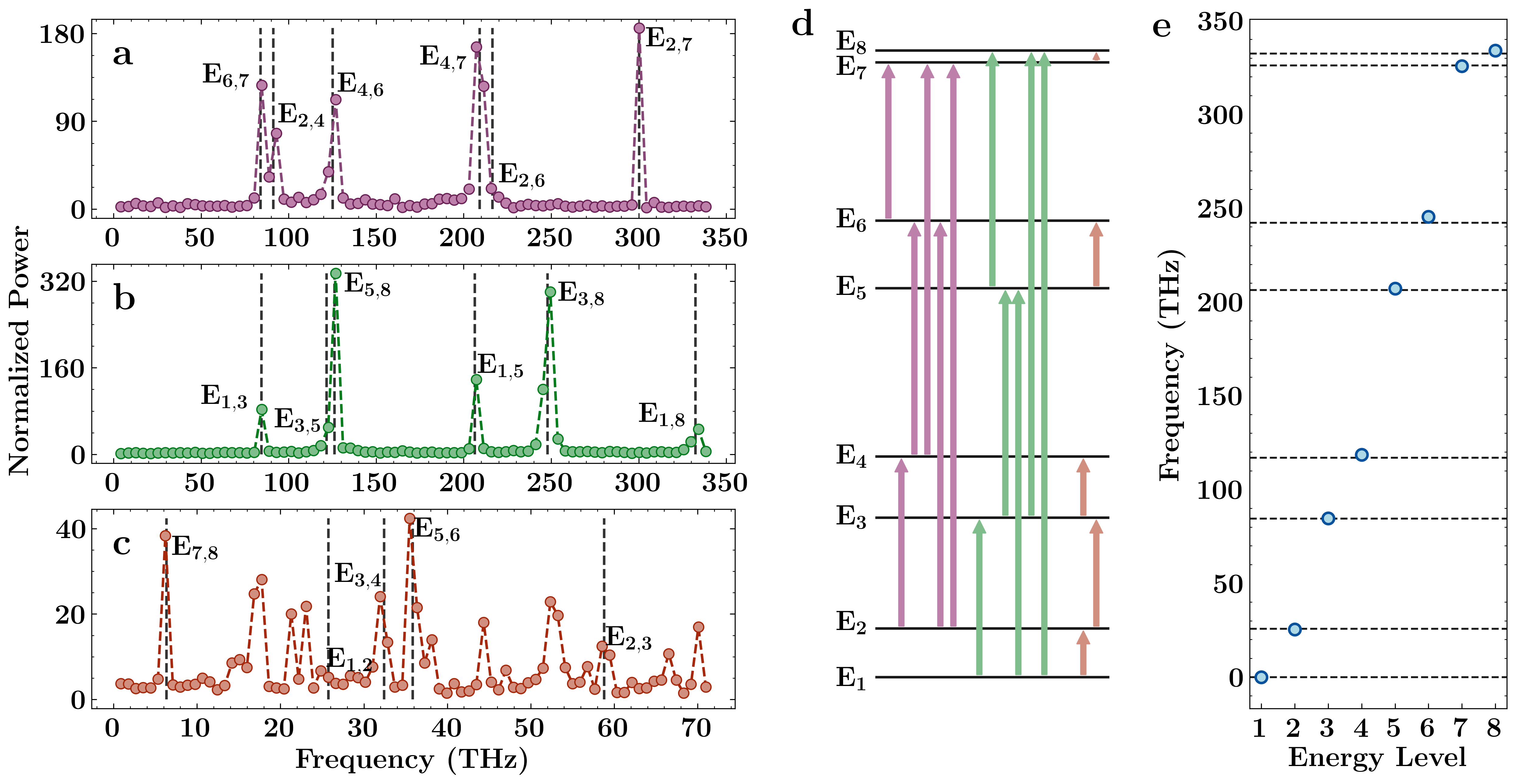}
 \caption{The figure presents the experimentally derived frequency and energy spectra of protonated 2,2'-bipyridine, focusing on the torsional degree of freedom (represented as $x_2$) associated with planar rotations of the pyridyl rings about the C-C bond. Leveraging the block diagonal structure of the Hamiltonian, each block is individually simulated using separate two-qubit ion-trap systems. The panel `a' illustrates the average Fourier transform resulting from quantum propagation of the upper block with 4 distict initial wavepackets (see \cref{Fig:FT_theta_upper}), while panel `b' (see \cref{Fig:FT_theta_lower} for more details) depicts the average Fourier transform obtained from the quantum propagation of the lower block along the $x_2$-direction starting with different initial wavepackets. Panel `c' exhibits the Fourier transform derived from the quantum dynamics of the full-Hamiltonian, elucidating the relative separation between the eigenstates of the lower and upper blocks. Furthermore, in \cref{Fig:ft_theta}d, frequencies extracted from panels `a' to `c' are color-coded by their respective spectra, facilitating the experimental determination of relative energies for all eigenstates. Additionally, \cref{Fig:ft_theta}e presents a comparison between the quantum-computed energy eigenstates of the nuclear Hamiltonian (represented by blue dots) and exact diagonalization results (depicted by dashed gray lines).}
   \label{Fig:ft_theta}
\end{figure}
Similarly, the Fourier transforms 
obtained from \cref{Fig:FT_theta_upper,Fig:FT_theta_lower,Fig:FT_theta_full} are utilized to derive the energy spectrum of the molecule, as depicted in \cref{Fig:ft_theta}. This figure encapsulates the experimentally derived frequency and energy spectra of protonated 2,2'-bipyridine, with a focus on the torsional degree of freedom ($x_2$) associated with planar rotations of the pyridyl rings about the C-C bond. Employing the block diagonal structure of the Hamiltonian, each block undergoes individual simulation using separate two-qubit ion-trap systems. Panel `a', also featured in \cref{Fig:FT_theta_upper}, illustrates the cumulative Fourier transform 
resulting from the quantum propagation of the upper block with 4 distinct initial wavepackets. Meanwhile, panel `b' (referenced in \cref{Fig:FT_theta_lower}) portrays the cumulative Fourier transform obtained from the quantum propagation of the lower block along the $x_2$-direction, initiated with different wavepackets. Panel `c', imported from \cref{Fig:FT_theta_full}, showcases the Fourier transform derived from the quantum dynamics of the full Hamiltonian, providing insights into the relative separation between the eigenstates of the lower and upper blocks. Further insight is provided in \cref{Fig:ft_theta}d, where frequencies extracted from panels `a' to `c' are color-coded according to their respective spectra, enabling the experimental determination of relative energies for all eigenstates. Additionally, \cref{Fig:ft_theta}e offers a comparison between the quantum hardware-computed energy eigenstates of the nuclear Hamiltonian (depicted by blue dots) and exact diagonalization results (represented by dashed gray lines).

\begin{figure}[t]
 	\centering
    \includegraphics[width=0.7\columnwidth]{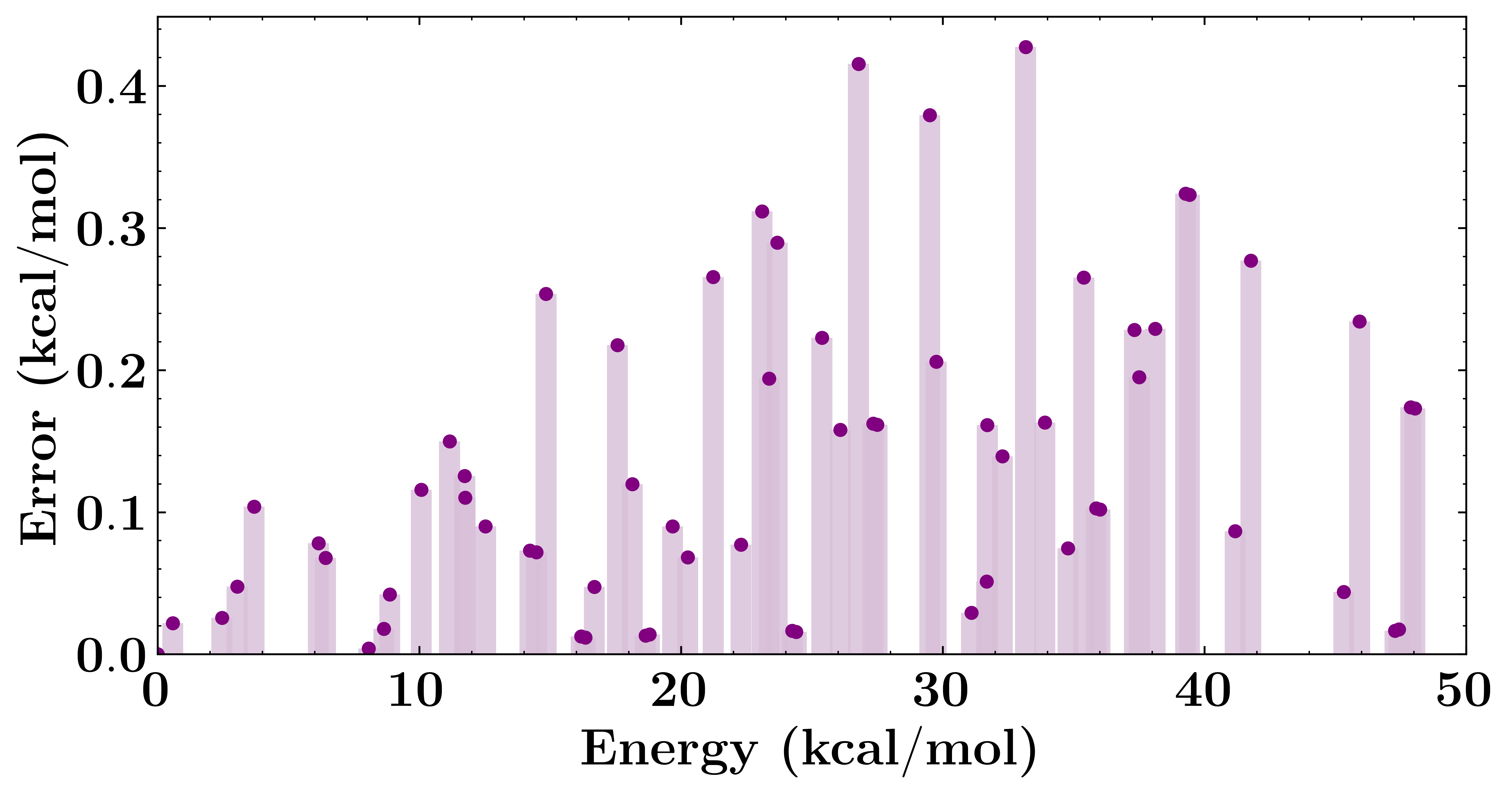}
  	\caption{The energy spectra from Figures \ref{Fig:ft_x}e and \ref{Fig:ft_theta}e are combined to obtain the vibrational eigenenergies corresponding to the two coupled modes for protonated 2,2'-bipyridine. 
   These states are compared with exact-diagonalization of the nuclear Hamiltonian with mean absolute error between the computational results is found to be 0.14 kcal/mol.
   }
  	\label{Fig:Energy2D}
\end{figure}
Next, the energy spectra in \cref{Fig:ft_x,Fig:ft_theta} are combined to obtain the vibrational energy levels corresponding to the two coupled degrees of freedom. 
The accuracy of these energies is compared with that of exact diagonalization results. As shown in \cref{Fig:Energy2D}, the mean absolute error is well within chemical accuracy. 
